\documentclass[fleqn,usenatbib]{mnras}

\usepackage[T1]{fontenc}

\DeclareRobustCommand{\VAN}[3]{#2}
\let\VANthebibliography\thebibliography
\def\thebibliography{\DeclareRobustCommand{\VAN}[3]{##3}\VANthebibliography}

\usepackage{graphicx}	
\usepackage{amsmath}	
\usepackage{amssymb}	

\usepackage{ae,aecompl}
\usepackage{bmpsize}
\usepackage{subfig}
\usepackage{makeidx}
\usepackage{physics}
\usepackage{booktabs,tabularx}
\usepackage{multirow}
\usepackage{siunitx}

\title[Particle Acceleration in MRI Turbulence]{Particle Diffusion and Acceleration in Magnetorotational Instability Turbulence}

\author[X. Sun \& X.-N. Bai]{
Xiaochen Sun,$^{1}$
Xue-Ning Bai$^{1,2}$\thanks{E-mail: xbai@tsinghua.edu.cn}
\\
$^{1}$Institute for Advanced Study, Tsinghua University, Beijing 100084, China\\
$^{2}$Department of Astronomy, Tsinghua University, Beijing 100084, China
}

\date{Accepted XXX. Received YYY; in original form ZZZ}

\pubyear{2021}

\begin{document}
\label{firstpage}
\pagerange{\pageref{firstpage}--\pageref{lastpage}}
\maketitle

\begin{abstract}
Hot accretion flows contain collisionless plasmas that are believed to be capable of accelerating particles to very high energies, as a result of turbulence generated by the magnetorotational instability (MRI). We conduct unstratified shearing-box simulations of the MRI turbulence in ideal magnetohydrodynamic (MHD), and inject energetic (relativistic) test particles in simulation snapshots to conduct a detailed investigation on particle diffusion and stochastic acceleration. We consider different amount of net vertical magnetic flux to achieve different disk magnetizations levels at saturated states, with sufficiently high resolution to resolve the gyro-radii ($R_g$) of most particles. Particles with large $R_g$ ($\gtrsim0.03$ disk scale height $H$) show spatial diffusion coefficients of $\sim30$ and $\sim5$ times Bohm values in the azimuthal and poloidal directions, respectively. We further measure particle momentum diffusion coefficient $D(p)$ by applying the Fokker-Planck equation to particle momentum evolution. For these particles, contribution from turbulent fluctuations scales as $D(p)\propto p$, and shear acceleration takes over when $R_g\gtrsim0.1H$, characterized by $D(p)\propto p^3$. For particles with smaller $R_g$ ($\lesssim0.03H$), their spatial diffusion coefficients roughly scale as $\sim p^{-1}$, and show evidence of $D(p)\propto p^2$ scaling in momentum diffusion but with large uncertainties.  We find that multiple effects contribute to stochastic acceleration/deceleration, and the process is also likely affected by intermittency in the MRI turbulence. We also discuss the potential of accelerating PeV cosmic-rays in hot accretion flows around supermassive black holes.
\end{abstract}




\section{Introduction}
\label{sec::intro}

Hot accretion flows around compact objects are formed when accretion rates are well below the Eddington accretion rate (see review by \citealp{2014ARA&A..52..529Y}). Such disks are optically thin and hence radiatively inefficient, and become geometrically thick as they are heated up by dissipation of accretion energy. Viscous dissipation in hot accretion flows also leads to rapid advection, and such flows are termed advection-dominated accretion flows (ADAF, \citealp{1977ApJ...214..840I,1994ApJ...428L..13N,1995ApJ...438L..37A}). They are the most common type of accretion disks, encompassing individual systems such as Sgr A$^*$ \citep{1995Natur.374..623N}, disk around the supermassive black hole (SMBH) in M87, and more generally in the hard states of X-ray binaries (XRBs, \citealp{1997ApJ...489..865E,2005ApJ...620..905Y} ) and low-luminosity active galactic nuclei (LLAGNs, e.g.,\citealp{2008ARA&A..46..475H,2014MNRAS.438.2804N}).

Due to low density and high temperature, plasmas in ADAFs near the central object are collisionless, where collision timescale among particles by Coulomb interactions well exceeds dynamical timescale \citep{1997ApJ...490..605M, 1999ApJ...520..298Q}. It naturally arises that particles may develop a non-thermal population and get accelerated to high energies. Observationally, presence of a non-thermal electron population is likely needed to fit the spectral energy distribution (SED) of Sgr A$^*$ across wide range of wavelengths (e.g., \citealp{1998Natur.394..651M, 2000ApJ...541..234O, 2003ApJ...598..301Y} ). Similar situations apply to the hard state of XRBs (e.g., \citealp{2014SSRv..183...61P}) and LLAGNs (e.g., \citealp{2013ApJ...764...17L}). It has also been proposed that protons can also be accelerated to very high energies ($\sim$PeV) that can produce energetic $\gamma$-rays and neutrinos and offer multi-messenger constraints to the physics of ADAFs (e.g., \citealp{2015ApJ...806..159K}). The generation of such PeV protons could potentially account for cosmic-ray energy spectrum up to the ``knee", especially during outburst of Sgr A$^*$ in the past \citep{2016Natur.531..476H,2017JCAP...04..037F}.

Particle acceleration in collisionless plasmas involves tapping the free energy from the bulk (thermal) plasmas into a small population of particles that becomes non-thermal, where the free energy can be kinetic in the case of first-order Fermi acceleration in shocks (e.g., \citealp{1978ApJ...221L..29B, 1978MNRAS.182..147B, 2008ApJ...682L...5S, 2013ApJ...765L..20C, 2015ApJ...809...55B}) and second-order Fermi acceleration in turbulence (e.g., \citealp{1949PhRv...75.1169F, 2012SSRv..173..535P, 2017PhRvL.118e5103Z, 2018PhRvL.121y5101C, 2019ApJ...879...53A}), or magnetic in the case of magnetic reconnection (e.g., \citep{2001ApJ...562L..63Z, 2014ApJ...783L..21S, 2014PhRvL.113o5005G}). In all these cases, particles gain energy in stochastic processes, and generally involve certain level of turbulence that are either self-generated or external.

In ADAFs, and in accretion disks in general, disk accretion is primarily driven by magneto-rotational instability (MRI, \citealp{1995ApJ...440..742H, 1998RvMP...70....1B}). The nonlinear saturation of the MRI generates and sustain vigorous turbulence that transports angular momentum radially outwards. There have been extensive numerical studies of the MRI over the past few decades, either local or global, mostly in the fluid framework with ideal magnetohydrodynamics (MHD), as we briefly discuss below.

Local simulations employ the shearing-box framework \citep{1965MNRAS.130...97G, 1995ApJ...440..742H}, which has the advantage of concentrating the resolution in a local disk patch without complications from global boundary conditions. Also, by ignoring radial gradients, turbulence is quasi-homogeneous in the simulation box which allows one to focus on the local microphysics. The main controlling parameter on the properties of the MRI is the strength of the net vertical magnetic field $B_{z0}$, characterized by its corresponding plasma beta (ratio of gas to magnetic pressure) at the midplane $\beta_0$. Being a weak field instability, the MRI requires $\beta_0\gtrsim10$ for most unstable wavelength to fit into the disk (e.g., \citealp{2013A&A...550A..61L}). Within this limit, turbulence becomes increasingly strong for stronger net vertical magnetic field \citep{2004ApJ...605..321S, 2007ApJ...668L..51P, 2013ApJ...767...30B}. We also note that while the shearing-sheet approximation applies mainly for thin disks with scale height $H\ll R$, the physics of the MRI remain similar even in thick disks as in ADAFs.

Global simulations of accretion disks can self-consistently capture the global disk structure and evolution, and for black-hole accretion disks, the use of general-relativistic MHD (GRMHD) to study ADAFs has become routine (e.g., \citealp{2003ApJ...599.1238D, 2003ApJ...589..444G}). Similar to the local simulations, the amount of vertical magnetic flux has been shown to be a defining characteristic of ADAF disks \citep{2012MNRAS.426.3241N}. In particular, disks with very little net vertical flux follows ``standard and normal evolution" (SANE), whereas disks with strong vertical magnetic flux evolves into ``magnetically arrested disks" (MAD, \citealp{2003ApJ...592.1042I, 2011MNRAS.418L..79T, 2012MNRAS.423.3083M, 2013MNRAS.436.3856S}). MAD disks are entirely magnetically-dominated in the saturated state, leading to powerful jet via the Blandford-Znajek mechanism \citep{1977MNRAS.179..433B}. A tight correlation between jet power and accretion luminosity implies that MAD disks are likely common around SMBHs \citep{2014Natur.510..126Z}.

In the recent years, there has been substantial interest in exploring the development and saturation of the MRI in the kinetic regime in both 2D \citep{2012ApJ...755...50R,2013ApJ...773..118H,2018ApJ...859..149I} and 3D \citep{2015PhRvL.114f1101H, 2016PhRvL.117w5101K}. These simulations generally show self-sustained MRI turbulence cascading to kinetic scales, the development of self-regulated pressure anisotropy, and magnetic reconnection that leads to the emergence of a non-thermal particle population. This particle injection process takes place at kinetic scales, followed by the development of a power-law tail in particle energy distribution. Nevertheless, kinetic simulations are highly computationally expensive, and generally only follow the initial stage of particle acceleration.

Upon injected, subsequent particle acceleration proceeds in the self-sustained MRI turbulence, which can be considered as second-order Fermi process (e.g., \citealp{1987PhR...154....1B}). Such processes have been studied by injecting test particles to driven or decaying MHD turbulence \citep{2004ApJ...617..667D, 2009ApJ...707..404L, 2012PhRvL.108x1102K,2012ApJ...758...78L, 2013ApJ...779..140X, 2014ApJ...791...71L}. More recent kinetic simulations of driven turbulence, mostly in the relativistic regime with a strong background field (with plasma $\beta$ for background field $\lesssim 10$, much stronger than those typical in the MRI simulations), demonstrate the development of the emergence of power-law non-thermal particle population \citep{2017PhRvL.118e5103Z, 2019PhRvL.122e5101Z}, with a two stage process of injection followed by stochastic/diffusive acceleration \citep{2018PhRvL.121y5101C, 2019ApJ...886..122C}, and the overall process can be well described by the Fokker-Planck equation (e.g., \citealp{2020ApJ...893L...7W}). On the other hand, the MRI turbulence is self-sustained without artificial driving, with turbulent field dominate over mean background field. As a different realization of MHD turbulence with specific application in accretion disks, measuring the corresponding particle diffusion coefficients will provide first-principle calibration for theoretical models of stochastic particle acceleration. 

When a particle gains more energy, its larger gyro-radius can further tap more free energy from orbital shear, resulting in shear acceleration \citep{1981SvAL....7..352B, 1988ApJ...331L..91E, 2006ApJ...652.1044R, 2013ApJ...767L..16O, 2019PhRvD..99h3006L, 2019Galax...7...78R}. Shear acceleration has been studied primarily in the context of jets, where large velocity gradient is present away from the jet axis (e.g., \citealp{2016ApJ...833...34R,2018ApJ...855...31W, 2017ApJ...842...39L}). It is related to second-order Fermi process in the sense that it requires particles to undergo spatial diffusion to experience background velocity shear, which enhances the rate of stochastic acceleration. In general, first-principle characterization of turbulence is essential to reveal the transition between the two regimes, and reveal their properties.

In this work, we aim to focus on stochastic particle acceleration in the MRI turbulence, encompassing both the second-order Fermi and shear acceleration regimes. In doing so, we inject test particles to snapshots of local high-resolution MHD simulations of the MRI turbulence, where the test particles can be considered as energetic particles injected from kinetic processes. By controlled experiments, we can measure the coefficients to the Fokker-Planck equation for different particle momenta, and analyze contributions from second-order Fermi and shear acceleration processes. Recently, a similar study has been conducted by \citet{2016ApJ...822...88K}. Our MRI simulations are run with several times higher resolution, so that we properly resolve the gyro-radii for most particle momenta. Moreover, we consider different levels of net vertical magnetic flux, covering the SANE and approaching the MAD regimes. With the specific realization of self-generated MRI turbulence, our work will corroborate kinetic simulations of stochastic particle acceleration in relativistic and/or low-$\beta$ regime to non-relativistic and more weakly magnetized regimes, and in the meantime, give detailed measurements of the interplay of particle acceleration in Fermi and shear processes.

This paper is organized as follows. In Section~\ref{sec:setup}, We describe numerical setup for both the MRI simulations and test particle simulations. Diagnostics of the MRI turbulence are analyzed in Section~\ref{sec:mri}. We study the particle diffusion in configuration space in Section~\ref{sec::diff_conf}. In Section~\ref{sec::evolute_mom}, we systematically analyze particle evolution in momentum-space by measuring coefficients in the Fokker-Planck equation, separating the contribution from shear. 
We discuss the physics of various stochastic acceleration mechanisms encountered in our simulations in Section~\ref{sec::stochastic_acc}. Our results are applied to astrophysical context, especially on particle acceleration in ADAF disks around SMBHs in Section~\ref{sec::discussion}, where we also compare results with previous works. We summarize and conclude in Section~\ref{sec::summary}.

\section{Numerical setup}
\label{sec:setup}

Our simulations consist of 3D local simulations of the MRI, and then test particle simulations in individual MRI simulation snapshots. These simulations are described in Section~\ref{sec::setup_mri} and \ref{sec::setup_par}, respectively.

\subsection{MRI simulations}
\label{sec::setup_mri}

We use the Athena MHD code \citep{2008ApJS..178..137S} to run the MRI simulations in the local shearing-box framework \citep{1995ApJ...440..742H}. It adopts a local reference frame corotating with some fiducial radius $r_0$ in the disk at angular velocity $\Omega_0$, dynamical equations are written in Cartesian coordinates by ignoring curvature and incorporating source terms of Coriolis force and tidal gravity. By convention, the coordinates $(x ,y ,z)$ refer to radial, azimuthal and vertical directions, where the fiducial radius corresponds to $x=0$. We denote gas density, velocity and pressure by $\rho$, $\mathbf{u}$ and $P$, respectively, and ${\mathbf B}$ represents magnetic field. We further decompose $\mathbf{u}=\mathbf{u}_0+\mathbf{u}'$ by subtracting background rotation velocity $u_0\equiv-q\Omega x$, where $q\equiv-d\ln\Omega/d\ln R$ is the shearing parameter. We adopt $q=3/2$ for Keplerian rotation. With these definitions, dynamical equations for $\mathbf{u}'$ can be written as \citep{2010ApJS..189..142S}
\begin{align}
	\frac{\partial \rho}{\partial t} +u_0\frac{\partial \rho}{\partial y} +\nabla \cdot \left(\rho \mathbf{u}'\right)=&0\ ,\\
    \frac{\partial \rho\mathbf{u}'}{\partial t}  +u_0\frac{\partial \rho \mathbf{u}'}{\partial y} &+\nabla\cdot\left(\rho\mathbf{u}'\mathbf{u}'+\sf{T}\right) \notag \\ = 
    \rho \Omega_0[2u'_y\hat{x}+ &(q-2)u'_x\hat{y}]\ ,\\
	\frac{\partial \mathbf{B}}{\partial t}=\nabla \times \left(\mathbf{u}'\times\mathbf{B}\right)&+\nabla\times(\mathbf{u}_0\times\mathbf{B})\  ,
\end{align}
with
\begin{equation}
{\sf T}\equiv(P+B^2/2){\sf I}+{\mathbf B}{\mathbf B}\ , 
\end{equation}
where ${\sf I}$ is the identity tensor.
We adopt an isothermal equation of state with $P=\rho c_s^2$, where $c_s$ is the isothermal sound speed. The disk scale height is thus given by $H\equiv c_s/\Omega_0$. Also note that the unit for magnetic field is such that magnetic permeability $\mu=1$ (i.e., magnetic pressure is $B^2/2$).
We focus on turbulence properties in the bulk disk around the midplane region, and neglect vertical gravity, so that our simulations are vertically unstratified.

Equations above are solved by methods described in \cite{2010ApJS..189..142S}, which employs an orbital advection technique \citep{2000A&AS..141..165M} that improves the efficiency and accuracy of the calculation.
The radial boundary conditions are shearing-periodic. Given the radial size of the simulation box $L_x$, it is given by
\begin{equation}
	\mathbf{A}(x,y,z,t) = \mathbf{A} (x \pm L_x, y \mp q \Omega_0 L_x t, z,t ),
\end{equation}
for any variables $\mathbf{A}(x,y,z,t)$. Boundary conditions in the $\hat{y}$ and $\hat{z}$ directions are periodic.

Our initial condition is given by a uniform density with $\rho=\rho_0$, threaded by a uniform vertical magnetic field $B_z=B_0$ parameterized by the plasma beta
\begin{equation}
	\beta_0= \frac{2\rho_0 c_s^2}{B_0^2}\ ,
\end{equation}
which is the only physical parameter in our simulations. In code units, we take $\rho_0=1$, $c_s=1$, and $\Omega_0=1$, thus $H=1$.

We run four simulations, varying the resolution and $\beta_0$, which are listed in Table~\ref{tab::mri}. Our simulation box size is fixed as $(L_x, L_y, L_z)=(4H, 8H, 2H)$, where the vertical domain size is chosen to represent the bulk of the midplane region while marginally compatible with the assumption of being unstratified. We choose $\beta_{0} = 10^2$, $10^3$, $10^4$ with a fiducial resolution of $(N_x, N_y, N_z) = (512, 512, 256)$ or $(512,768,256)$. These runs are named as B2, B3 and B4, respectively. For the case of  $\beta_{0} = 10^4$, we run one additional simulation with nearly twice the resolution with $(N_x, N_y, N_z) = (960, 960, 480)$, named as B4-hires. The large horizontal box size and high resolution offer us sufficiently large dynamical range for test particle simulations. In fact, the resolution chosen here ($128-240$ cells per $H$ in $x$ and $z$) allows us to well resolve the most unstable MRI wavelength ($\sim 9.18\beta_0^{-0.5} H$, \citealp{1995ApJ...440..742H}) in all cases.

In our simulations, the MRI turbulence saturates in 10-20 orbits. We take snapshots after $t=150\Omega_0^{-1}$ at a constant time interval of $\Delta t=12\Omega_0^{-1}$ or $15\Omega_0^{-1}$(Table~\ref{tab::mri}). Each snapshot is treated as an independent realization of the MRI turbulence of given parameters, for which we conduct particle simulations on top of its electromagnetic field.

Two points are worth clarifying regarding our simulations. Firstly, MHD simulations of the MRI assume the gas can be described as collisional fluid, despite we aim at the collisionless regime. The dispersion relation for collisionless MRI is found to be different but qualitatively similar to the MHD counterpart \citep{2002ApJ...577..524Q, 2003ApJ...596.1121S}. Incorporating sub-grid prescriptions of kinetic effects into MHD simulations, the general properties of the MRI turbulence are also found to be qualitatively similar \citep{2007ApJ...671.1696S, 2015MNRAS.454.1848R, 2017MNRAS.466..705S, 2017ApJ...844L..24R, 2017MNRAS.470.2367C, 2018MNRAS.478.5209C, 2020MNRAS.494.4168D}. Therefore, the properties of the MRI turbulence beyond kinetic scale, to a first approximation, can be reasonably taken from ideal MHD simualtions of the MRI turbulence, which serves the purpose of this paper. Secondly, the shearing box approximation is strictly applicable to thin disks, whereas ADAF disks are thick. Nevertheless, main properties of the MRI turbulence remain similar, as long as the MRI is properly resolved (e.g., 
\citealp{2011ApJ...738...84H}). 

\begin{table*}
    \centering
    \begin{tabular}{ccccc}
        \toprule
        Name & B2 & B3 & B4 & B4-hires \\
        \midrule
        $(N_x,N_y,N_z)$ & (512,768,256) & (512,768,256) & (512,512,256) & (960,960,480) \\
        Snapshots$[\Omega_0^{-1}]$ &$150 - 300, \Delta t = 15$ & $150 - 300, \Delta t = 15$ & $156 - 360,  \Delta t = 12$ &  $156 - 336,  \Delta t = 12$ \\
        $\alpha_{\text{Rey}} $ & $2.51\times 10^{-1}$ & $5.20\times 10^{-2}$ & $1.20\times 10^{-2}$ & $1.49\times 10^{-2}$ \\
        $\alpha_{\text{Max}}$ & $1.01$ & $2.90\times 10^{-1}$ & $5.75\times 10^{-2}$ & $8.12\times 10^{-2}$ \\
        $\alpha^\dagger $ & $1.26$ & $3.41\times 10^{-1}$ & $6.95\times 10^{-2}$ & $9.62\times 10^{-2}$ \\
        $\langle \rho {u'_x}^2 /2 \rangle [\rho_0 c_s^2]$ & $4.33\times 10^{-1}$ & $9.78\times 10^{-2}$ & $2.20\times 10^{-2}$ & $2.86\times 10^{-2}$ \\
        $\langle \rho {u'_y}^2 /2 \rangle [\rho_0 c_s^2]$ & $3.96\times 10^{-1}$ & $1.22\times 10^{-1}$ & $2.59\times 10^{-2}$ & $3.68\times 10^{-2}$ \\
        $\langle \rho {u'_z}^2 /2 \rangle [\rho_0 c_s^2]$ & $1.23\times 10^{-1}$ & $4.23\times 10^{-2}$ & $1.09\times 10^{-2}$ & $1.43\times 10^{-2}$ \\
        $\langle B^2_x /2 \rangle [\rho_0 c_s^2]$ & $5.63\times 10^{-1}$ & $1.29\times 10^{-1}$ & $2.19\times 10^{-2}$ & $3.31\times 10^{-2}$ \\
        $\langle B^2_y /2 \rangle [\rho_0 c_s^2]$ & $1.41$ & $5.65\times 10^{-1}$ & $1.07\times 10^{-1}$ & $1.53\times 10^{-1}$ \\
        $\langle B^2_z /2 \rangle [\rho_0 c_s^2]$ & $2.32\times 10^{-1}$ & $6.77\times 10^{-2}$ & $1.13\times 10^{-2}$ & $1.82\times 10^{-2}$ \\
        $\overline{B} [\sqrt{\rho_0c_s^2}]$ & $2.10$ & $1.23$ & $5.30\times 10^{-1}$ & $6.39\times 10^{-1}$ \\
        \bottomrule
    \end{tabular} \\
    \footnotesize{$ ^\dagger$The $\alpha$ parameter, contributed by Reynolds and Maxwell stress tensor,
    $\langle \rho u'_x u'_y - B_x B_y \rangle/(\rho_0 c_s^2)$.
    }
    \caption{Parameters in the MRI simulations and the resulting turbulence quantities.}
    \label{tab::mri}
\end{table*}

\subsection{Test particle simulations}
\label{sec::setup_par}

We inject cosmic-ray (CR) particles into snapshots of the MRI simulations, and solve the equation of motion of individual kinetic particles
\begin{align}
	\frac{\dd \mathbf{p}}{\dd t} &= e \left(\mathbf{E}+ \frac{\mathbf{v}}{c} \times \mathbf{B} \right)\ , \notag \\
	\mathbf{p} &= \gamma m \mathbf{v}\ , \label{equ::equ_of_motion}
\end{align}
where $\mathbf{v}$, $\mathbf{p}$, $m$, $e$ and $\gamma$ denote particle velocity, momentum, mass, charge and Lorentz factor respectively. We focus on relativistic ions with $\gamma\gg1$ and hence $v\approx c$, while the results apply equally to relativistic electrons.
The electric field $\mathbf{E}$ is obtained from the MRI simulation snapshot based on the frozen-in condition (preventing direct acceleration in parallel electric field, e.g., in magnetic reconnection):
\begin{equation}\label{eq:frozen}
	\mathbf{E} + \frac{\mathbf{u}}{c} \times \mathbf{B} = 0\ .
\end{equation}

In addition, to isolate the contribution of Keplerian shear from the MRI, we also consider the putative ``shear-subtracted'' turbulence (See Section~\ref{sec::evolute_mom}), where,
\begin{equation}\label{eq:frozen2}
	\mathbf{E'} + \frac{\mathbf{u'}}{c} \times \mathbf{B} = 0\ .
\end{equation}

We use the Boris integrator \citep{boris1972proceedings} to push particles, with velocity and $B$ field interpolated to particle positions by the standard triangular-shaped cloud (TSC) scheme \citep{birdsall2004plasma}. We then compute $\mathbf{E}$ at particle position from Equ.~\ref{eq:frozen} or ~\ref{eq:frozen2}, which guarantees that there is no spurious acceleration by parallel electric field.

From the MRI simulations, we use $\overline{ B}\equiv\sqrt{\langle B^2\rangle}$ to denote the mean strength of the magnetic field in the saturated state, where $\langle\cdot\rangle$ denotes averages over time and volume. With this definition, we can non-dimensionize the particle simulations as follows.
The role of $e$ and $m$ are on particle dynamics can be fully encapsulated in to a single factor
\begin{equation}
    \omega_0\equiv e\overline{B}/mc\ ,
\end{equation} 
namely, the cyclotron frequency, as can be seen by reorganizing Equation~\ref{equ::equ_of_motion} as
\begin{equation}
	\frac{\dd}{\dd t}\bigg(\frac{\gamma\mathbf{v}}{\omega_0}\bigg) = 
	\left(\mathbf{v}- \mathbf{u}\right) \times \frac{\mathbf{B}}{\overline{B}} \ .
\end{equation}

Similarly, particle momentum is characterized by its mean gyro radius
\begin{equation}
    R_{g0}=\frac{pc}{e\overline{B}}=\frac{\gamma v}{\omega_0}\approx\frac{\gamma c}{\omega_0}\ ,
\end{equation}
where the last approximate equality applies for relativistic particles.

In our simulations, we inject mono-energetic particles with momentum $p=p_0$, corresponding to gyro radius of $R_{g0}$. For relativistic particles, the equation of motion is fully specified by an additional choice of the speed of light $c$. For non-relativistic particles, we need to specify the corresponding velocity $v_0$ to the gyro radius. Thus, our particle simulation parameters are given by the pair $(R_{g0}, c)$ or $(R_{g0}, v_0)$ for relativistic/non-relativistic cases. 

There are several constrains on the choice of parameters based on our assumptions. First, the use of MRI simulation snapshot requires that particle gyro-frequency $\omega$ be much higher than disk orbital frequency. In other words,
\begin{equation}
	\omega\equiv 
	\omega_0/\gamma\gg \Omega_0\ . \label{equ::gyrofreq}
\end{equation}
We also note that the ratio of $\omega/\Omega_0$ is on the same order of the ratio of Lorentz force to Coriolis force or tidal gravity, thus they can be dropped in integrating particle equations of motion. A related condition is that particle gyro radius must be well constrained within the disk, or $R_g \lesssim H$. For typical disk conditions and particle energies, we show in Section~\ref{sec::acc_ADAF} that these conditions are always satisfied. 

Second, the choice of speed of light (essentially particle velocity, which is $v_0$ in the non-relativistic case) must be much larger than any MHD velocities in the simulation box, which consist of turbulent motion and shear motion. As the MRI turbulence is mostly sub-thermal, this means we must have\footnote{More rigorously, we may write it as $c\gg q\Omega_0N_{\rm cross}L_x+c_s$, with $N_{\rm cross}$ being number of times a particle cross the radial boundary in one direction, because each crossing effectively means the particle moves to a neighboring copy of the simulation box that is a distance $L_x$ further away.}
\begin{equation}
    c\gg q\Omega_0L_x+c_s\ .\label{equ::light_relation}
\end{equation}
Physically, this choice corresponds to how far the fiducial radius $r_0$ in the shearing box is from the black hole. Clearly, larger $c$ corresponds to a location that is further away.

We generally choose the initial $R_{g0}$ from $0.004H$ ($0.002H$ for run B4-hires) to $0.2H$, and we set $c=50c_s$ by default, but also vary it to $100c_s$. For each MRI simulation snapshot, we inject $2^{17}\approx1.3\times10^5$ particles at fixed momentum $p_0$ at random positions and directions, and track the evolution of their trajectories and momenta. The results from all snapshots of the same MRI simulations are then combined for further analysis.

Time in the simulations can then be conveniently normalized by particle gyro time
\begin{equation}
    T_0=\frac{2\pi R_{g0}}{c}=\frac{2\pi\gamma_0}{\omega_0}\ ,
\end{equation}
where $c$ should be replaced by $v_0$ for the non-relativistic case. Because $R_{g0}\lesssim H$ and $c\gg c_s$, we have $\Omega_0T_0=2\pi(R_g/H)(c_s/c)\ll1$. We typically run the particle simulations for $T=125T_0\lesssim \Omega_0^{-1}$. We set integration timestep to be $\Delta t\lesssim 0.1T_0$ and that a particle does not move by more than one MHD cell, and we have verified that this choice is sufficiently small to guarantee numerical convergence of simulation results.

Boundary conditions for charged particles are mostly same as those in MHD simulations, except that the shearing periodic boundary condition which needs to  incorporate  relativity \citep{2012ApJ...755...50R,2016ApJ...822...88K}. For particles crossing the radial boundary, the Lorentz transformation is applied to particle momenta:
\begin{align}
	p'_y&=\Gamma_{\rm box}\left(p_y-\beta_{\rm box} \varepsilon / c \right),\notag \\
	p'_x&=p_x,\qquad p'_z=p_z,\notag \\
	\beta_{\rm box}&=\mp \frac{q \Omega_0 L_x}{c},\qquad
	\Gamma_{\rm box}=(1-\beta_{\rm box}^2)^{-1/2}. \label{equ:shear_bound_par}
\end{align}
where the $\pm$ sign corresponds to particles crossing from either sides and $\varepsilon$ represents the energy of each particle.

We further comment on our choice of particle gyro radii. The smallest $R_g$ we choose marginally fit to the grid resolution. The maximum gyro radii of $0.2H$ is chosen for two reasons. First, they can easily escape from the disk and hence in reality particles can hardly reach such high energies. Second, the boundary condition above implicitly requires that particles undergo sufficient scattering that their distribution function become isotropic in the background shear flow. Otherwise, such transformations would introduce unphysical accelerations as particles cross radial boundaries (like the runaway particles in \citet{2016ApJ...822...88K}). We confirm that with $R_g\lesssim0.2H$, particles are sufficiently isotropized, and hence do not suffer from spurious accelerations.

\section{MRI turbulence}
\label{sec:mri}

\begin{figure}
    \centering
    \includegraphics[width=28em]{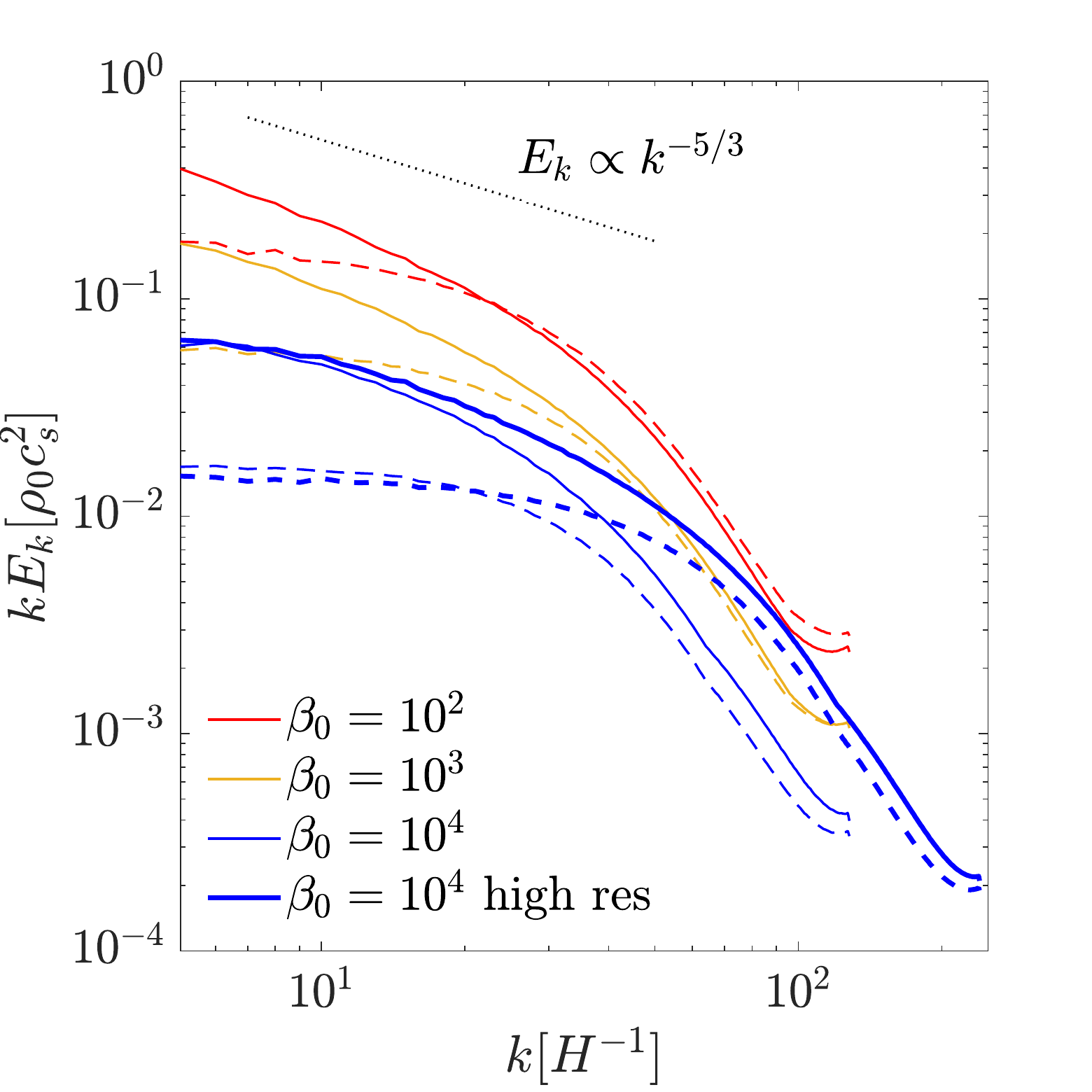}
    \caption{The compensated power spectra density (PSD, $kE_k$) of the magnetic (solid) and kinetic (dashed) energy densities from each MRI simulation run. Different runs are indicated by color and line width (see legend). The dotted line scales as the Kolmogorov power spectrum, where $E_k \propto k^{-5/3}$.}
    \label{fig::mri_power_spectrum}
\end{figure}

In this section, we describe the main properties of the MRI turbulence from our simulations, following standard analysis procedures. 

The mean energy density of MRI turbulence increases with decreasing $\beta_0$, and the turbulence is anisotropic with $\langle \rho u'^2_y /2 \rangle \gtrsim \langle \rho u'^2_x /2 \rangle > \langle \rho u'^2_z /2 \rangle$ and $\langle B^2_y /2 \rangle > \langle B^2_x /2 \rangle > \langle B^2_z /2 \rangle$, where angle brackets represent spatial and time averages. 
The turbulent energy density in the higher resolution run slightly exceeds that in the lower resolution run, as run B4-hires better resolves the most unstable mode than run B4 \citep{1998RvMP...70....1B}. Besides the energy, the Reynolds stress $\alpha_\text{Rey}=\langle \rho u'_x u'_y \rangle/\left( \rho_0 c_s^2 \right)$ and the Maxwell stress $\alpha_\text{Max}=- \langle B_x B_y \rangle/\left( \rho_0 c_s^2 \right)$ characterize the strength of MRI turbulence in transporting angular momentum. The Maxwell stress is always greater than the Reynolds stress by a factor of $\sim5$, and correlates with the square of mean field strength $\overline{B}$ (e,g., \citealp{1995ApJ...440..742H,2011ApJ...736..144B}).

We also examine the shell-integrated power spectrum density(PSD), averaged over all snapshots, defined as
\begin{equation}
    E_k \equiv  4\pi k^2 \tilde{E}_\mathbf{k}, 
\end{equation}
where $\tilde{E}_\mathbf{k}$ denotes the average of kinetic or magnetic energy density over constant $k=|\mathbf{k}|$ in the Fourier space, with proper remapping procedures to account for shearing motion and boundary conditions, following \cite{1995ApJ...440..742H} and \citet{2010ApJ...713...52D}. We show the results in terms of $kE_k$ in Figure~\ref{fig::mri_power_spectrum}, which has a unit of energy density (naturally normalized to $\rho_0c_s^2$). The magnetic PSD of these runs are generally found to be slightly steeper than the Kolmogorov slope of $-5/3$ at low $k$, while the kinetic energy PSDs are smaller and flatter, consistent with previous findings \citep{2016MNRAS.457L..39W}. The PSD slope
steepens at larger $k$ resulting from numerical dissipation. Higher magnetization (lower $\beta_0$) leads to stronger PSD, as expected. The higher-resolution run B4-hires shows consistent normalization in its PSDs with those in run B4 at low $k$, indicating convergence in the bulk energetics, but it also shows extended inertial range with a magnetic PSD closer to the $k^{5/3}$ scaling than in run B4. Such high resolution is likely needed to properly capture the inertial range in the MRI turbulence.

\begin{figure}
    \centering
    \includegraphics[width=28em]{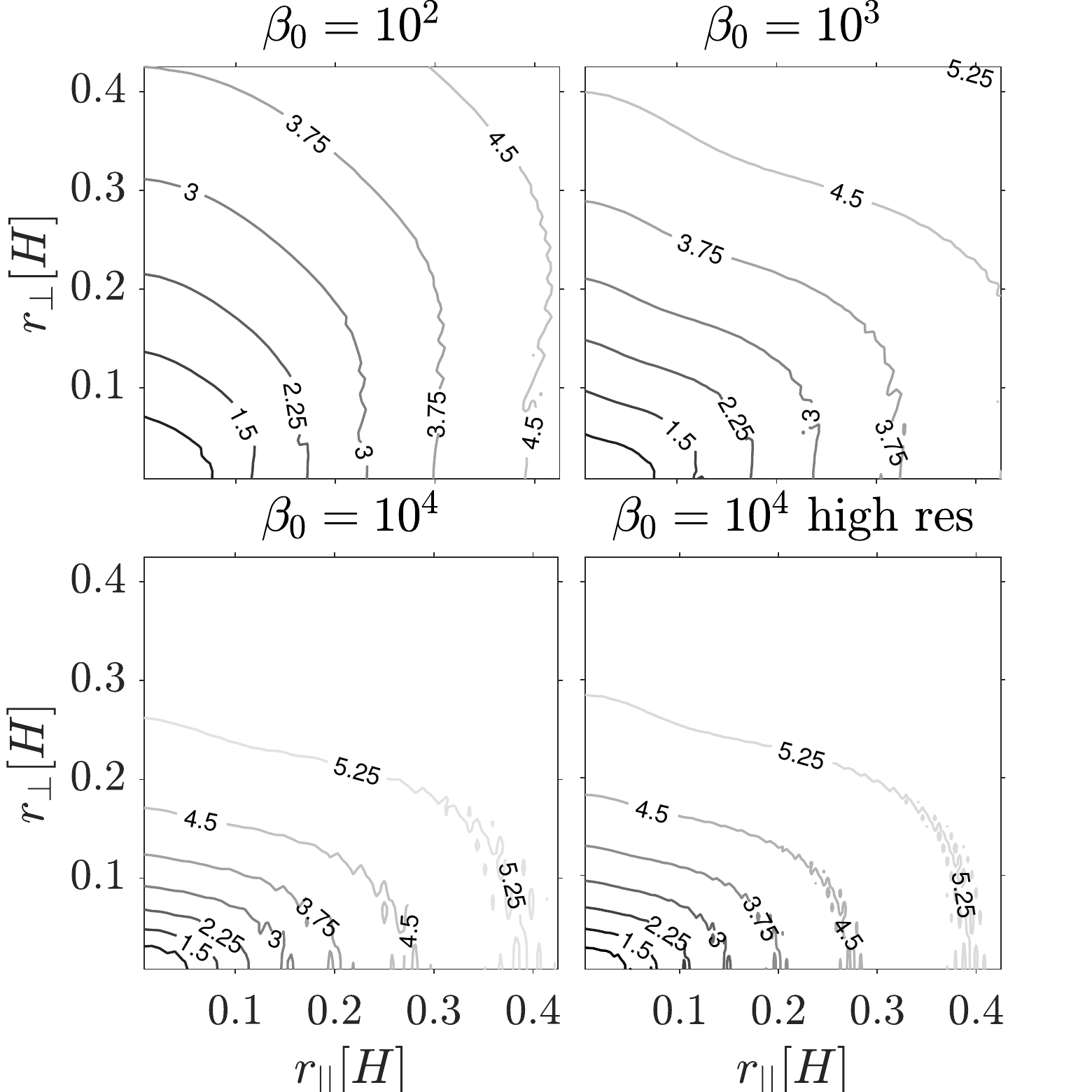}
    \caption{Contour plots of the second-order normalized structure function $SF_2(r_\parallel , r_\perp)$ (\ref{eq:SF2}) for magnetic fields from each of the MRI simulations, which can be considered as an illustration of eddy shapes relative to the local mean field direction at different scales.}
    \label{fig::mri_eddy}
\end{figure}
\begin{figure}
    \centering
    \includegraphics[width=28em]{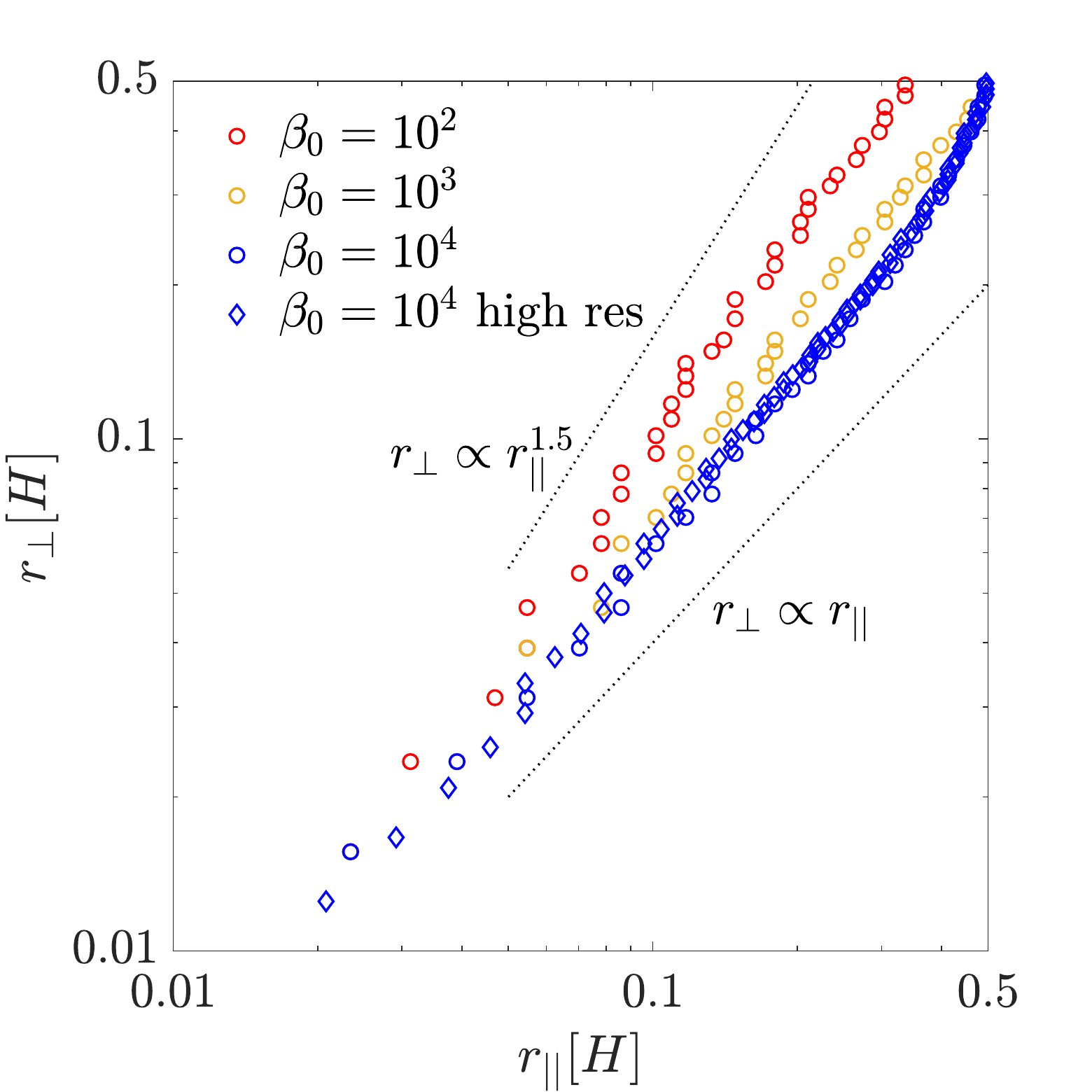}
    \caption{Anisotropy of turbulent eddies from all our MRI simulation runs (indicated by different colors and symbols, see legend). Data points are measured from the intercept values of $r_{\parallel}$ and $r_\perp$ in the contour plots of Figure~\ref{fig::mri_eddy}, respectively. The dotted line indicates the slope expected from critical balance.}
    \label{fig::mri_anisotropy}
\end{figure}

MHD turbulence is well known to be anisotropic (e.g., \citealp{1995ApJ...438..763G, 2000ApJ...539..273C, 2003MNRAS.345..325C}). We further characterize the anisotropy of the MRI turbulence, by evaluating the second-order structure function for magnetic field
\begin{equation}
   SF_2(r_\parallel , r_\perp)\equiv  \langle |\mathbf{B}(\mathbf{x+r})-\mathbf{B}(\mathbf{x}) |^2\rangle\ ,
\end{equation}
where angle bracket averages over $\mathbf{x}$, and $\mathbf{r}=r_\parallel\hat{\mathbf r}_\parallel+r_\perp\hat{\mathbf r}_\perp$, with $\hat{\mathbf r}_\parallel$, $\hat{\mathbf r}_\perp$ denoting unit vectors parallel and perpendicular to the local mean field $\mathbf{B}_L$. 
There can be different ways of measuring $\mathbf{B}_L$, especially given that the fluctuating field in the MRI turbulence well exceeds background mean field. Here we adopt a three point second-order normalized structure function devised by \cite{2009ApJ...701..236C}, and we further normalize it by the mean field as follows
\begin{equation}\label{eq:SF2}
    SF_2(r_\parallel , r_\perp) = \frac{\langle |\mathbf{B}\left(\mathbf{x}+\mathbf{r}\right) - 2\mathbf{B}\left(\mathbf{x}\right)+\mathbf{B}\left(\mathbf{x}-\mathbf{r}\right)|^2 \rangle}
    {\langle{B^2}\rangle},
\end{equation}
where
\begin{equation}
    r_\parallel = \mathbf{r} \cdot \frac{\mathbf{B}\left(\mathbf{x}+\mathbf{r}\right) + \mathbf{B}\left(\mathbf{x}\right)+\mathbf{B}\left(\mathbf{x}-\mathbf{r}\right)}{|\mathbf{B}\left(\mathbf{x}+\mathbf{r}\right) + \mathbf{B}\left(\mathbf{x}\right)+\mathbf{B}\left(\mathbf{x}-\mathbf{r}\right)|}, r_\perp = \sqrt{|\mathbf{r}|^2 - r_\parallel^2}. \notag
\end{equation}
We show contours of the structure function in Figure~\ref{fig::mri_eddy}, which depicts the eddy's shapes. Note that the local mean magnetic field could only be well defined within typical eddy size of $|\mathbf{r}| \lesssim H$. The level of anisotropy can be characterized by the intercept of the contours with the two axis, which are shown in Figure~\ref{fig::mri_anisotropy}. We see that in runs B3 and B4, eddies are nearly isotropic on scales of $\sim H$, and become more elongated along the local field towards smaller scales. Interestingly, eddies in B2 is more elongated in the perpendicular direction at large scales, and the eddy shape then reverses at smaller scales. Critical balance of \cite{1995ApJ...438..763G} suggests $r_\perp \propto r_\parallel^{3/2}$. In our simulations, we find that the results are between $r_\perp \propto r_\parallel$ and $r_\perp \propto r_\parallel^{3/2}$, where run B2 is closer to the scaling from critical balance.
Comparing the result between run B4 and B4-hires, we see that level of anisotropy agrees well between different resolutions. 

\begin{figure}
    \centering
    \includegraphics[width=28em]{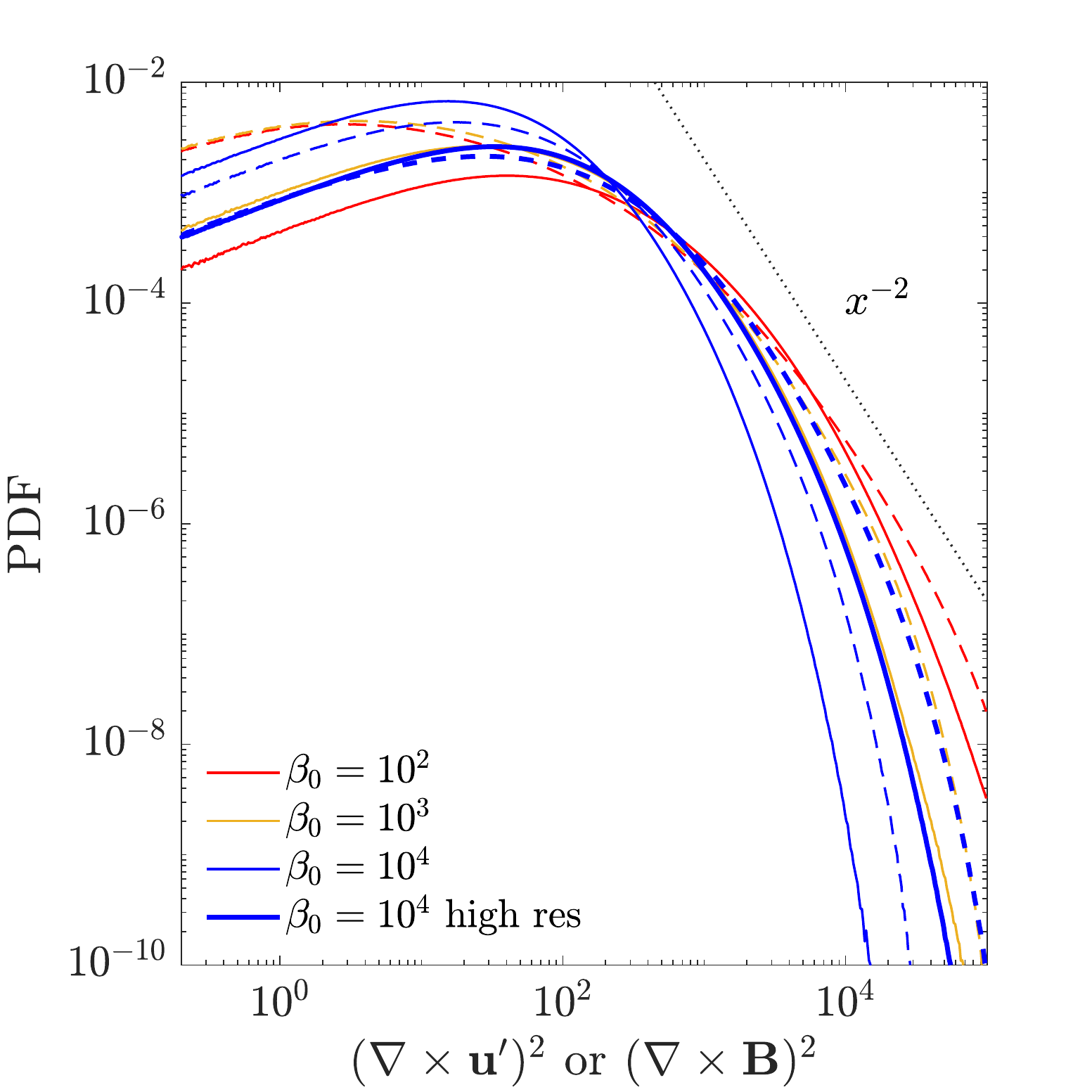}
    \caption{The probability density functions for the square of vorticity (solid) and the square of current density (dashed) in code units. Different runs are indicated by color and line width (see legend). The dotted line shows the asymptotic scaling of a Lorentzian profile $x^{-2}$.}
    \label{fig::intermittency}
\end{figure}

While there is no explicit dissipation in our ideal MHD simulations, it is still useful to 
examine the structures that are associated with energy dissipation, which can be related to intermittency (scale-dependent inhomogeniety). In the presence of explicit viscosity and resistivity, rate of kinetic and magnetic energy dissipation is proportional to the square of
vorticity $\nabla \times \mathbf{u}'$ and current density $\nabla \times \mathbf{B}$.
We show the probability density functions (PDF) of these quantities in Figure~\ref{fig::intermittency} from all our simulations. The PDFs for all four runs decay after reaching a peak, and they have more extended wings for simulations with stronger magnetization and higher resolution, and hence generating more small-scale structures. The PDFs of $\left( \nabla \times \mathbf{B} \right)^2$ are always higher than those of $\left( \nabla \times \mathbf{u}' \right)^2$ in the wings, suggesting that magnetic energy dissipation dominates. For comparison, we also show the an $x^{-2}$ scaling for a Lorentzian-like distribution. While the decaying part of the PDFs matches this scaling at certain ranges, the overall slope decays continuously.
We note that the PDFs presented here can only be considered as a crude way of characterizing dissipation in the simulations. We do observe that in saturated state, the system is characterized of small-scale current-sheet like structures that occupy very small fraction of the simulation volume (as in several other works, e.g., \citealp{2011ApJ...736..144B, 2014ApJ...791...62M}). \citet{2017MNRAS.467.3620Z} found that the PDF for the dissipation rates of these structures follows a $x^{-2}$ scaling.
Implications of this intermittency will be revisited in Section \ref{sec::direct_acc}.

\section{Diffusion in configuration space}
\label{sec::diff_conf}

\begin{figure*}
    \centering
    \subfloat[]{
    \includegraphics[width=21em]{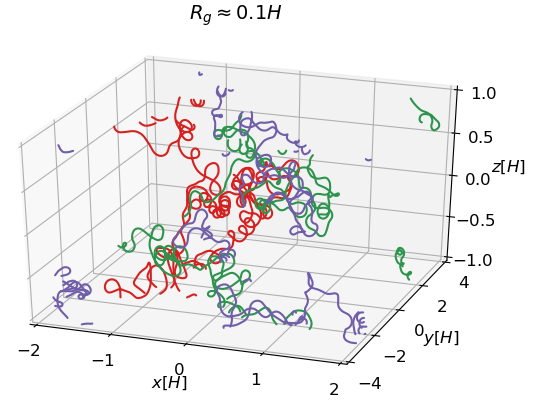}
    \label{fig::traj_rel_mri_6}}
    \subfloat[]{
    \includegraphics[width=21em]{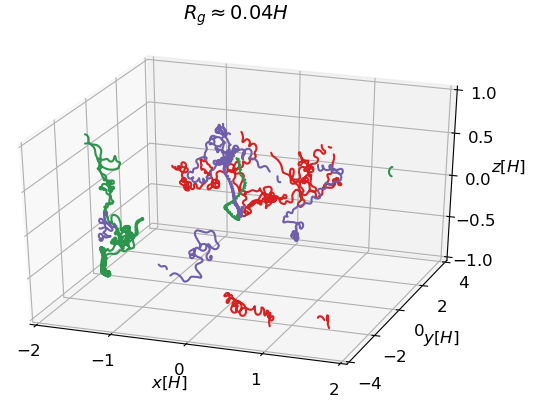}
    \label{fig::traj_rel_mri_2}}
    \subfloat[]{
    \includegraphics[width=21em]{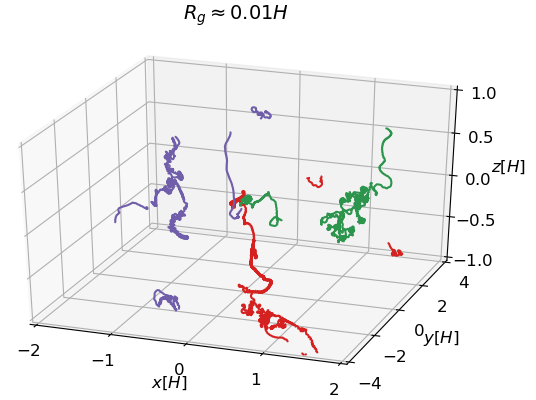}
    \label{fig::traj_rel_mri_0.5}}
    \caption{Trajectories of representative CR particles in a B4-hires snapshot over the same time intervals (50,125,500$T_0$ for $R_g=0.1,0.04,0.01H$). Three particles with similar initial gyro-radii are shown in each panel, indicated by different colors. 
    }
    \label{fig::traj_mri}
\end{figure*}

\begin{figure}
    \centering
    \includegraphics[width=28em]{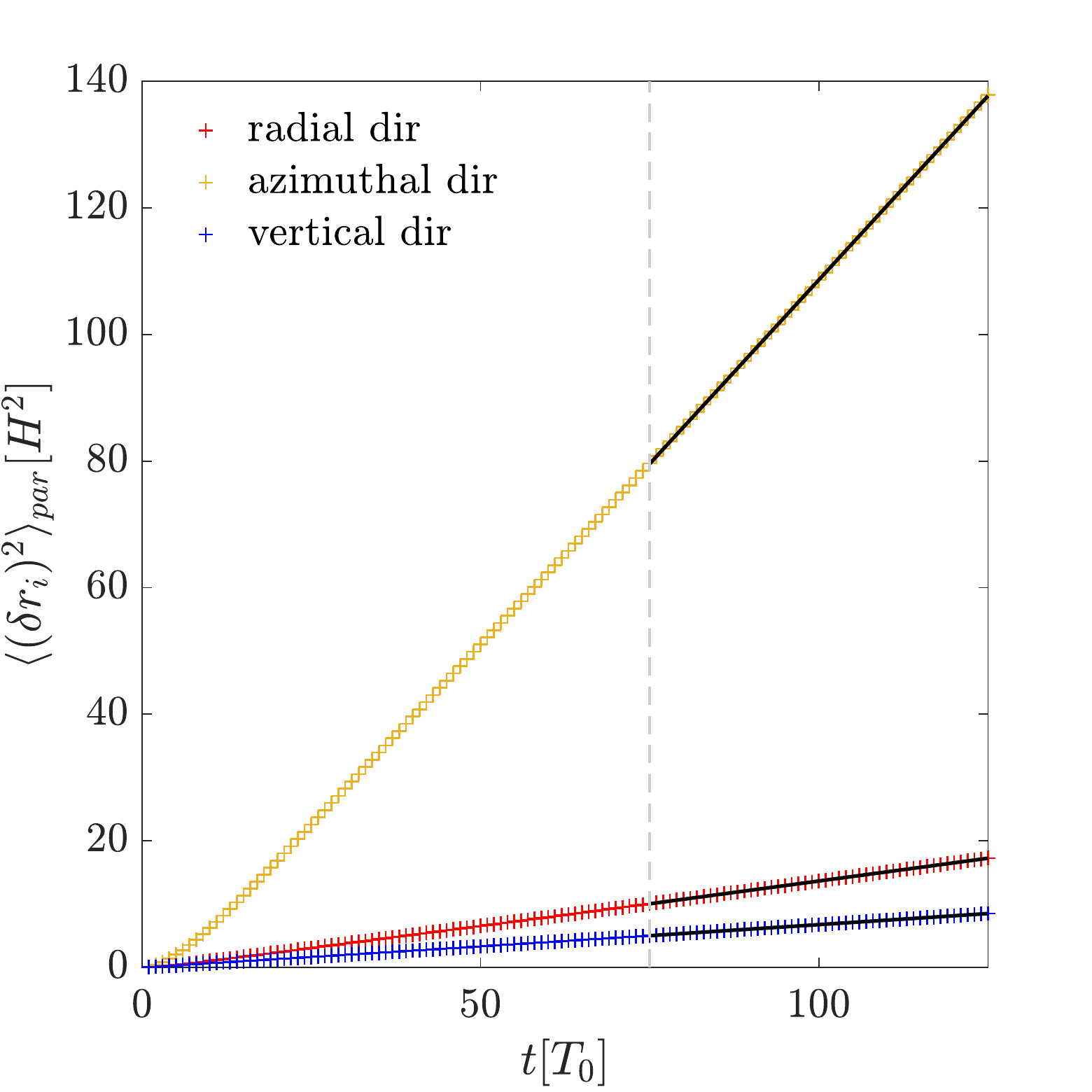}
    \caption{Illustration of relativistic particle diffusion in configuration space shown as time evolution of mean square displacement in three directions (see the legend), with particle gyro radius $R_{g0}=0.1H$ in run B4-hires. Linear fitting is performed after 75$T_0$ to derive the diffusion coefficients.
    }
    \label{fig::diffusion_conf_with_t_rel}
\end{figure}

In this section, we discuss diffusion of relativistic CR particles in configuration space.
In Figure \ref{fig::traj_mri}, we show typical trajectories in run B4-hires for particles with three different gyro radii over the same time intervals. Clearly, all particles move stochastically. Particles with small gyro radii ($R_g\approx0.01-0.04H$, Figure~\ref{fig::traj_rel_mri_0.5},\ref{fig::traj_rel_mri_2}) largely follow guiding-center motion along turbulent magnetic field lines. Among them, particles with larger gyro radii become less strongly tied to field lines. More energetic particles ($R_g=0.1H$, Figure~\ref{fig::traj_rel_mri_6}) exhibit more stochastic trajectories reaching wider range of locations over the same time interval.

More quantitatively, we measure the displacement of each particle, $\delta \mathbf{r} = \mathbf{r}(t) - \mathbf{r}(0)$, as a function of time. In Figure \ref{fig::diffusion_conf_with_t_rel}, we show the mean square displacements in each direction for particles with $R_{g0} \approx 0.1H$ in run B4-hires. 
There is clearly a linear relation $\langle \left(\delta r_i\right)^2 \rangle_\text{par} \propto t$ for each direction $i$ with tight correlation (correlation coefficient $>0.997$). Therefore, particles exhibit normal diffusion (random walk) in configuration space, from which we can fit the diffusion coefficient as
\begin{equation}\label{eq:Dconf}
 \langle \left(\delta r_i\right)^2 \rangle_\text{par}=2D_{{\rm conf},i}t\ .
\end{equation}
for each direction $i$. 
From the fitting in Figure \ref{fig::diffusion_conf_with_t_rel}, we see that $D_{{\rm conf},y} \gg D_{{\rm conf},x} > D_{{\rm conf}, z}$. This is in line with the fact that $\langle B_y^2 \rangle > \langle B_x^2 \rangle > \langle B_z^2 \rangle$ (Table~\ref{tab::mri}, also see \citealp{2016ApJ...822...88K}).
With azimuthal field being the most dominant field component, particles travel much more freely along the $y$ direction than in the poloidal plane.

\begin{figure}
    \centering  
    \includegraphics[width=28em]{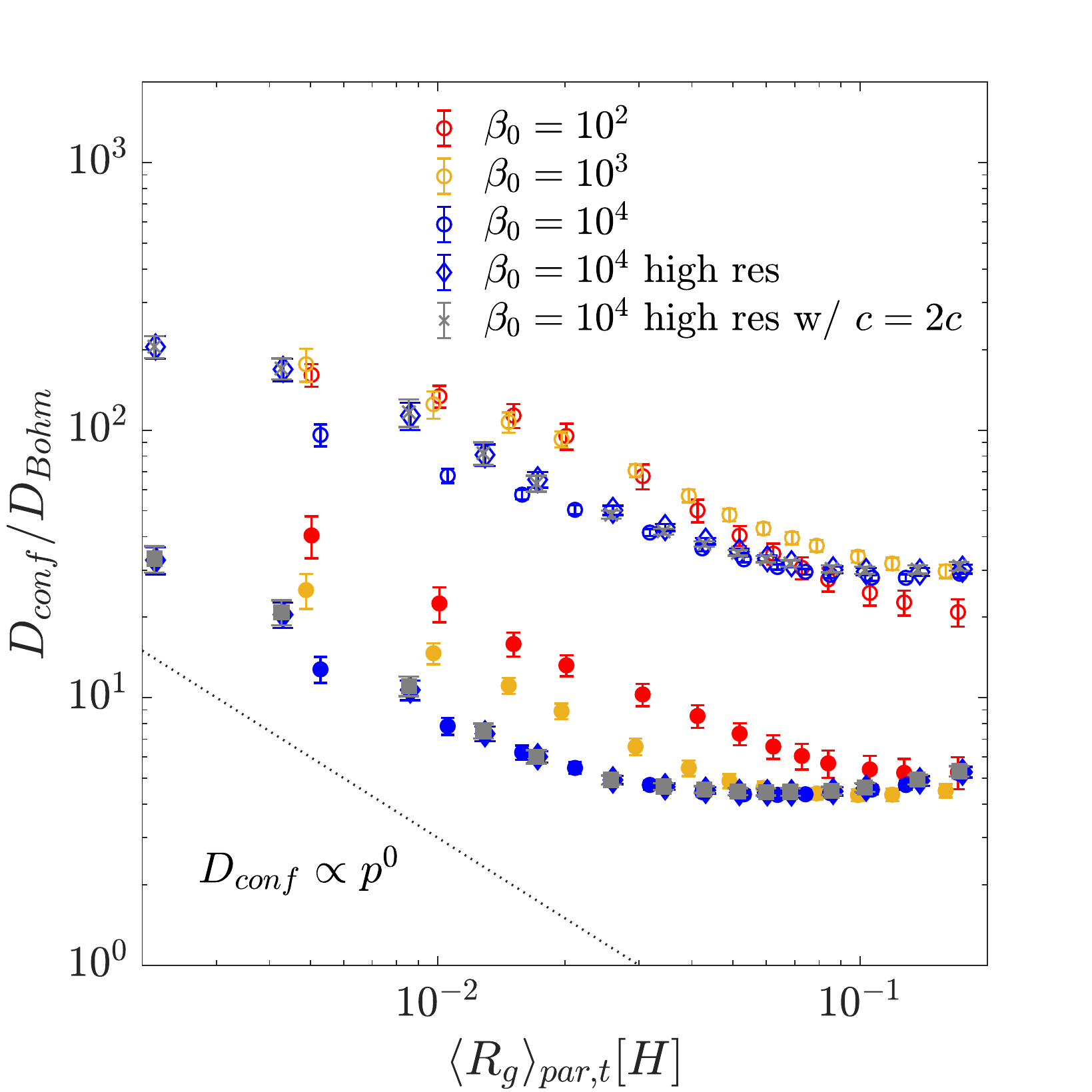}
    \caption{Diffusion coefficients for relativistic particles in the configuration space as a function of mean gyro radii $\langle R_{g} \rangle_\text{par, t}$ after $t_{\rm begin}$. The value has been normalized by the Bohm diffusion coefficient $D_\text{Bohm}$. Different colors represent  MRI simulations with different $\beta_{0}$, and a different symbol is used for the higher-resolution run B4-hires and cases with $c' = 100 c_s=2c$.
    Open symbols refer to total diffusion coefficient $D_{\rm conf}$, summed from all directions, and filled ones represent $D_{\rm conf,pol}$, summed from radial and vertical directions.}
    \label{fig::diffusion_conf_with_p_rel}
\end{figure}

The total diffusion coefficient is $D_{\rm conf} = D_{{\rm conf},x} + D_{{\rm conf},y} + D_{{\rm conf},z}$, and its poloidal component is $D_{{\rm conf}, {\rm pol}} = D_{{\rm conf},x} + D_{{\rm conf},z}$. It is customary to normalize the diffusion coefficient to that in the Bohm limit. For particles with mean gyro radius $R_{g0}$, it is given by
\begin{equation}
	D_{\rm Bohm}\approx\frac{v R_{g0}}{3}\ ,
\end{equation}
with $v\approx c$ for relativistic particles. It is considered as the minimum diffusion coefficient possible where particles get efficiently scattered over one gyro time.

In measuring $D_{\rm conf}$,
we start to fit the relation Equ.~\ref{eq:Dconf} after a time $t_{\rm begin}$ that satisfies
\begin{equation}\label{eq:tstart}
    t_{\rm begin} > \frac{D_{\rm conf}}{D_{\rm Bohm}} T_0\ .
\end{equation}
This is to avoid artificial features from initial injection (on top of our randomization injection procedures) where particles are free-streaming. 
In practice, we choose $t_{\rm begin} = 75 T_0$, for which we find Equ.~\ref{eq:tstart} is satisfied for most particle simulations. Although not obvious in Figure \ref{fig::diffusion_conf_with_t_rel}, we do find this procedure necessary in some cases.

\begin{figure*}
    \centering
    \subfloat[]{
    \includegraphics[width=21em]{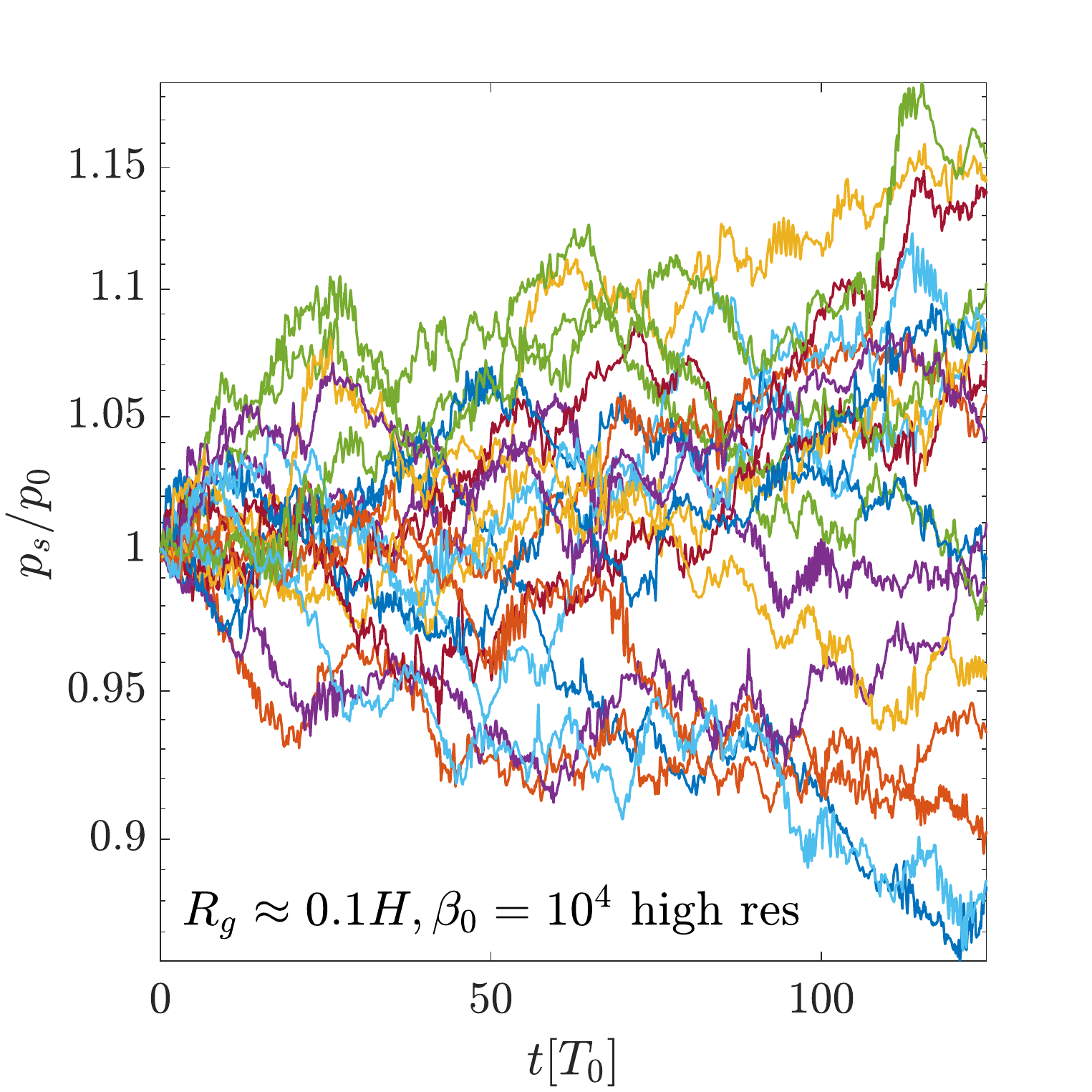}
    \label{fig::energy_hist_0.1_b1e4_hires}}
    \subfloat[]{
    \includegraphics[width=21em]{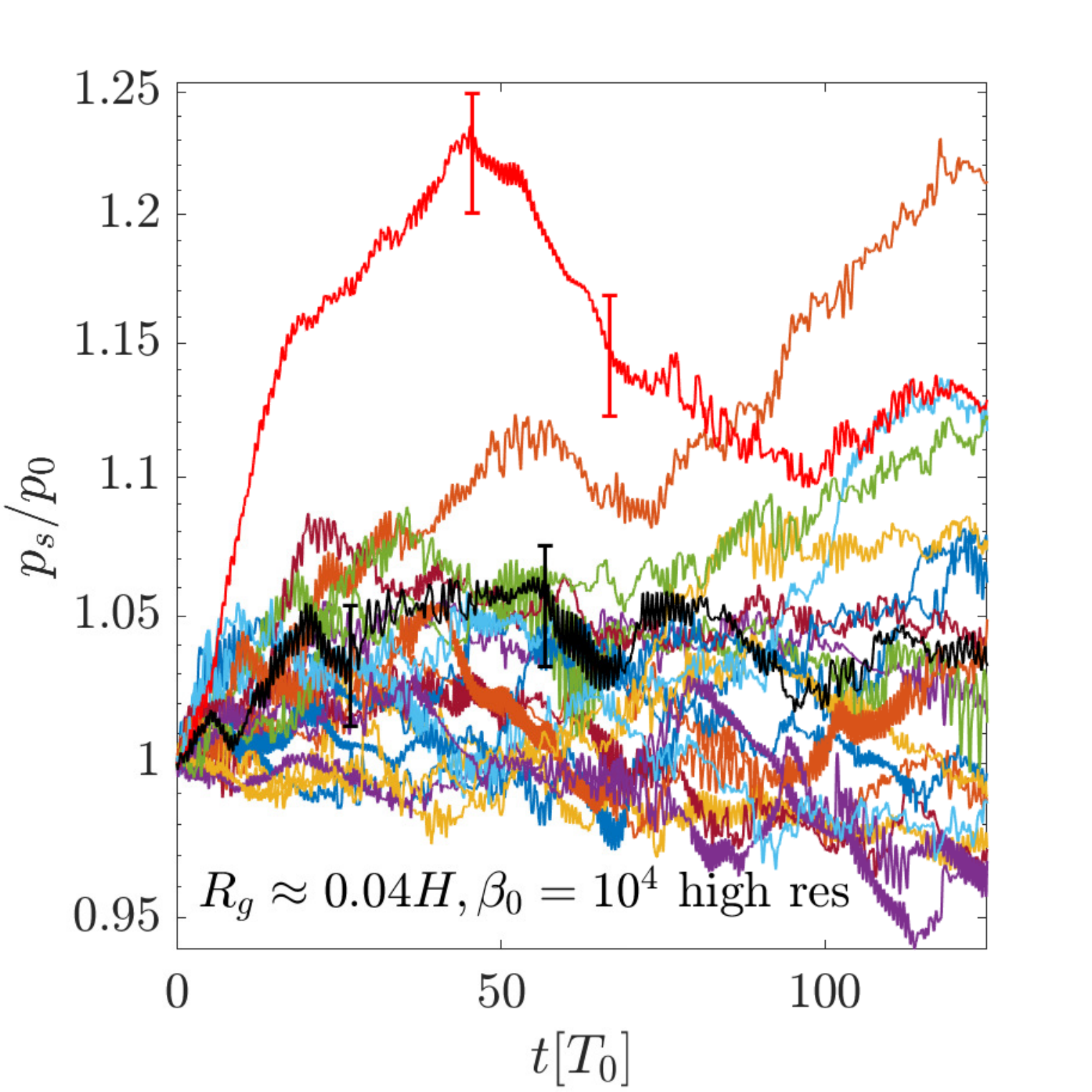}
    \label{fig::energy_hist_0.04_b1e4-hires}}
    \subfloat[]{
    \includegraphics[width=21em]{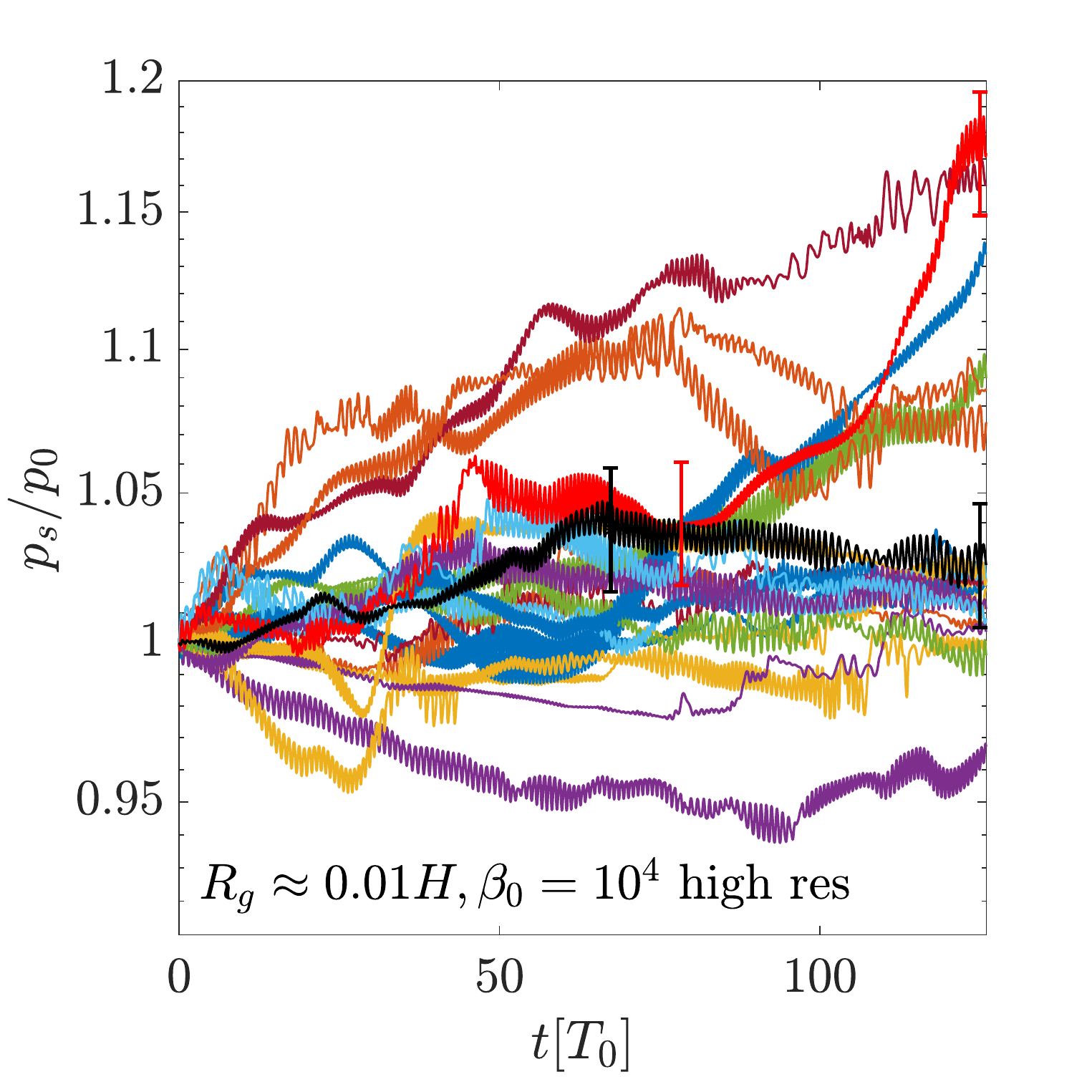}
    \label{fig::energy_hist_0.01_b1e4-hires}}
    \caption{Momentum magnitude histories of representative CR particles in a B4-hires snapshot from beginning to the end (125 gyro-period). In each panel, twenty particles whose gyro radii are all close to $0.1H$ (left), $0.04H$ (middle) and $0.01H$ (right), indicated by different colors. The momentum $p_s$ is defined in the shear frame (see Section~\ref{sec::result_mri}). The lines with error-bar-like markers (in red or black) in the middle and right panels will be further analyzed in Figure~\ref{fig::energy_hist_anomalous} and Figure~\ref{fig::energy_hist_norm} in Section~\ref{sec::direct_acc}.}
    \label{fig::energy_history}
\end{figure*}

In Figure \ref{fig::diffusion_conf_with_p_rel}, we show the measured diffusion coefficients $D_{\rm conf}$ for all particle simulations. 
They are shown as a function of the mean gyro radius, and averaged over all particles and from different MRI simulation snapshots (denoted as $\langle R_g\rangle_\text{par, t}$). It can be slightly different from $R_{g0}$ as particle mean gyro-radii slowly evolve with time. 
The error bars are estimated from the standard error among diffusion coefficients measured from different snapshots.

For particles with $R_g\gtrsim0.03H$, the total $D_{\rm conf}\sim30 D_{\rm Bohm}$, whereas in the poloidal plane, particles have much shorter mean free path, with $D_{\rm conf, pol}\sim 5D_{\rm Bhom}$. 
For particles with smaller gyro radii ($R_g\lesssim0.03H$), we see that their diffusion coefficients are generally larger,
close to a $D_{\rm conf}/D_{\rm Bohm}\sim R_{g}^{-1}$ scaling, or $D_{\rm conf}\sim{\rm const}$. We also see that in Bohm units, the results are largely independent of the choice of speed of light $c$.
These results are consistent with the fact that particles largely follow guiding-center motion along the field lines, and spatial diffusion is largely due to field line wandering \citep{1966ApJ...146..480J, 1974ApJ...193..231E}.

Despite very different levels of turbulence among different MRI simulations, the differences in spatial diffusion coefficients, when expressed as a function of $R_{g}$, are not as pronounced, typically within a factor of 2-3. Runs B2 and B3 generally show larger $D_{\rm conf}$, which likely reflects that magnetic field lines are more rigid in these runs despite yielding stronger turbulence. Moreover, run B4-hires also yields very similar diffusion coefficients to those in run B4,
an indication of convergence in the MRI properties. 

Results presented in this section are based on the MRI turbulence. We have also measured the coefficients in the ``shear-subtracted'' turbulence (See Section~\ref{sec::evolute_mom}). The resulting diffusion coefficients are almost identical to those in the MRI turbulence, being slightly larger by less than 10\%.

\section{Evolution in momentum space}
\label{sec::evolute_mom}

Generally, evolution of particle momentum in a turbulent flow is described by the Fokker-Planck advection-diffusion equation \citep{2004ApJ...610..550P}
\begin{equation}
    \frac{\partial f\left(p, t\right)}{\partial t} = - \frac{1}{p^2} \frac{\partial}{\partial p} \left[A(p) p^2 f \right] + \frac{1}{p^2} \frac{\partial}{\partial p} \left[p^2 D\left(p\right)\frac{\partial f}{\partial p}\right], \label{equ::mom_FP}
\end{equation}
where the momentum distribution function $f(\mathbf{p}, t)$ is considered uniform and isotropic (which we confirm from our simulations). The first term on the right hand side is the advection term, describing direct acceleration/deceleration with coefficient $A(p)$, and the second term describes momentum diffusion with coefficient $D(p)$. With an isotropic distribution function, we define
$n\left(p,t\right)\equiv 4 \pi p^2 f\left(p,t\right)$, and Equation~\ref{equ::mom_FP} becomes\footnote{We note this is different from used by \cite{2020ApJ...893L...7W}, where $A$ in their definition corresponds to $A + 2D/p$ in this paper.}:
\begin{equation}
    \frac{\partial n}{\partial t} = \frac{\partial}{\partial p} \left(D \frac{\partial n}{\partial p}\right) - \frac{\partial}{\partial p} \left[\left(A + \frac{2D}{p} \right) n \right]. \label{equ::final_FP}
\end{equation}

In our numerical setup, particles initially share the same momentum $p_0$, with $n$ being a $\delta$-function around $p=p_0$. As long as particle momenta do not evolve substantially, we can regard $A$ and $D$ as constants in each particle run, which can be obtained from linear fitting the mean $\langle p \rangle_\text{par}$ and variance $\left(\langle p^2 \rangle_\text{par} - \langle p \rangle^2_\text{par}\right)$ of momenta as a function of time, where we expect that over a time interval $\Delta t$
\begin{align}
    \frac{\Delta \left(\langle p^2 \rangle_\text{par} - \langle p \rangle^2_\text{par}\right)}{2 \Delta t} &= D \left(p_0\right), \\
    \frac{\Delta \langle p \rangle_\text{par}}{ \Delta t} - \frac{D}{p}\left(2 + \frac{\partial \ln D}{\partial \ln p}\right)  &= A\left(p_0\right).
\end{align}
For the same reason as in the previous section, our fitting starts from time $t_{\rm begin} = 75T_0$ and, instead of using initial $p_0$, we consider the measured results apply to $p=\langle p\rangle_{{\rm par},t}$, averaged across all the particles from $t_{\rm begin}$ to the end of the simulation. Note that particle momentum used here are transformed to the frame of local Keplerian flow, described in Section \ref{sec::result_mri}.

\begin{figure}
    \centering
    \includegraphics[width=28em]{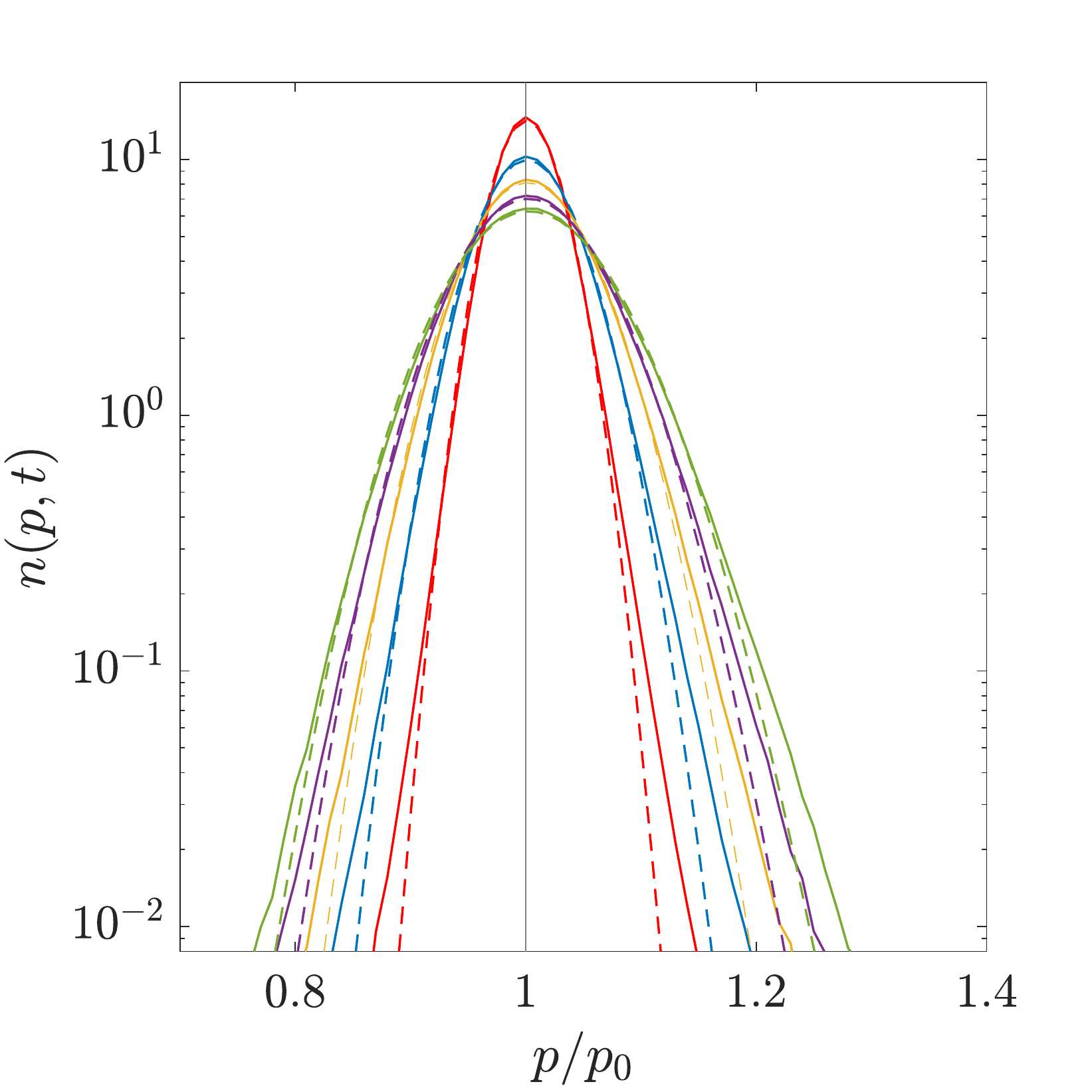}\\
    \includegraphics[width=28em]{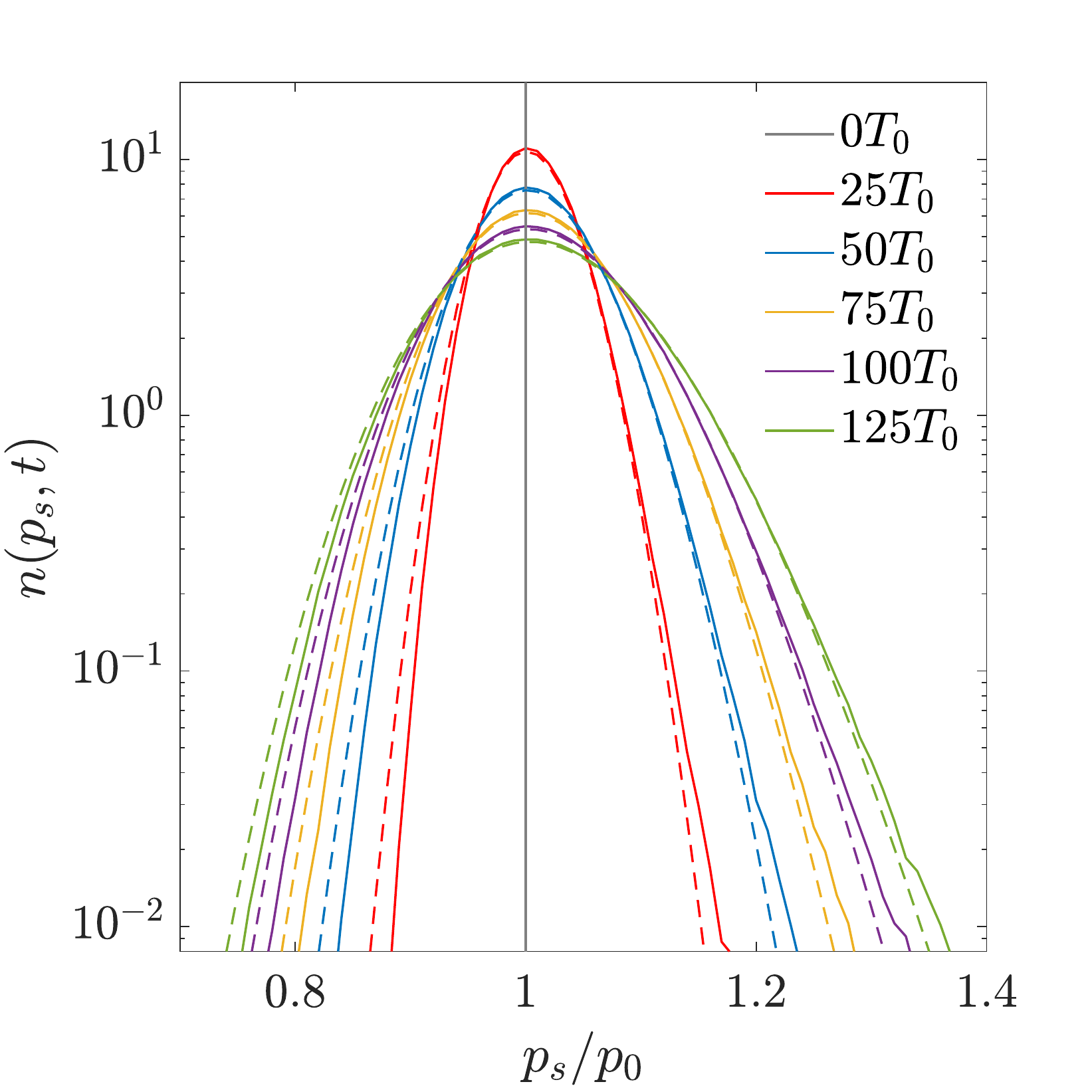}
    \caption{Evolution of momentum distributions for relativistic particles in run B4-hires. The initial gyro radius is close to $0.1H$. Different times are represented by different colors, as indicated by the legend, in units of the mean gyro period $T_0$. The upper panel is for particles in the ``shear-subtracted" turbulence and the lower panel is for particles in full MRI turbulence. Solid lines are simulation data while dashed lines correspond to numerical solutions of Equation~\ref{equ::final_FP}, with $A_\text{Fermi}=A_\text{shear}=0$, $D_\text{Fermi} / p_0^2 = 1.62 \times 10^{-5} (p/p_0) T_0^{-1}$, $D_\text{shear} / p_0^2 = 1.15 \times 10^{-5} (p/p_0)^3 T_0^{-1}$ (See Figure~\ref{fig::coefficient_rel} and Figure~\ref{fig::coefficient_mri_rel})}.
    \label{fig::energy_distribution}
\end{figure}

As an example, in Figure~\ref{fig::energy_history}, we show energy histories of several particles randomly chosen from our simulations. Here, particle energy (equivalent to momentum) is measured in the shear frame, given in Equation~\ref{eq:pshear} later in this section. In general, particles stochastically gain or lose energy, exhibiting random-walk behaviors as expected, particularly for particles with $R_g\approx0.1H$. 
Gathering all particles, we show in Figure~\ref{fig::energy_distribution} how the energy distribution $n(p)$ evolves with time with initial $R_{g}\approx 0.1H$ in run B4-hires. The width of the distribution function increases with time, which we find to be approximately as $\sqrt{t}$, consistent with diffusive transport.
We further show the numerical solution of Equation~\ref{equ::final_FP}, with the coefficients measured in the following subsections. The results reasonably match simulation results, especially near the central region of the distribution function. There are small deviations in the large and small ends of the distribution function, which are more prominent in the shear-subtracted case (see definition below), though more in-depth study of such deviations is beyond the scope of this work.

For particles with smaller $R_g$, while most of them similarly exhibit random-walk behaviors, some may occasionally experience more systematic acceleration or deceleration over tens of orbital periods and lead to further spread, as seen in Figure~\ref{fig::energy_hist_0.04_b1e4-hires} and Figure~\ref{fig::energy_hist_0.01_b1e4-hires}, which we will discuss in more detail in Section \ref{sec::direct_acc}. Moreover, as these particles largely follow guiding-center motion, their energy tends to oscillate over gyro-time, as seen in Figure~\ref{fig::energy_hist_0.01_b1e4-hires}. This is due to non-zero inductive electric field when energy is not measured in the co-moving frame with the gas. While this can be mitigated by procedures as described in \citet{2020ApJ...893L...7W}, we choose to stay simple especially that we have already done one frame transformation (to the shear frame). Moreover, the oscillation amplitude is on the order of $\delta v/c\approx c_s/c\ll1$ and barely affects our results. 

To disentangle the shear acceleration and the second-order Fermi acceleration in the Fokker-Planck equation, for each particle simulations that we have described, we perform an additional particle simulation where we subtract the background flow velocity from the velocity field, i.e., use $\mathbf{u}'$ instead of $\mathbf{u}$ to construct the electric field (see Equation ~\ref{eq:frozen2}). With this "shear-subtracted" turbulence, there is no background shear, and particle acceleration is expected to proceed as in the standard second-order Fermi process, and we denote the corresponding coefficients as $A_{\rm Fermi}(p)$ and $D_{\rm Fermi}(p)$. Coefficients obtained from the original simulations (with background shear included, full MRI turbulence) are denoted as $A_{\rm MRI}$ and $D_{\rm MRI}$. From the same example in Figure \ref{fig::energy_distribution}, we can see that the width of the distribution function is narrower in shear-subtracted turbulence (dashed lines), implying additional contribution shear acceleration plays a substantial role in energizing such particles.

In Figure~\ref{fig::coefficient_rel}, we show results for measured coefficients of $A(p)$ and $D(p)$ for different particle momenta (in terms of $\langle R_{g} \rangle_\text{par,t}$) from all simulation runs for both shear subtracted and full MRI turbulence. The coefficients are normalized as $D / \langle p \rangle^2_\text{par, t}$ and $A / \langle p \rangle_\text{par, t}$, which have a unit of time$^{-1}$, expressed in $\Omega_0$. Thus, these normalized coefficients represent the characteristic rate at which particles get accelerated or diffuse in momentum space. These results are discussed in more detail in the following subsections.

\subsection{Results in the ``shear-subtracted" turbulence}
\label{sec::result_tur}

\begin{figure*}
    \centering
    \subfloat[]{
    \includegraphics[width=28em]{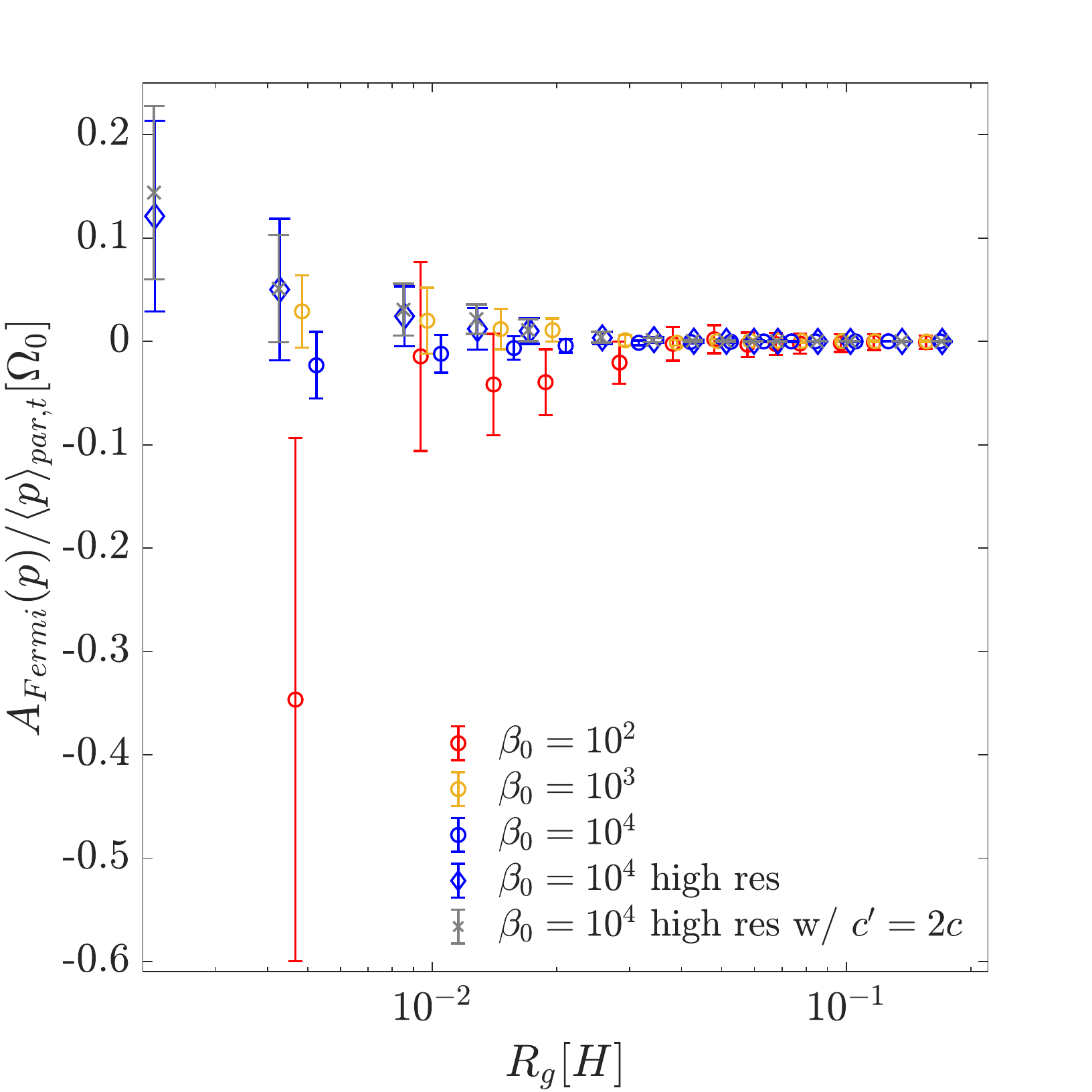}
    \label{fig::A_rel_tur}}
    \subfloat[]{
    \includegraphics[width=28em]{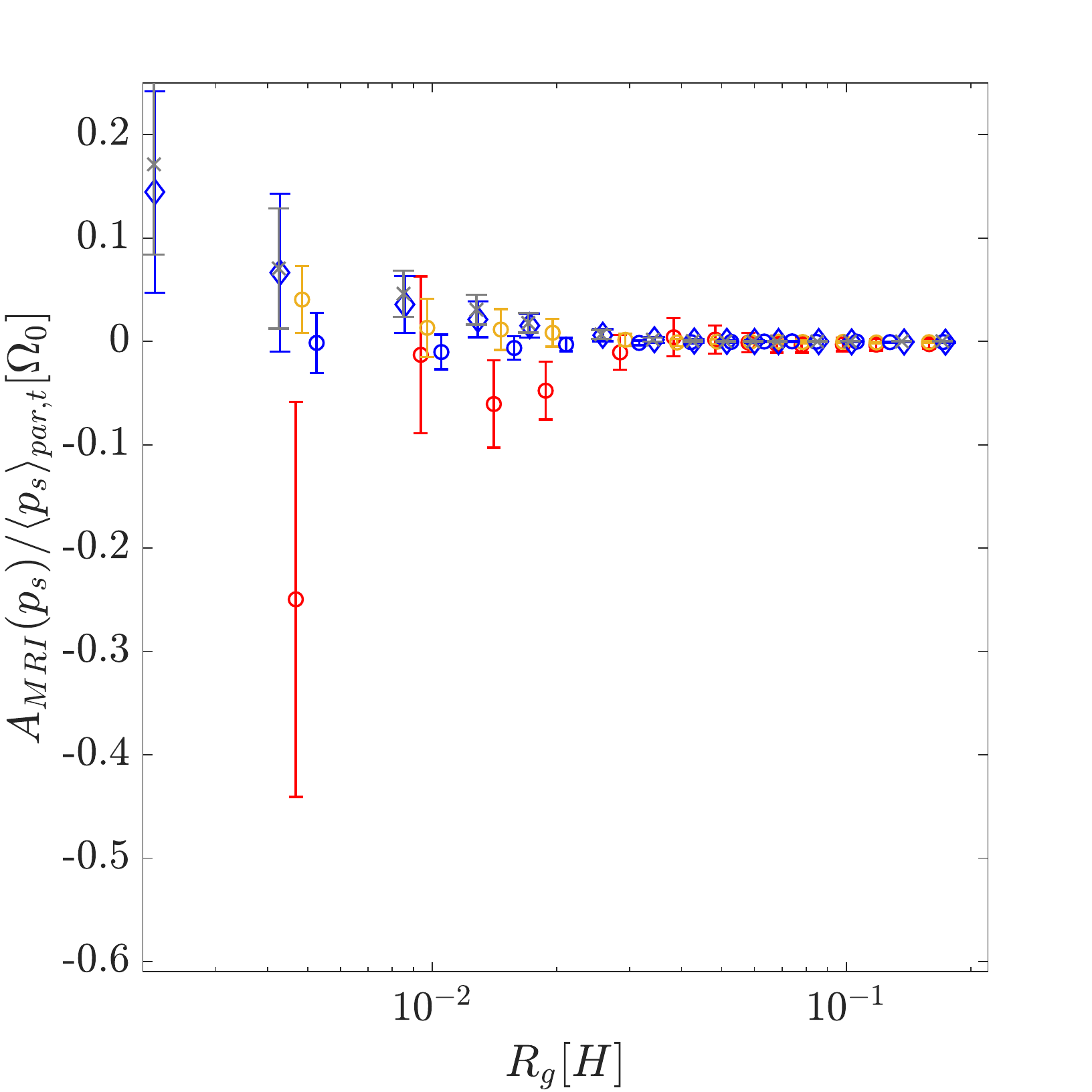}
    \label{fig::A_rel_mri}} \\
    \subfloat[]{
    \includegraphics[width=28em]{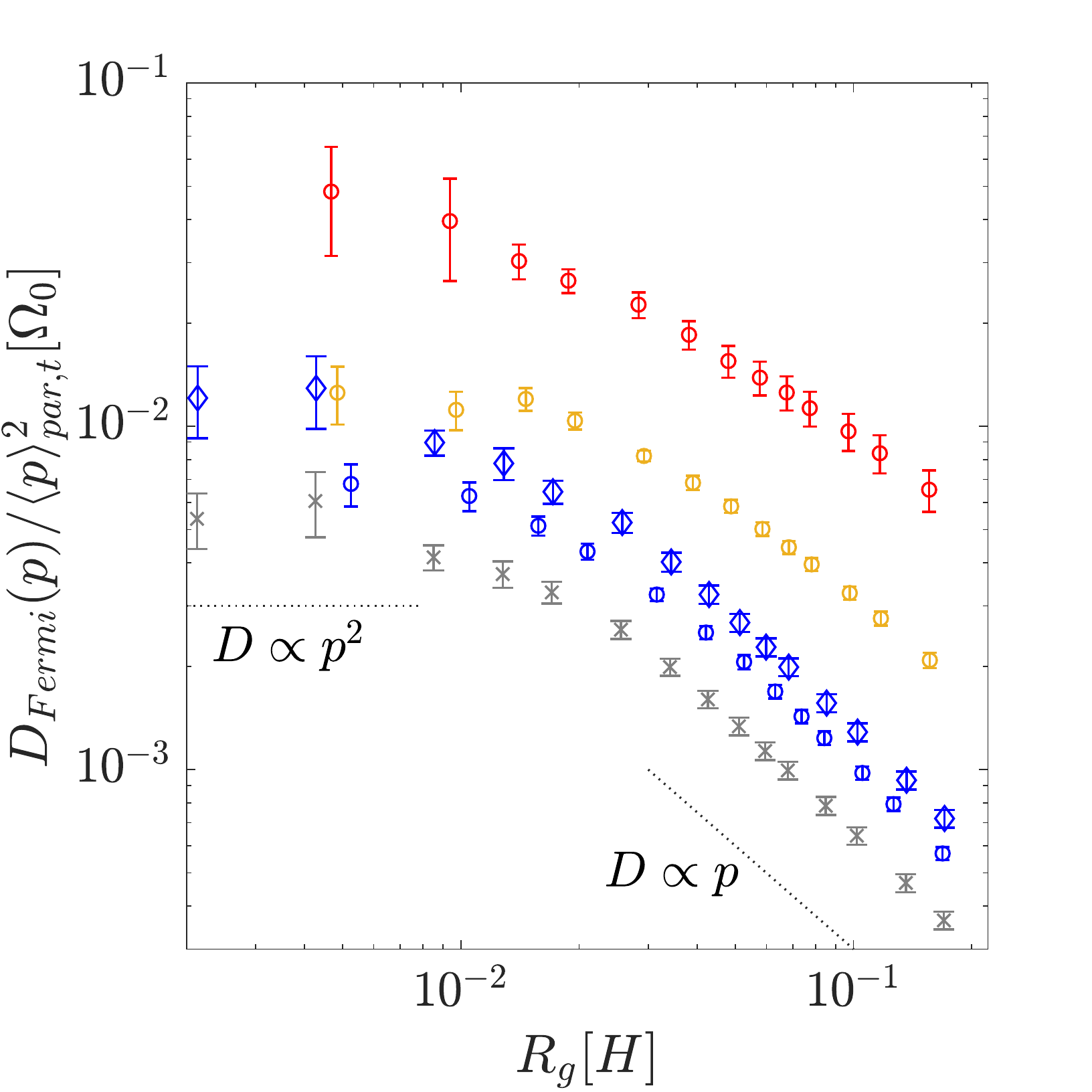}
    \label{fig::D_rel_tur}}
    \subfloat[]{
    \includegraphics[width=28em]{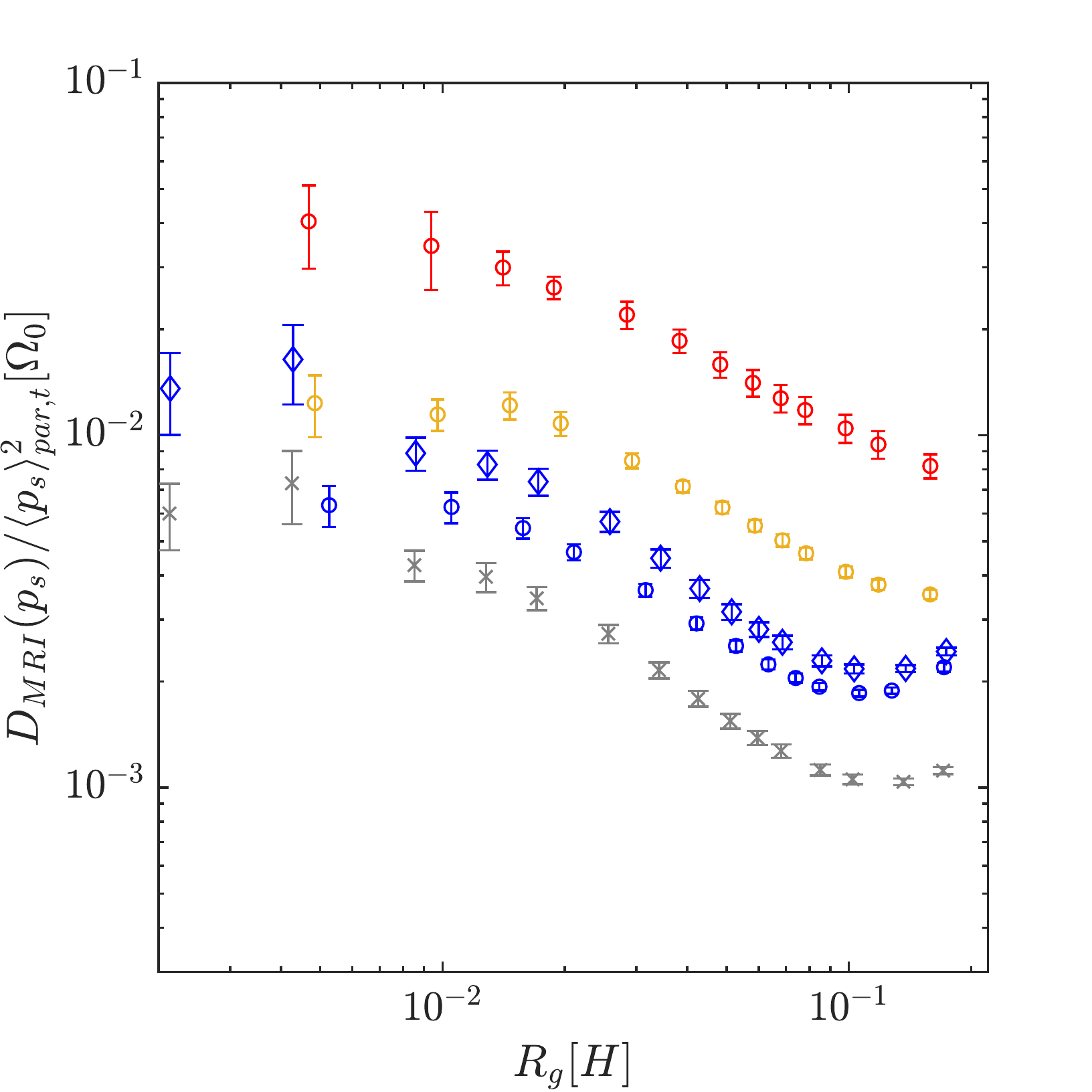}
    \label{fig::D_rel_mri}}
    \caption{The direct acceleration coefficients $A(p)$ (upper) and the diffusion coefficients $D(p)$ (lower) for relativistic particles in shear-subtracted turbulence (left) and full MRI turbulence (right). They are shown as a function of particle gyro radius, $\langle R_{g} \rangle_\text{par,t}$ (momentum). Color stands for $\beta_0$ of background MHD simulations, while symbol shape represents resolution or numerical speed of light (see legend). Error bars are estimated from the standard deviation among results obtained from different MRI simulation snapshots.}
    \label{fig::coefficient_rel}
\end{figure*}

We first study the diffusion coefficient $D_{\rm Fermi}(p)$ in shear-subtracted turbulence. In Figure~\ref{fig::D_rel_tur}, we see that for $R_g \gtrsim 0.03H$, diffusion coefficient roughly scales as $D_{\rm Fermi} \propto p$, or as shown in the plot, that the mean rate of momentum diffusion scales as $D/p^2\propto p^{-1}$. Increasing the speed of light by a factor of 2 reduces $D_{\rm Fermi}$ by about the same factor in our code units. In this case, we can write $D/p^2\approx C(H/R_g)(c_s/c)\Omega_0= C\omega(c_s/c)^2$, with $C$ being a constant.
A rough fitting yields $C$ being about $6 \times 10^{-3} $ for run B4 and B4-hires, $1.5 \times 10^{-2}$ for run B3, and $4 \times 10^{-2}$ for run B2.

This scaling is interesting in that the rate of momentum diffusion, and hence stochastic acceleration, is proportional to particle gyro-frequency. The resulting maximum particle energy particle would increase according to $d\ln p/dt\sim C\omega(c_s/c)^2\propto 1/p$, or that the maximum Lorentz factor increases linearly in time at the rate 
\begin{equation}
    \frac{d\gamma}{dt}\approx C\bigg(\frac{c_s}{c}\bigg)^2\omega_0\ ,\quad {\rm for}\ R_g\gtrsim0.03H\ .
\end{equation}
Despite that maximum particle energy only increases linearly in time, its overall rate is substantial, as $\omega_0\gg\omega$ for ultra-relativistic particles, and it becomes increasingly efficient in hotter environment (e.g., closer to the black hole in the disk).

The scaling of $D_{\rm Fermi}(p)/p^2$ with $p$ flattens towards smaller $R_g\lesssim0.03H$. In other words, diffusion rate is less than $C\omega(c_s/c)^2$ mentioned earlier. We note that our MRI simulations barely resolve particle orbits in this regime, and we may not be able to reliably measure the coefficients. 
If we tentatively assume the curves for small $R_g$ shown in Figure \ref{fig::D_rel_tur} is exactly flat, or $D_{\rm Fermi}\propto p^2$. This yields $d\ln p/dt\sim C'\Omega_0(c_s/c)$, or that particle energy grows exponentially as
\begin{equation}
    \frac{dp_{\rm max}}{dt}\sim \exp[C'\Omega_0t(c_s/c)]\ \quad {\rm for}\ R_g\lesssim0.03H\ ,
\end{equation}
where we may take $C'$ to be about 0.01 for runs B3 and B4, while it could reach $\sim0.05$ for run B2.
We note that the $D(p)\propto p^2$ behavior is also observed in recent kinetic simulations of relativistic turbulence \citep{2019ApJ...886..122C,2020ApJ...893L...7W}.
Despite maximum particle energy grows exponentially, it is measured in disk dynamical time $\Omega_0^{-1}$ (rather than gyro-time) with a relatively small coefficients. Therefore, the overall rate of acceleration is only modest.

At a given $R_g$, the diffusion rate in runs B3 and B2 are typically higher than those in run B4 by a factor around 1-3 and 5-10, respectively. Runs with different $\beta_0$ give different normalized diffusion coefficients, but the difference is not substantial when shown as a function of $R_g$ (instead of $p$) as we do in Figure \ref{fig::coefficient_rel}, especially between runs B3 and B4. If we draw $D_{\rm Fermi}$ directly as a function of $p$, because particles with the same $R_g$ are more energetic in lower-$\beta_0$ runs, contrast in $D_{\rm Fermi}$ for particles with same momentum will be much more significant between in lower-$\beta_0$ and higher-$\beta_0$ runs. 

Results for the other coefficient, $A_{\rm Fermi}$ measuring direct acceleration, are shown in Figure~\ref{fig::A_rel_tur}. We see that it approaches zero for $R_g \geq  0.03H$ in all runs. 
Towards smaller $R_g$, $A_{\rm Fermi}$ turns to a relatively large positive values in B3 and B4-hires, while tends to be negative in run B2 and B4 (where large or small is seen by comparing $A/p$ and $D/p^2$).
Besides, when we increases the numerical speed of light two times, $A_{\rm Fermi}$ is largely unchanged. As the coefficients deviate from zero, the error bar, obtained from comparing results different MRI simulation snapshots, also becomes large. There is additional uncertainty in evaluating $A_{\rm Fermi}$ as it involves evaluating the derivative $d\ln D_{\rm Fermi}/d\ln p$. The large uncertainty with potentially anomalous non-zero $A_{\rm Fermi}$ may be attributed to the intermittency in the MRI turbulence, which will be further discussed in Section~\ref{sec::direct_acc}. On the other hand, despite the deviation, $A(p)$ is still consistent with zero within $2\sigma$ limit in all cases. 

\subsection{Results in the MRI turbulence}
\label{sec::result_mri}

In full MRI turbulence, we first define particle momentum in the shear frame, $\mathbf{p}_s$, as also in \citet{2016ApJ...822...88K}, done by a Lorentz transformation according to background shearing motion
\begin{align}\label{eq:pshear}
    p_{s, y}&=\Gamma_s\left(p_y-\beta_s \varepsilon / c \right),\notag \\
    p_{s,x}&=p_x\ ,\qquad p_{s,z}=p_z\ ,\notag\\
    \varepsilon_s &=\Gamma_s\left(\varepsilon-\beta_s c p_y \right) ,\notag\\
	{\rm where}\quad \beta_{s}&=-\frac{q \Omega_0 x}{c},\quad
	\Gamma_{s}=(1-\beta_{s}^2)^{-1/2}\ .
\end{align}
The transformation is compatible with the shearing-perordic boundary conditions (Equation~\ref{equ:shear_bound_par}), and ensures that particles can smoothly cross radial boundaries without artificial accelerations.

From our particle simulations, we first confirm that momentum in the shear frame remains isotropic in $f(\mathbf{p_s})$, consistent with \citet{2016ApJ...822...88K}. Therefore, we can apply the same Equation~\ref{equ::final_FP} with $p$ replaced by $p_s$. 
We have already seen in Figure~\ref{fig::energy_distribution} that for $R_g \approx 0.1$ particles, $n(p_s)$ in the MRI case (solid lines) is broader due to additional contribution from shear. In the right panels of Figure \ref{fig::coefficient_rel}, we show more detailed measurements of the acceleration and momentum diffusion coefficients.
Overall, the measured direct acceleration coefficients remain similar to those in the shear subtracted counterparts, consistent with no contribution from shear. Below we discuss the momentum diffusion coefficients.

In runs B4 and B4-hires, particles with $R_g\lesssim0.05H$ show largely identical momentum diffusion coefficients with those in the shear-subtracted cases. Beyond this range, another component rises steeply, and eventually takes over at $R_g\gtrsim0.1H$. This corresponds to additional contribution from shear acceleration, 
which we will discuss in detail in Section~\ref{sec::shear}. 
For run B3, we see that shear acceleration starts to take over at larger $R_g$. For run B2, the larger value of $D_{\rm MRI}$ than the counterpart $D_{\rm Fermi}$ again suggests shear acceleration also operates.

\section{Physics of Stochastic Acceleration}
\label{sec::stochastic_acc}

In this section, we discuss the physics behind the results obtained in the previous sections, focusing on the acceleration mechanisms.

\subsection{The second-order Fermi acceleration}
\label{sec::2nd_Fermi}

In the absence of shear, and phenomenologically, the momentum diffusion coefficient in isotropic turbulence for second-order Fermi acceleration can be written as \citep{1987PhR...154....1B}
\begin{equation}
	\frac{D_{\rm Fermi}\left(p\right)}{p^2} \approx \frac{\langle u'^2 \rangle}{3vL_{\rm scatter}}\ , \label{equ::diff_fermi_origin}
\end{equation}
where $\langle\delta u'^2\rangle$ represents turbulent velocity squared, $v=c$ for relativistic particles, and $L_{\rm scatter}$ is the scattering mean free path in the turbulence. 
On the other hand, analogous to the concept of Bohm diffusion, we can write the diffusion coefficients in configuration space as
\begin{equation}
	D_{\rm conf}(p) \approx \frac{v L_{\rm scatter}}{3}\ .
\end{equation}
We have seen in Figure~\ref{fig::coefficient_rel} that for relatively large $R_g \gtrsim 0.03H$, $D_{\rm Fermi}/p^2\propto p^{-1}$, suggesting that $L_{\rm scatter}\propto p$. This is consistent with the results from configuration diffusion (Figure~\ref{fig::diffusion_conf_with_p_rel}), where in the same parameter range $L_{\rm scatter}\propto R_g\propto p$. At smaller $R_g$, our results from configuration space indicate that $L_{\rm scatter}$ becomes larger than expected from the same proportionality, as turbulent energy at such scales diminishes. This is also reflected in momentum space diffusion, as $D_{\rm Fermi}$ is smaller than extension from the power law at large $R_g$. In fact, this scaling $D_{\rm Fermi}\propto p^2$ is also consistent with
the quasi-linear theory from Goldreich-Sridhar spectrum of fluctuations \citep{2000PhRvL..85.4656C, 2020PhRvD.102b3003D}.
Combining the two equations above, we arrive at
\begin{equation}
	\frac{D_{\rm Fermi} D_{\rm conf}}{p^2} \approx \frac{1}{9} \langle u'^2 \rangle. \label{equ::norm_diff_Fermi}
\end{equation}
This is a handy relation that is independent of $L_{\rm scatter}$ and $c$. 

\begin{figure}
    \centering
    \includegraphics[width=28em]{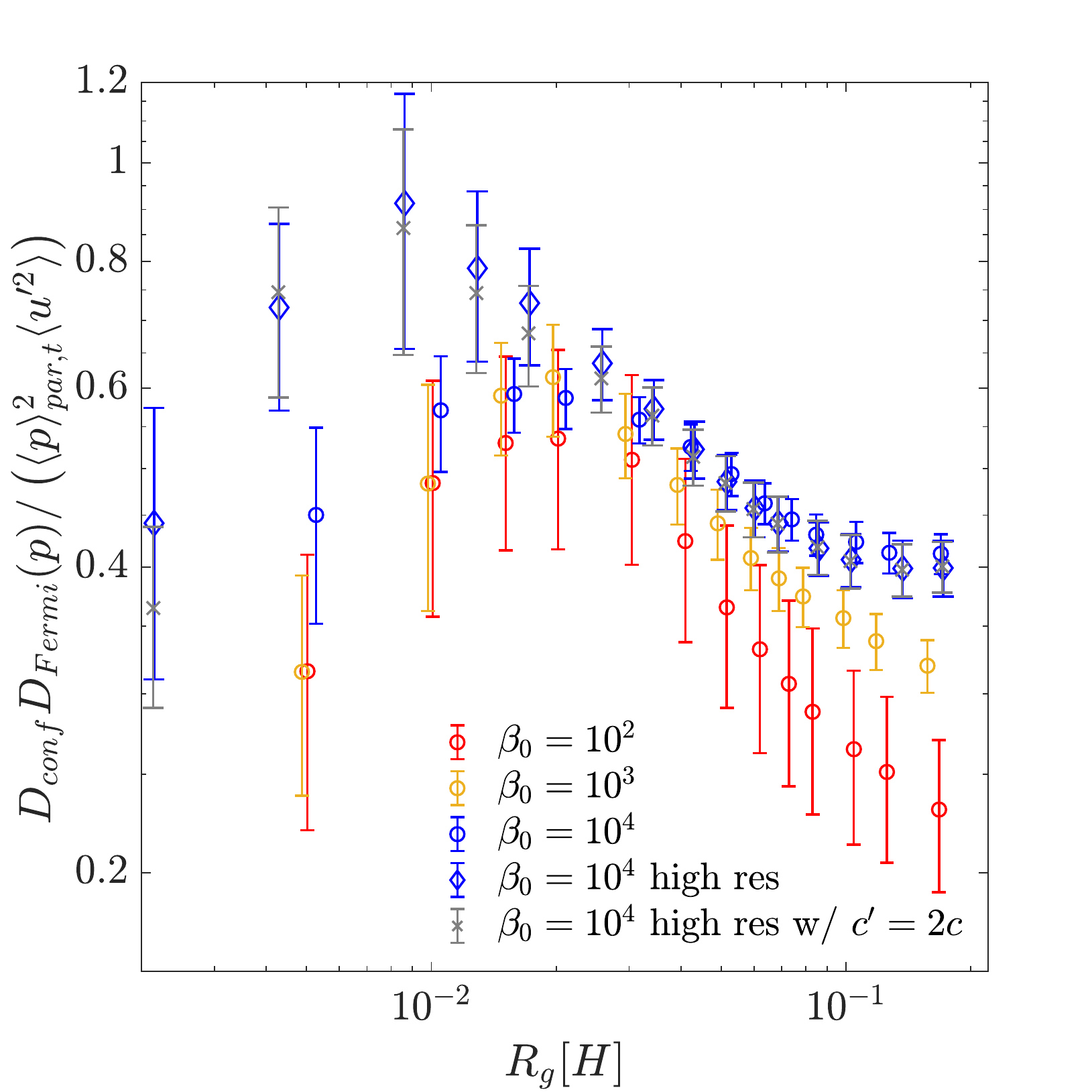}\\
    \includegraphics[width=28em]{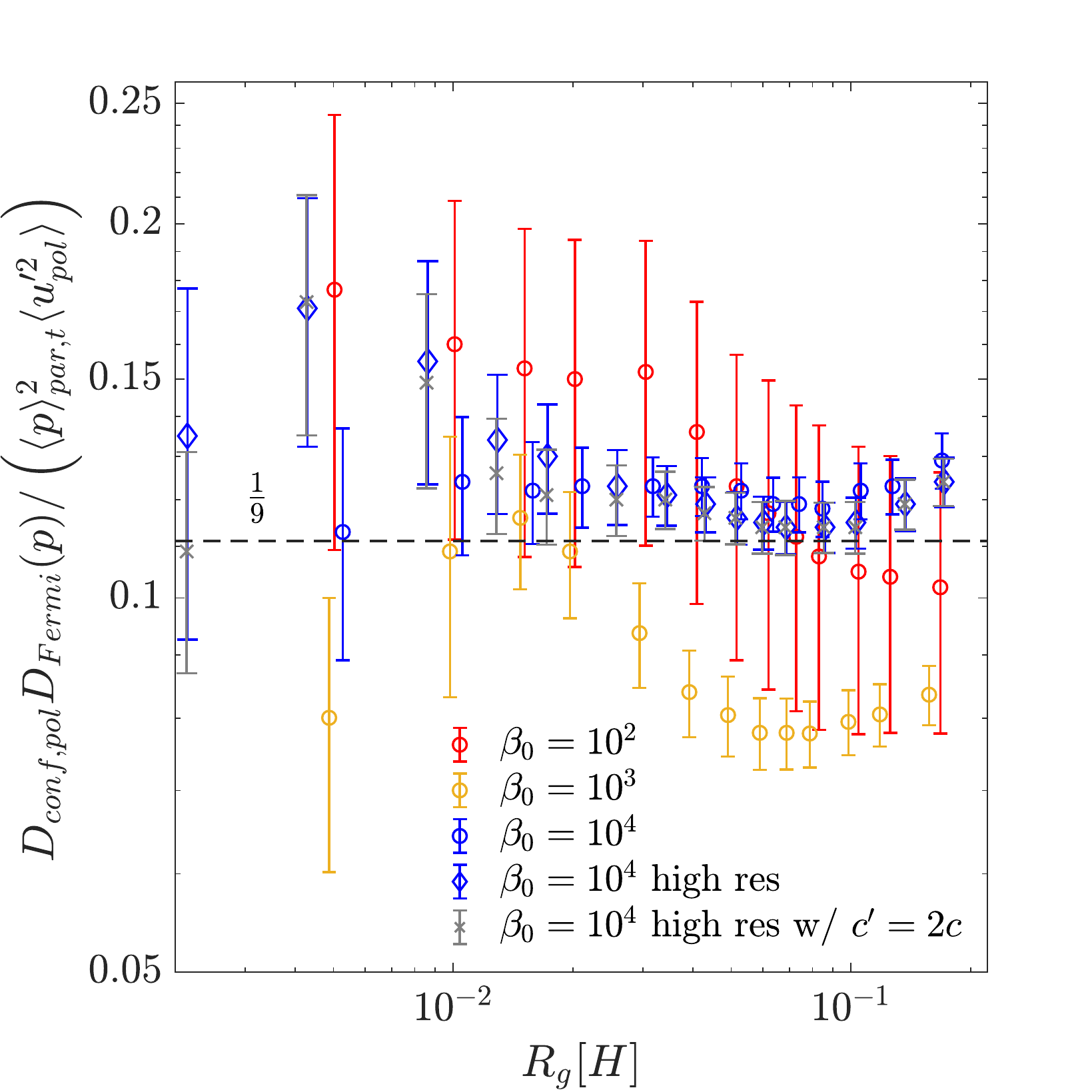}
    \caption{The normalization for diffusion coefficients $D_{\rm conf}D_{\rm Fermi}/(p^2\langle u'^2\rangle)$ (upper) and $D_{\rm conf, pol}D_{\rm Fermi}/(p^2\langle u'^2_\text{pol}\rangle)$ (lower) for relativistic particles in shear-subtracted turbulence, to test the validity of Equation~\ref{equ::norm_diff_Fermi}. Results are shown as a function of $\langle R_{g} \rangle_\text{par,t}$ (momentum).  Different colors stand for the $\beta_0$ value of the MRI simulations, and the shapes indicate different resolution and numerical speed of light.}
    \label{fig::D_norm_tur_rel}
\end{figure}

To verify to what extent it is valid, we show in Figure~\ref{fig::D_norm_tur_rel} the combination of $D_{\rm conf} D_{\rm Fermi} / \left( p^2 \langle u'^2\rangle\right)$ for the shear-subtracted turbulence. For consistency, in this subsection we applied $D_{\rm conf}$ measured in the ``shear-subtracted'' turbulence.
We see that all results are larger than the predicted value $1/9$ by a factor of up to 10. We attribute this to the anisotropic nature of the MRI turbulence. We have already seen that particle mean free path is much longer in the toroidal ($\hat{y}$) direction ($D_{{\rm conf},y}$ being dominant). In fact, as also shown in Figure~\ref{fig::D_norm_tur_rel}, if we replace $D_{\rm conf}$ by its poloidal component $D_{\rm conf, pol}$, and replace $\langle u'^2\rangle$ by $\langle {u'^2_{\rm pol}}\rangle^2$, then the results show good agreement with expectation within uncertainties. There is also some dependencies on the turbulence structure (different runs) and $R_g$, but such dependencies are weak and deviations are generally within a factor of 2. Overall, this result suggests that momentum diffusion in the MRI turbulence is largely due to scattering in the poloidal plane. Based on this result, we can define
\begin{equation}
    L_{\rm scatter, pol}=R_g\frac{D_{\rm conf, pol}}{D_{\rm Bohm}}\ ,
\end{equation}
which can be estimated from Figure \ref{fig::diffusion_conf_with_p_rel}, and it can serve as a convenient indicator to characterize the efficiency of diffusion in both configuration and momentum spaces.

\subsection{Acceleration of particles with small $R_g$}
\label{sec::direct_acc}

Our measurements of $D_{\rm Fermi}$ and $D_{\rm MRI}$ bare large uncertainties for particles with smaller $R_g\lesssim0.03H$, and sometimes exhibit nonzero direct acceleration coefficient (despite still compatible to zero). These particles, as seen in Figure~\ref{fig::traj_mri}, primarily follow guiding center motion.
Hereby we focus on the momentum evolution of these particles, discuss the mechanisms behind their statistical acceleration and deceleration, and seek for clues behind the uncertainties.
We have also seen from Figure~\ref{fig::coefficient_rel} that for particles with such small radii, effect of shear is largely negligible. Therefore, in our analysis below, we neglect the contribution from the Keplerian shear flows for better transparency. The role of shear will be discussed in the next subsection.

\begin{figure*}
    \centering
    \includegraphics[width=18em]{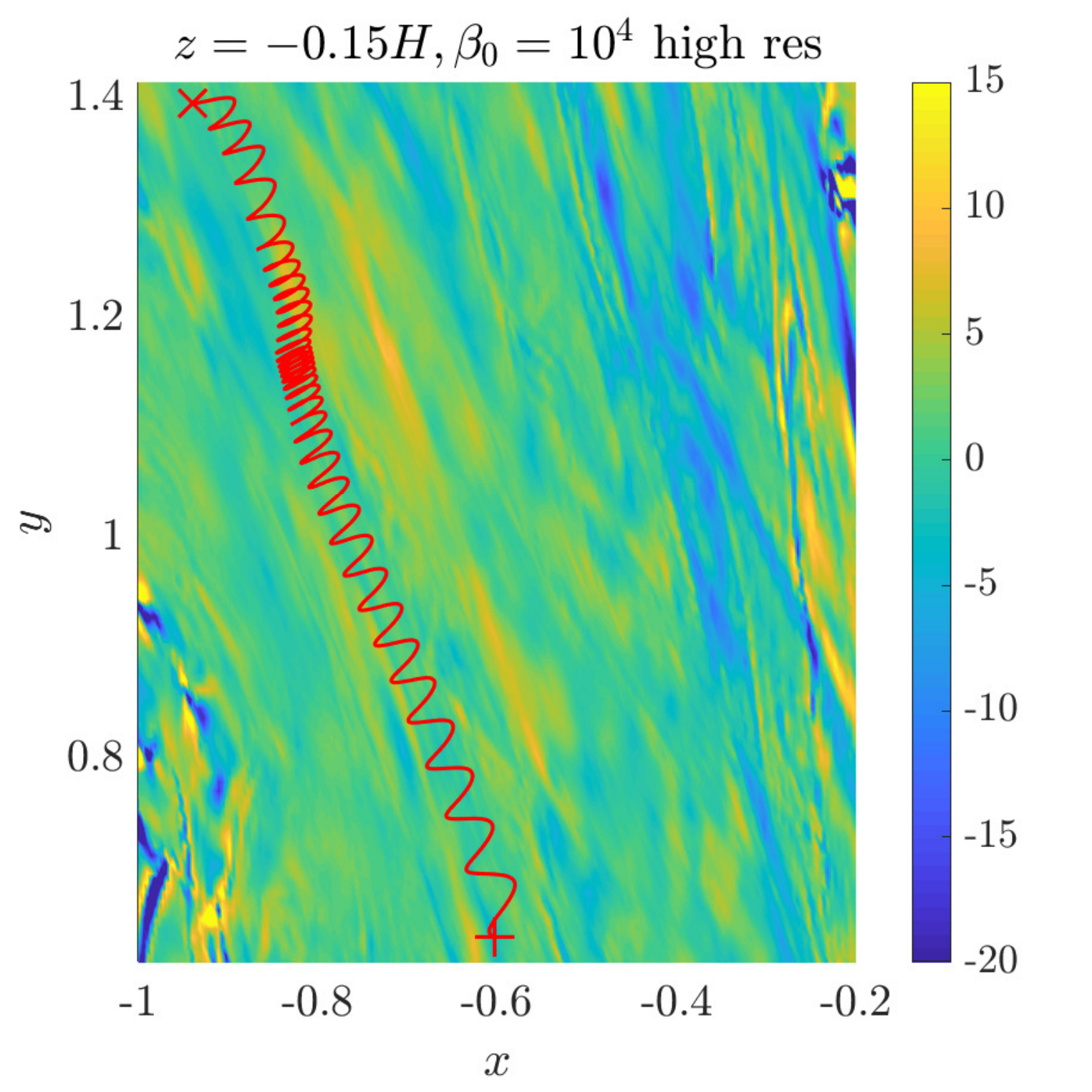}
    \includegraphics[width=18em]{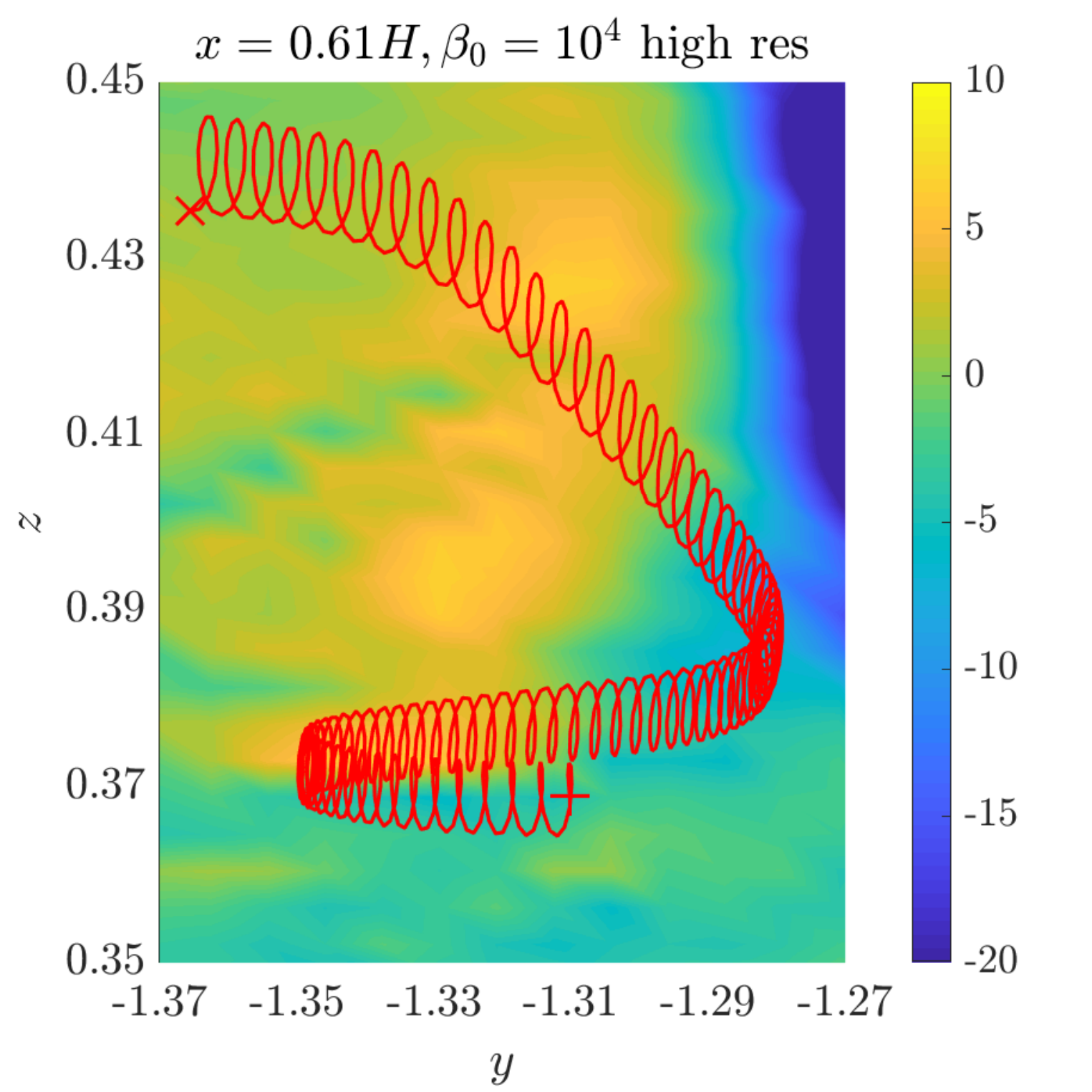}
    \includegraphics[width=18em]{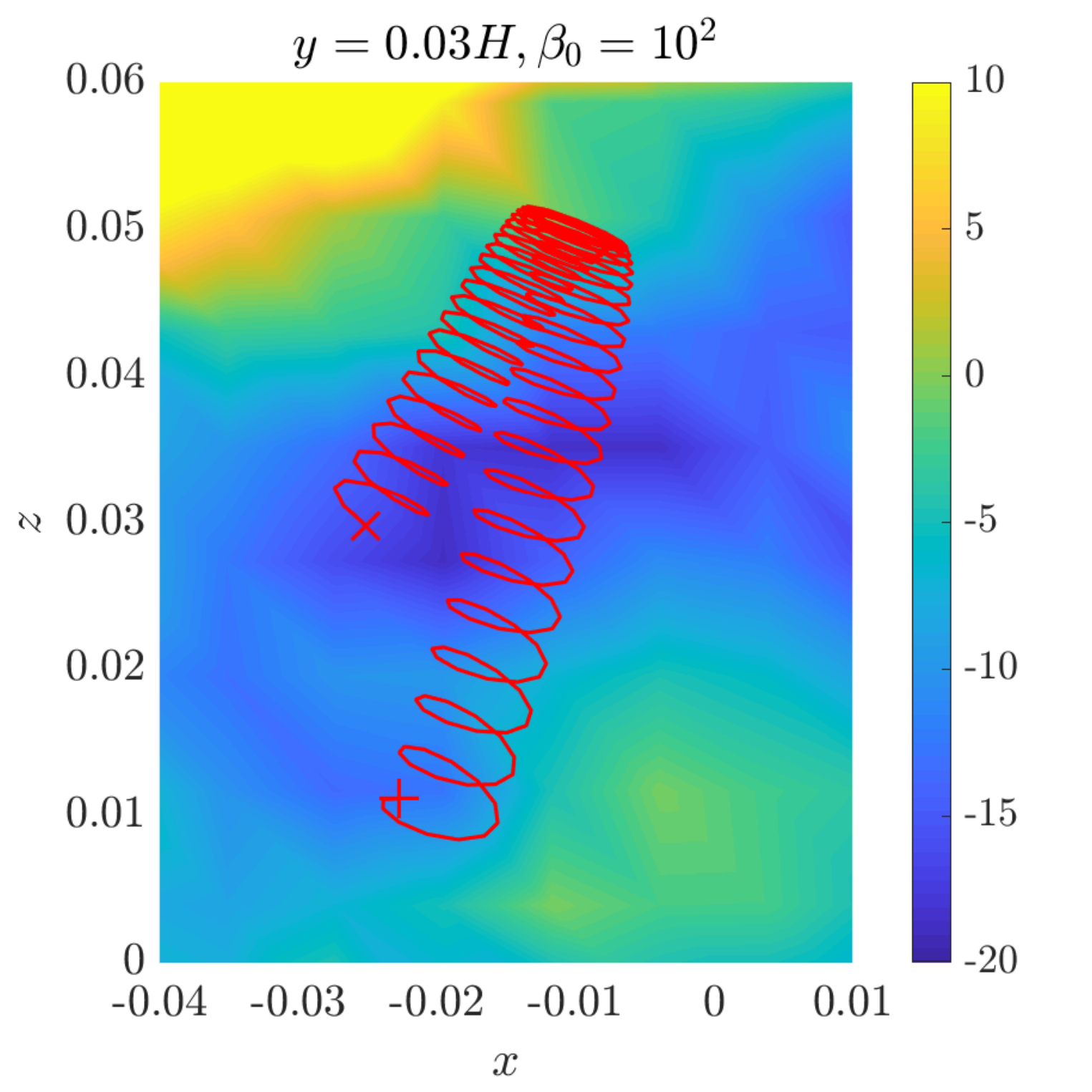}\\
    \subfloat[]{
    \includegraphics[width=21em]{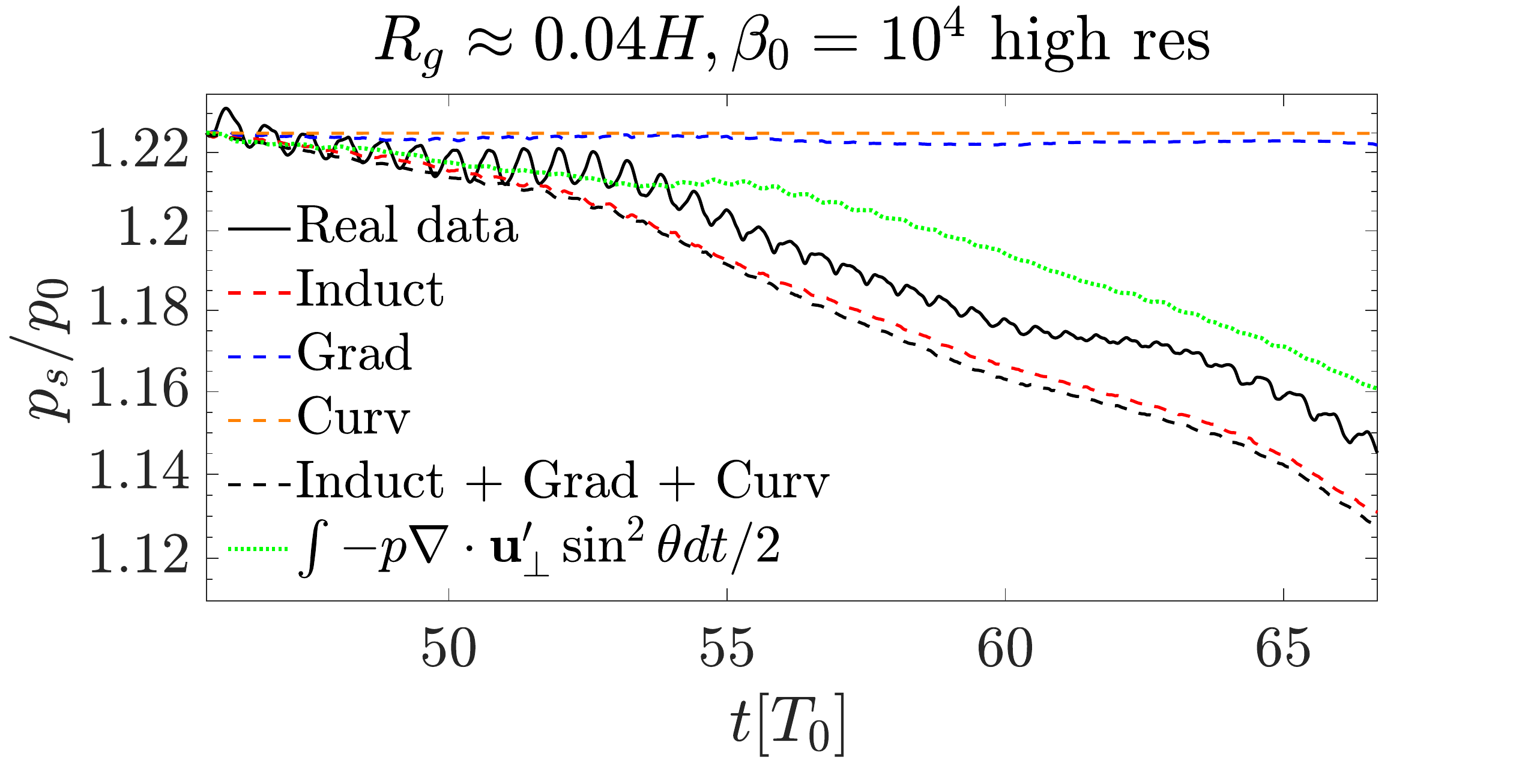}
    \label{fig::anomalous_1}}
    \subfloat[]{
    \includegraphics[width=21em]{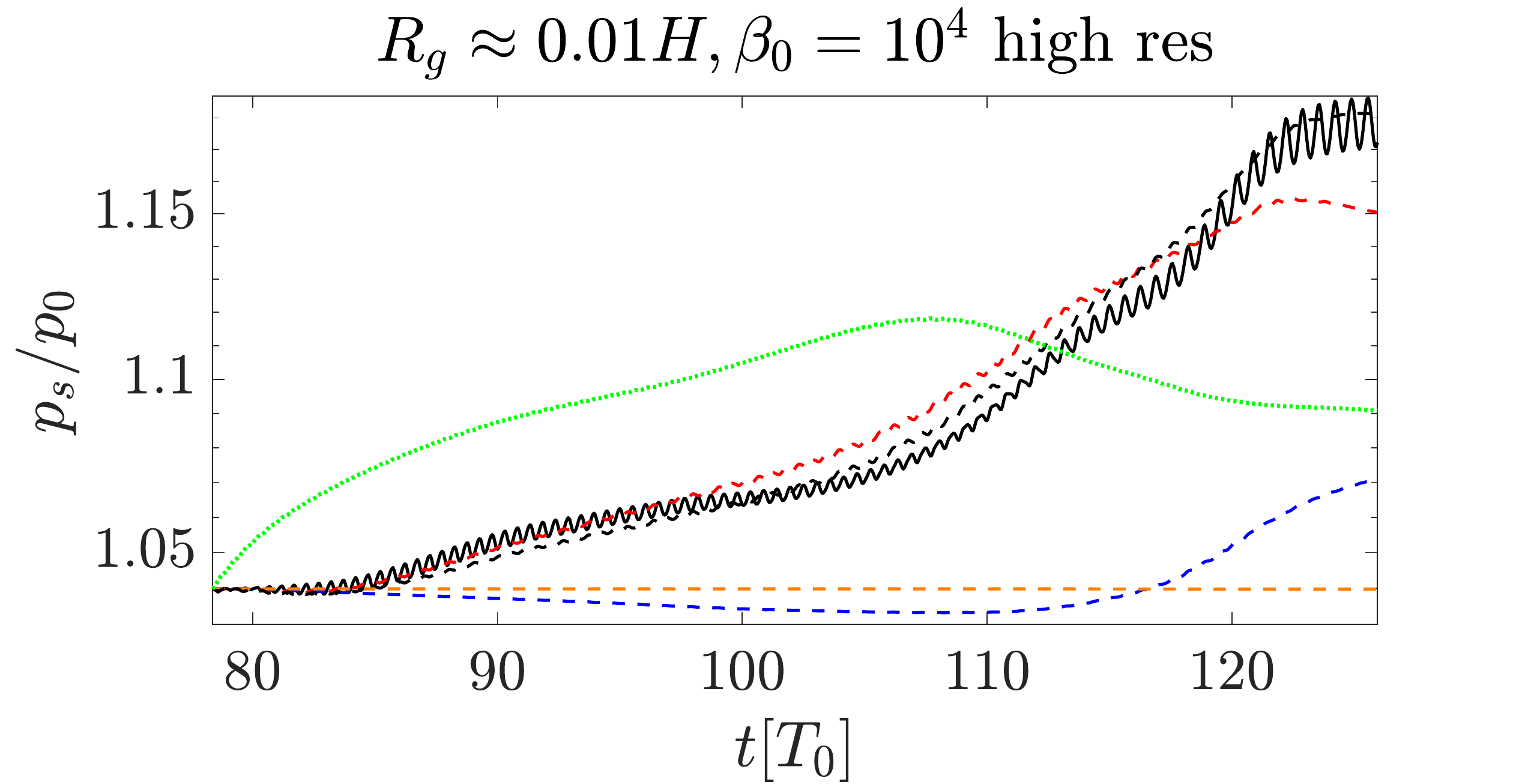}
    \label{fig::anomalous_2}}
    \subfloat[]{
    \includegraphics[width=21em]{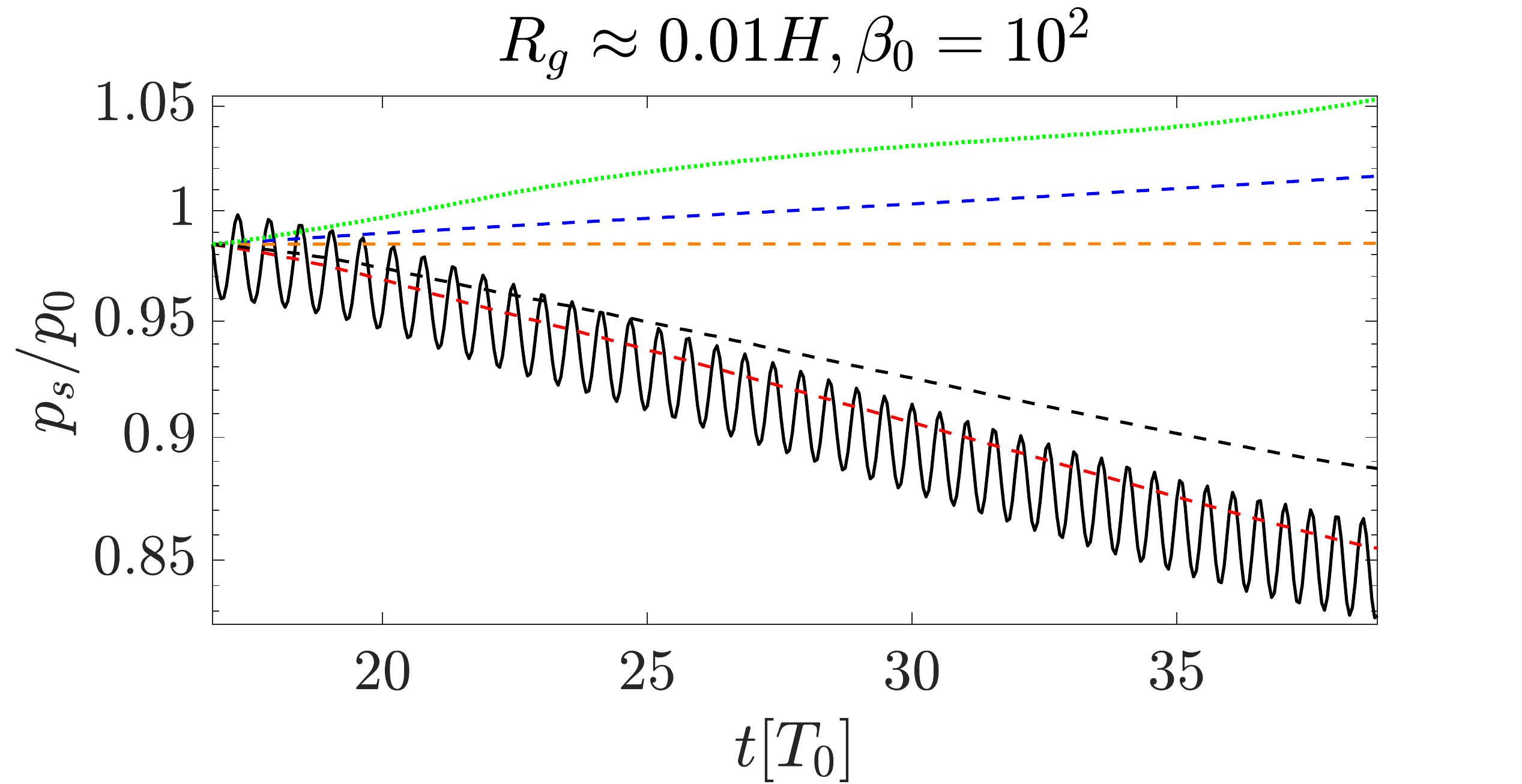}
    \label{fig::anomalous_3}}
    \caption{Three representative particles that show systematic acceleration/deceleration. The top panels are these particles' trajectories (red lines) with the $\nabla \cdot \mathbf{u}'_\perp $ profile at the corresponding slices (note these slices are in different directions). The symbol $+$ denotes the starting point while $\times$ is for the ending point. The bottom panels are the energy history (black solid lines) over the same time intervals. We also show the expected energy evolution assuming contribution only from induction effect (red dashed lines), the gradient-B drift (blue dashed lines) and the curvature drift (orange dashed lines) respectively. The black dashed lines are the predicted energy history incorporating contributions of all three mechanisms. The green dotted lines show contribution from compressibility, $\int - \nabla \cdot \mathbf{u}'_\perp p \sin^2 \theta  \dd t/ 2$ (See Equation~\ref{eq:decomp_ind}). The left two energy histories in B4-hires correspond to the red lines wedged between error-bars in Figure~\ref{fig::energy_hist_0.04_b1e4-hires} and Figure~\ref{fig::energy_hist_0.01_b1e4-hires} respectively.}
    \label{fig::energy_hist_anomalous}
\end{figure*}

Particles following guiding center motion along mean magnetic fields also undergo drift motion, which we denote as $\mathbf{v}_g$ (perpendicular to mean field to lowest order). Particles can get accelerated/decelerated due to the work done by the (perpendicular) electric field $\mathbf{E}'\cdot\mathbf{v}_g$ (in shear frame).  In addition, particles with small $R_g$ approximately respond adiabatically to changes in mean field. The net effect to relativistic particles can be written as
\begin{align}
    \frac{dp_sc}{dt}\approx e \mathbf{E}' \cdot \mathbf{v}_g + \frac{p_\perp^2}{2\gamma m B} \frac{\partial B}{\partial t},
\end{align}
where 
$_\perp$ denotes the perpendicular component relative to the local mean magnetic field. More complete expressions can be found in, e.g., \citet{1963RvGSP...1..283N} and \citet{2015ApJ...800...88S}, while we only keep the dominant terms given that background flows are non-relativistic.

We anticipate drift motion to be dominated by gradient-B drift and curvature drift \footnote{The polarization drift velocity is negligible, smaller than gradient-B drift or curvature drift with the order of magnitude $u/c$}, given by (e.g., \citealp{fitzpatrick2014plasma})
\begin{equation}
    \mathbf{v}^{\left(1\right)}_g \approx \frac{\mathbf{b}}{\omega} \times \left(\frac{v^2_\perp}{2B} \nabla B + v^2_\parallel \mathbf{b} \cdot \nabla \mathbf{b}\right)\ ,
\end{equation}
where $\mathbf{b} \equiv\mathbf{B}/B$ is a unit vector along the direction of magnetic field and $_\parallel$ denotes the component parallel to $\mathbf{b}$. 
The change in field strength $B$ is due to induction effect, given by
\begin{equation}
    \frac{\partial B}{\partial t}\approx\mathbf{b}\cdot(\nabla\times\mathbf{E}')\ ,
\end{equation}
which naturally arises when background gas undergoes compression or expansion perpendicular to the local mean field.
Combining the above, we have
\begin{align}
    \frac{1}{p_s}\frac{dp_sc}{dt}\approx [ & \sin^2 \theta \frac{\mathbf{u'}_\perp}{2B} \cdot \nabla B + \cos^2 \theta \left( \mathbf{b} \cdot \nabla \mathbf{b} \right) \cdot \mathbf{u'}_\perp 
    \notag \\ & - \sin^2 \theta \frac{c\mathbf{b}}{2B} \cdot \left( \nabla \times \mathbf{E}' \right) ].
    \label{equ::par_acc}
\end{align}
We see that besides particle pitch angle $\theta$, contributions from turbulence is entirely encapsulated into three terms in the bracket, the gradient-B drift acceleration (Grad, $\mathbf{u_\perp}'\cdot \nabla B / \left(2B\right)$), the curvature drift acceleration (Curv, $\left( \mathbf{b} \cdot \nabla \mathbf{b} \right) \cdot \mathbf{u'}_\perp $) and the induction effect (Induct, $-c\mathbf{b} \cdot \left(\nabla \times \mathbf{E}' \right) / \left(2B\right)$). 
Physically, stochastic particle acceleration in MHD turbulence is commonly associated with (but not limited to) transit time damping (TTD, see \citealp{2008ApJ...673..942Y, 2018ApJ...868...36X,2012ApJ...758...78L}) and Fermi type-B acceleration (e.g., see \citealp{2013ApJ...777..128L,2014ApJ...791...71L}).
TTD results from with particles interacting with moving magnetic mirrors that arise from fast modes in compressible MHD turbulence, while Fermi type-B is associated with particles traveling along moving field lines. There is no clear cut which of the three (Grad, Curv and Induct) terms contribute to which of these mechanisms, but one can expect that acceleration/deceleration from curvature drift is more likely linked to Fermi type-B, while induction effect and gradient-B drift are more likely connected to TTD.

Acceleration by the induction effect is also known as betatron acceleration. It can be further decomposed that contains more physically intuitive terms
\begin{align}
    -\frac{c\mathbf{b}}{2B} \cdot \left( \nabla \times \mathbf{E}' \right) =&  \frac{1}{2} \left[-\nabla \cdot \mathbf{u}'_\perp + \mathbf{b} \cdot \left(\mathbf{b} \cdot \nabla\right) \mathbf{u}'_\perp - \mathbf{b} \cdot \left(\mathbf{u}_\perp' \cdot \nabla\right) \mathbf{b}\right] \notag \\ &- \frac{\mathbf{u'}_\perp}{2B} \cdot \nabla B\ .\label{eq:decomp_ind}
\end{align}
In particular, the first term corresponds to compression/expansion perpendicular to magnetic field. Flux freezing in converging/diverging flows enhances/reduces local magnetic field, creating an inductive field leading to betatron acceleration.
The last term exactly cancels the gradient-B drift term in Equation (\ref{equ::par_acc}). In our later analysis, we will also show separate contribution from compressible effect.

Previously in Figure \ref{fig::energy_history}, we have seen that for particles with large $R_g$, particles stochastically gain or lose energy akin to a random walk, while for particles with smaller $R_g$, some particles (especially those that achieve largest and lowest energies) tend to experience periods of systematically energy gain or loss over many tens of gyro periods. For the time being, we tentatively call such particles ``anomalous", with examples found in the red lines in Figure \ref{fig::energy_hist_0.04_b1e4-hires} and \ref{fig::energy_hist_0.01_b1e4-hires}. While we consider other particles as ``normal", as an example, the black lines in the same plots.
We speculate that the presence of such ``anomalous particles" to be the source of large uncertainties in our estimate of $D_{\rm Fermi}$, $D_{\rm MRI}$, and non-zero $A_{\rm Fermi}$, $A_{\rm MRI}$.

\begin{figure}
    \centering
    \includegraphics[width=28em]{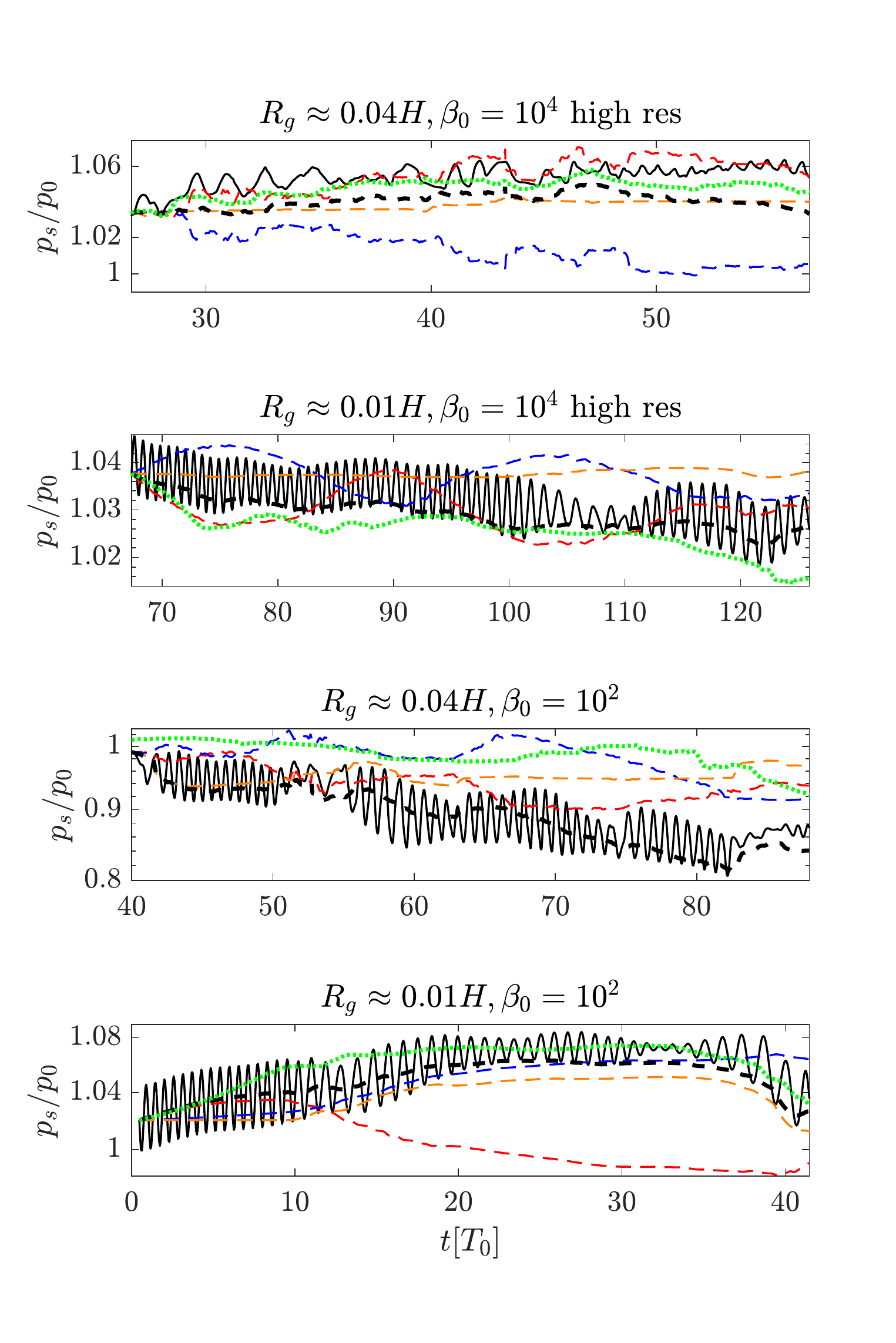}
    \caption{Energy histories (black solid lines) of four particles whose energy largely vary stochastically.
    We also show the expected energy evolution assuming contribution only from induction effect (red dashed lines), the gradient-B drift (blue dashed lines) and the curvature drift (orange dashed lines) respectively, and the green lines indicate contribution from compressbility. The black dashed lines are the predicted energy history incorporating contributions of all three mechanisms. Legend is not shown but is the same as that in Figure \ref{fig::energy_hist_anomalous}. The upper two energy histories in B4-hires correspond to the black lines wedged between error-bars in Figure~\ref{fig::energy_hist_0.04_b1e4-hires} and Figure~\ref{fig::energy_hist_0.01_b1e4-hires} respectively.}
    \label{fig::energy_hist_norm}
\end{figure}

To further understand the underlying source of particle acceleration and diffusion, we select a few such ``anomalous" particles with small $R_g$ drawn from Figure \ref{fig::energy_history} as well as in run B2, and analyze their energy evolution, studying contributions from Grad, Curv and Induct. The results are shown in Figure \ref{fig::energy_hist_anomalous}. For comparison, we conduct the same exercise for some ``normal" particles and show results in Figure \ref{fig::energy_hist_norm}. Overall, we find that we can reasonably well reproduce the energy history as long as $R_g$ is sufficiently small (notable deviation occurs mainly for $R_g=0.04H$ particles but the trend is well captured). Below, we separately discuss the results for these two groups of particles.

For those particles continuously accelerated/decelerated, we see that the induction effect dominates the energy evolution. In some cases, particles undergo reflection in configuration space during the acceleration/deceleration (e.g. Figure~\ref{fig::anomalous_2} and Figure~\ref{fig::anomalous_3}), analogous to mirror interactions.
For other cases, they mainly reside in compression/expansion flows (e.g. Figure~\ref{fig::anomalous_1}), where $\nabla \cdot \mathbf{u}'_\perp$ is consistently negative or positive.
These regions are associated with the spiral density waves generated in the MRI turbulence \citep{2009MNRAS.397...52H,2009MNRAS.397...64H}.

For ``normal" particles, as shown in Figure~\ref{fig::energy_hist_norm}, we find that the three mechanisms all contribute their energy evolution (in different ways), and the sum of three acceleration mechanisms very well reproduces the mean energy history for particles with small $R_g$. We observe an anti-correlation between gradient-B drift and induction effects, especially for the top three panels, anticipated from the decomposition (Equation~\ref{eq:decomp_ind}) if the last term dominates. As a result, their contributions largely cancel. On the other hand, contribution from $\nabla \cdot \mathbf{u}'_\perp$, seen in the green curves, matches more closely to the energy evolution (except for the third panel).

\begin{figure*}
    \centering
    \includegraphics[width=66em]{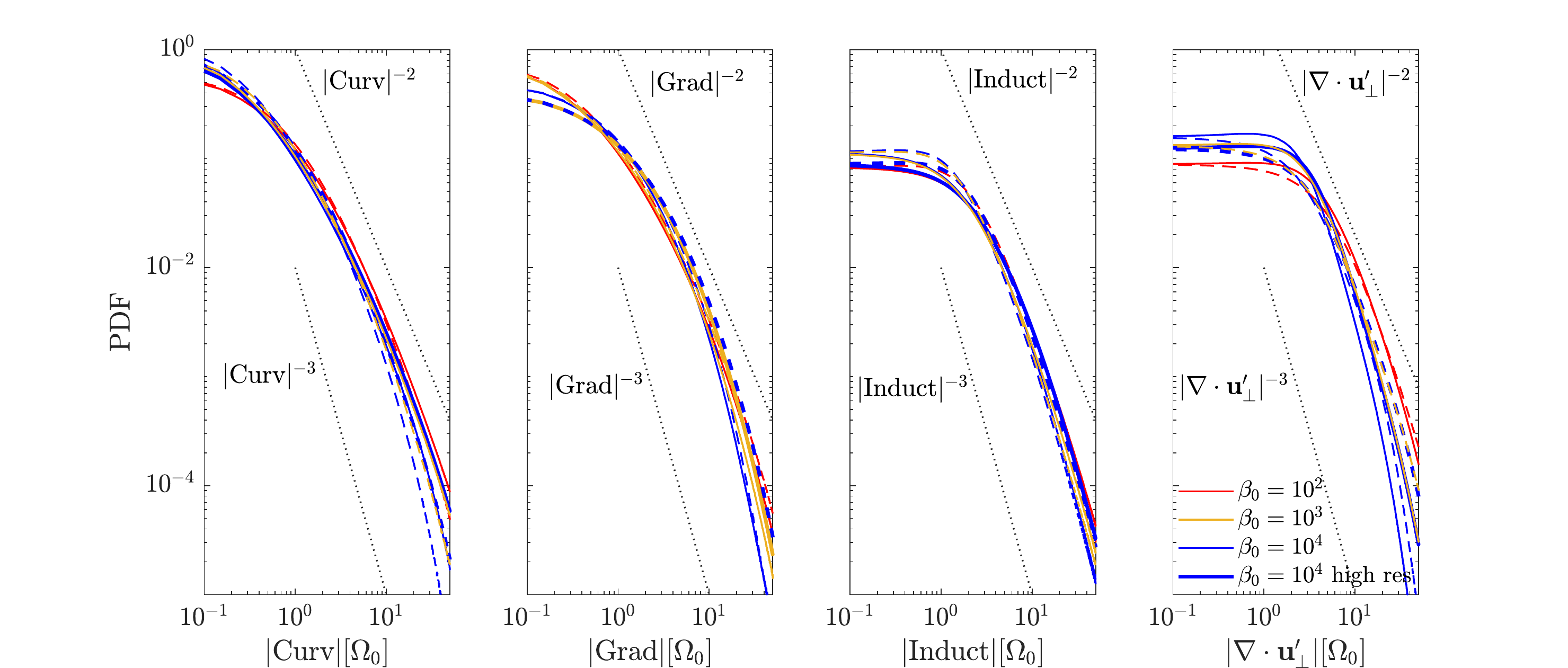}
    \caption{The probability density functions (PDF) for the curvature drift acceleration (Curv), the gradient-B drift acceleration (Grad), the induction effect (Induct), and $\nabla \cdot \mathbf{u}'_\perp$, respectively. Color and thickness indicates different MRI simulations (see legend). For each simulation, the solid and dashed lines correspond to the part with positive and negative values, where the negative parts are flipped to take absolute values to enhance contrast.}
    \label{fig::pdf_mri_acc}
\end{figure*}

Having seen that the three terms largely govern particle energy evolution, we can draw information about the strength of individual turbulent ``kicks" to accelerate/decelerate particles (related to $D_{\rm Fermi}$) by studying the statistical properties on the strength of these terms. 
As particles are approximately uniformly distributed in space and pitch angle,
we have computed these terms (Grad, Curv and Induct in Equation (\ref{equ::par_acc})) from individual cells in our MRI simulations. Additionally, based on Equation~\ref{eq:decomp_ind}, we have also computed $\nabla \cdot \mathbf{u}'_\perp$ representing contribution from compressibility. In Figure~\ref{fig::pdf_mri_acc}, we show the probability density functions (PDFs) of these three terms (in code units). 
We see that the negative and positive sides of the PDF are largely symmetric, but not exact. Both sides of the PDFs have extended tails that are close to a power law with an index between $-2$ and $-3$ for runs B2, B3 and B4-hires. Such PDFs are ill-conditioned in the sense that it does not have a well-defined variance, and that the law of large numbers fails. We thus speculate this is likely related to the large uncertainties in our measured diffusion and acceleration coefficients $D_{\rm Fermi}(p)$ at small $R_g$. Because the calculation of $A_{\rm Fermi}(p)$ depends on $D_{\rm Fermi}$ and its gradient in $p$, its non-convergence to zero is also likely one of the consequences. In other words, particle energy gain and loss are strongly affected by rare events, rather than the accumulation of small random kicks.

The result discussed above is closely related to the intermittency in the MRI turbulence, as discussed in Section \ref{sec::result_mri}. In the context of MHD, intermittency in turbulence is commonly studied focusing on the scale-dependent inhomogeneities in turbulent dissipation \citep{2006ApJ...640L.175B,2015ApJ...807...39C, 2017MNRAS.466.3918M,2020arXiv201000699S}. At microphysical level, it is expected to lead to intermittent injection of energetic particle. Here, our results show that subequent acceleration of particles can also be strongly affected by intermittency in turbulence.
Nevertheless, an important caveat is that our simulations lack explicit dissipation, and future studies are needed to address the detailed microphysical mechanisms of particle acceleration and more definitely assess the role intermittency in the MRI turbulence.

\subsection{Shear acceleration}
\label{sec::shear}

\begin{figure}
    \centering
    \includegraphics[width=25em]{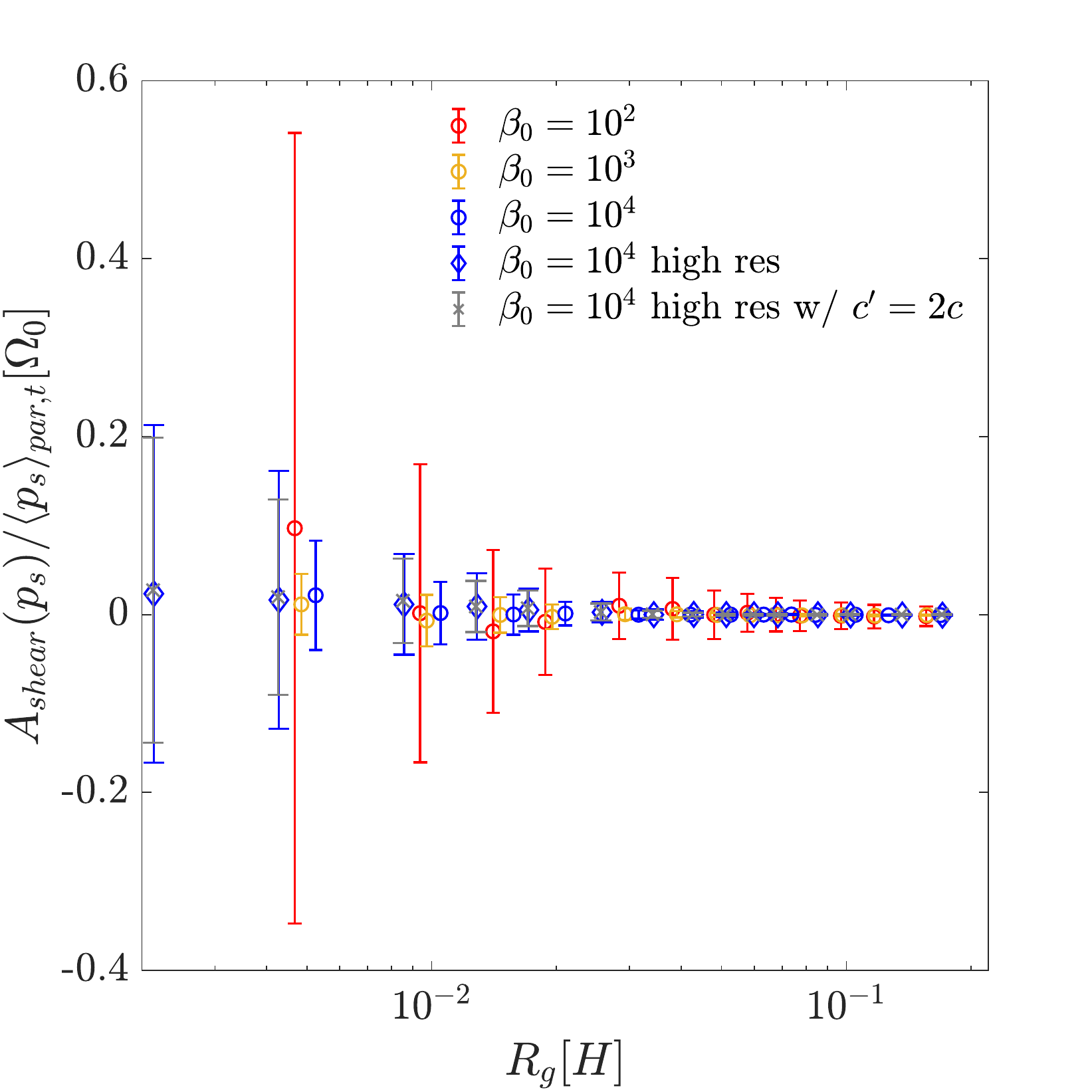}\\
    \includegraphics[width=25em]{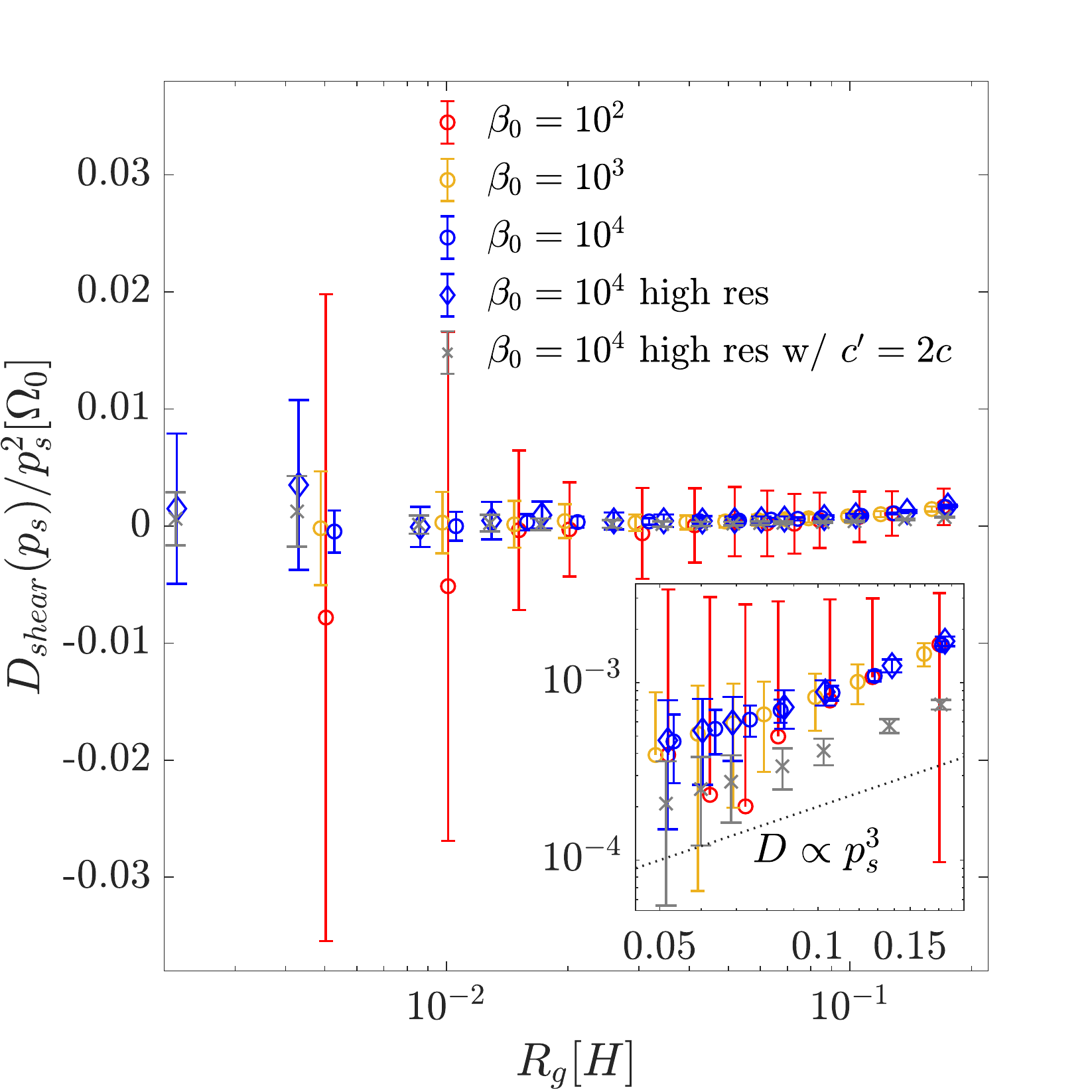}
    \includegraphics[width=25em]{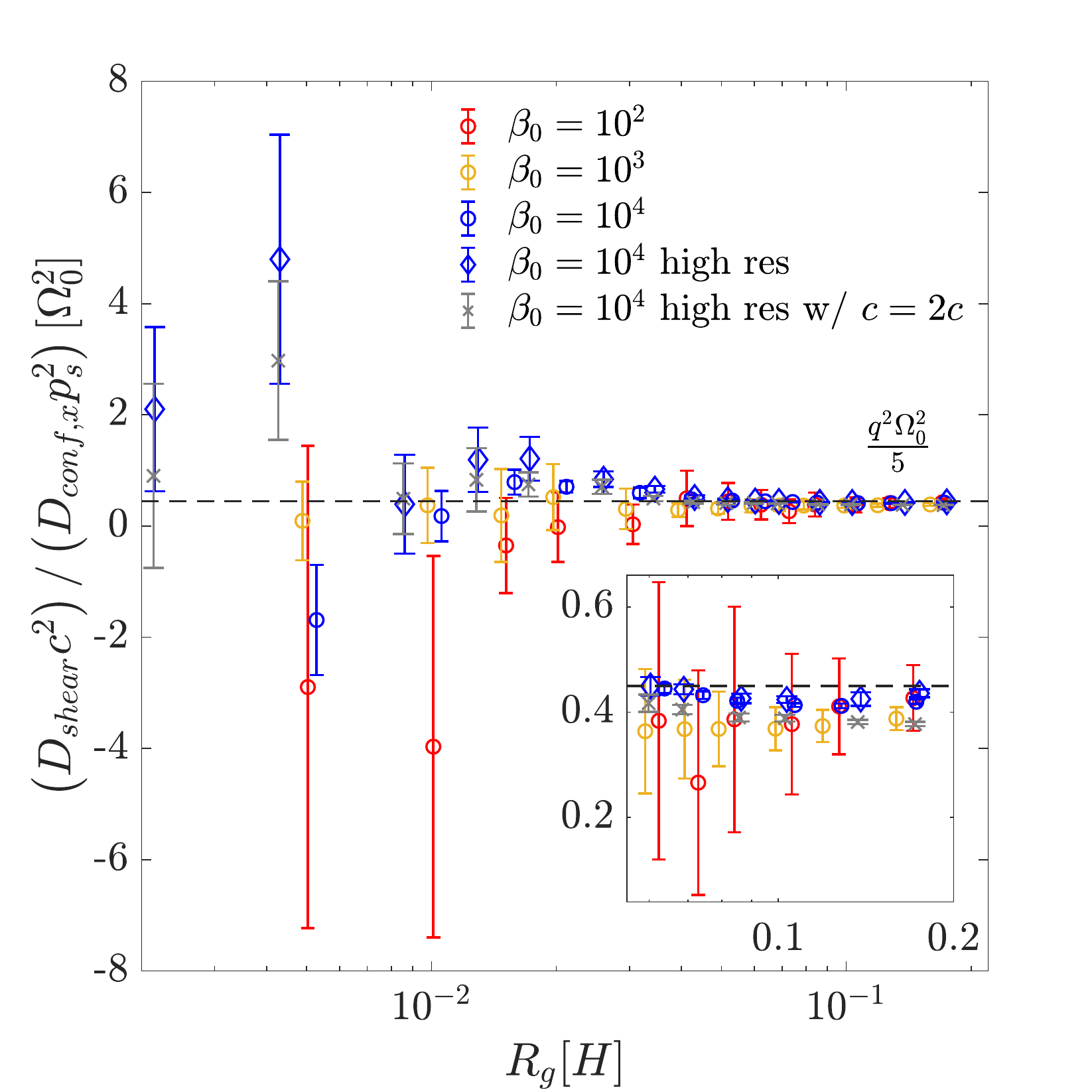}
    \caption{Acceleration (top) and diffusion (middle) coefficients for relativistic particles due to contributions from shear, shown as functions of particle gyro radii $R_g$. The bottom panel shows a comparison with relation (\ref{equ:shear_comb}). Results from different simulations are marked by different colors and symbols (see legend).}
    \label{fig::coefficient_mri_rel}
\end{figure}

We assume contribution from shear to the Fokker-Planck equation is additive so that $A_{\rm MRI} = A_{\rm Fermi}+A_{\rm shear}$ and $D_{\rm MRI} = D_{\rm Fermi}+D_{\rm shear}$, where $A_{\rm shear}$ and $D_{\rm shear}$ denote the contributions from the shear flow.
With the linear assumption, we therefore subtract the left and right panels of Figure~\ref{fig::coefficient_rel}, and show in Figure~\ref{fig::coefficient_mri_rel} the inferred values of $A_{\rm shear}$ and $D_{\rm shear}$. 

Based on the assumption of isotropic turbulent scattering, \cite{1988ApJ...331L..91E} derived that $A_{\rm shear}=0$ for simple planar shear as in the MRI. 
Given the shear rate of $q\Omega_0$ in our simulations, the shear diffusion coefficient can be estimated as \citep{1988ApJ...331L..91E,1989ApJ...340.1112W,2019PhRvD..99h3006L,2019Galax...7...78R}\footnote{The diffusion coefficient based on \cite{2019PhRvD..99h3006L} is four times that given by $D_{\rm shear}$\citep{1988ApJ...331L..91E} which we adopt, but the difference could be readily absorbed to $\tau$.
}.
\begin{equation}
    D_{\rm shear} = \frac{q^2 \Omega_0^2 p_s^2 \tau}{15}\ , \label{equ::shear_Earl}
\end{equation}
where $\tau$ represents the relaxation time of turbulent scattering. In our case, we can write $\tau\approx L_{{\rm scatter}, x}/v=L_{{\rm scatter}, x}/c$. We choose to use the $x$-component of $L_{\rm scatter}$ because it is scatterings in the radial dimension that is responsible for shear acceleration (see below). 
We can divide $D_{\rm shear}/p_s^2$ by $D_{{\rm conf}, x}$, thus canceling $L_{{\rm scatter},x}$ \citep{2016ApJ...822...88K}, and obtain \citep{1988ApJ...331L..91E}
\begin{equation}\label{equ:shear_comb}
    \frac{D_{\rm shear}}{D_{\rm conf, x} p_s^2} = \frac{q^2 \Omega_0^2}{5v_s^2}\ .
\end{equation}
We will test these analytic results below.

From the top panel of Figure~\ref{fig::coefficient_mri_rel}, we see that while our measured $A_{\rm shear}$ values are not exactly zero, especially for particles with small $R_g$, they are consistent with zero well within the $1\sigma$ uncertainties.
We therefore treat $A_{\rm shear}=0$ and mainly focus on $D_{\rm shear}$ below. 

Given the numerical speed of light, we see that all three runs (B2, B3, B4) tend to yield the same $D_{\rm shear}/p_s^2$ at a given $R_g\gtrsim0.05H$, with a clear scaling relation that $D_{\rm shear}/p_s^2\propto p\propto R_g$. This is consistent with $L_{{\rm scatter},x}\propto R_g$, or $D_{{\rm conf},x}/D_{\rm Bohm}\sim$const for such particles, as we have already discussed in previous sections.
The values of $D_{\rm shear}/p_s^2$ are subject to large uncertainties at smaller $R_g$, largely due to the values being small compared to uncertainties obtained from $D_{\rm MRI}$. On the other hand, based on the behavior of $L_{{\rm scatter},x}$, we expect the curve of $D_{\rm shear}/p_s^2$ to be close to flat, and remain at a value much smaller than $D_{\rm Fermi}$.
We also see that doubling the numerical speed of light reduces $D_{\rm shear}$ by half, which is expected due to the reduction of $\tau=L_{{\rm scatter}, x}/c$. 
In the bottom panel of Figure~\ref{fig::coefficient_mri_rel}, we show the ratio given by Equation (\ref{equ:shear_comb}). It varies strongly when $R_g\lesssim0.03H$, but approaches a constant value consistent with the predicted value of $q^2\Omega_0/5=0.45$ within uncertainties. Overall, our simulation results are broadly consistent with the general theory of shear acceleration.

Now we further investigate the microphysics behind shear acceleration (see \citealp{1988ApJ...331L..91E,1989ApJ...340.1112W,2019PhRvD..99h3006L,2019Galax...7...78R}). To tap the free energy from shear, particles must scatter back and forth in the radial direction. During this process, particles experience varying a background velocity, and gain or lose energy to adapt to the local velocity field. For example, we expect that a particle moving radially outward $\delta x>0$ will experience a negative azimuthal velocity from background shear. If the particle has a positive azimuthal velocity $v_{s,y}>0$ in the shearing frame, it tends to gain energy because the isotropization process can ``reflect" $v_{s,y}$ with an additional boost from relative shear, just as a ball bouncing off an approaching wall. Other cases can be analyzed similarly, and overall we expect a positive correlation between $\delta x \times v_{s,y}$, and energy (and hence magnitude of momentum) gain in the shearing frame $\delta p_s$.

\begin{figure}
    \centering
    \includegraphics[width=28em]{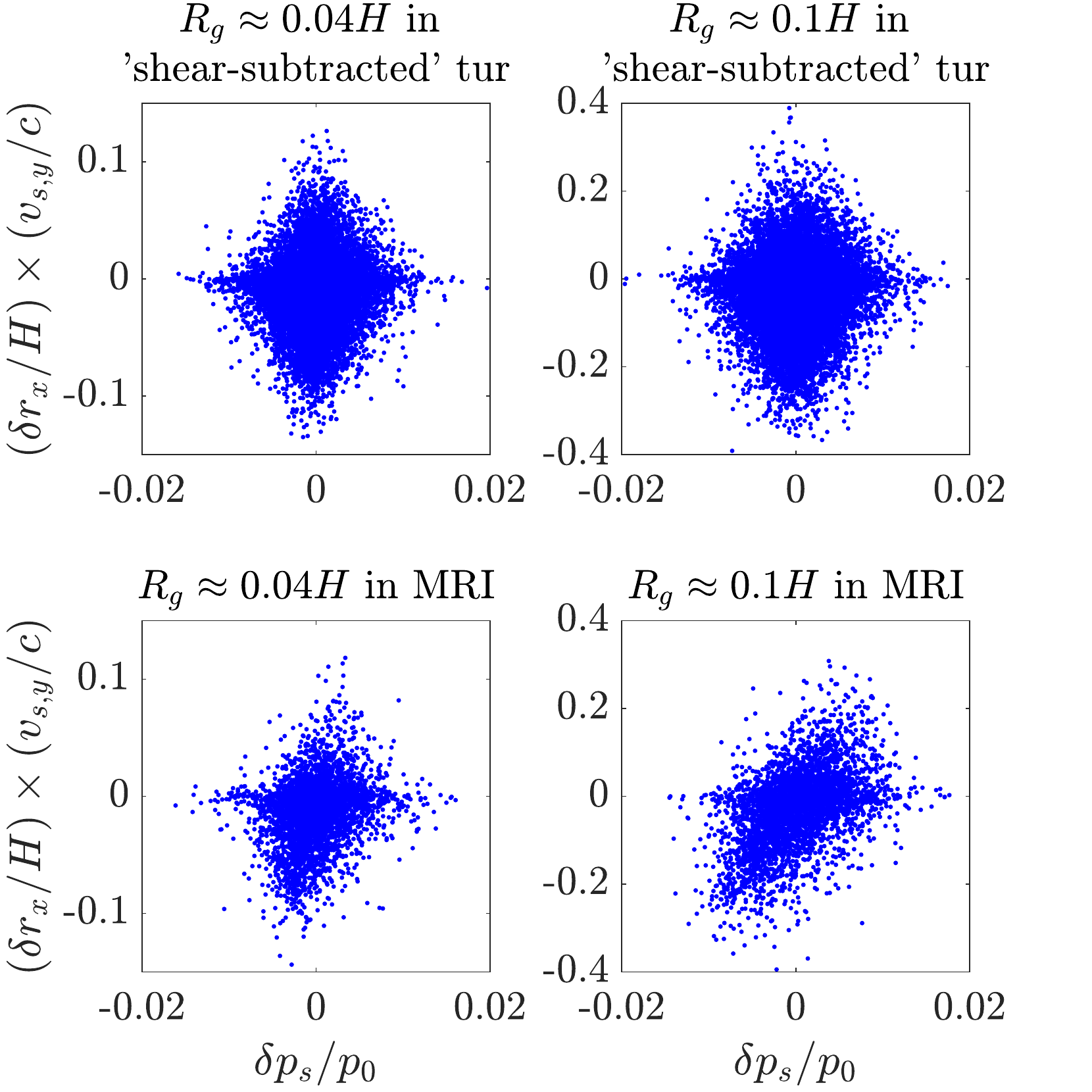}
    \caption{Scatter plots for the relation between momentum changes of particles over a time interval $\delta t=2T_0$ and radial displacements $\delta x$ multiplied by mean azimuthal velocity $v_{s,y}$ over this time interval (normalized to dimensionless units), for relativistic particles in run B4-hires. The left two panels are for particles with gyro radii  $R_g\approx0.04H$, and the right ones for $R_g \approx 0.1H$. The upper panels are for particles in the ``shear-subtracted" turbulence and the lower ones are for particles in full MRI turbulence.}
    \label{fig::corr_traj}
\end{figure}
\begin{figure}
    \centering
    \includegraphics[width=28em]{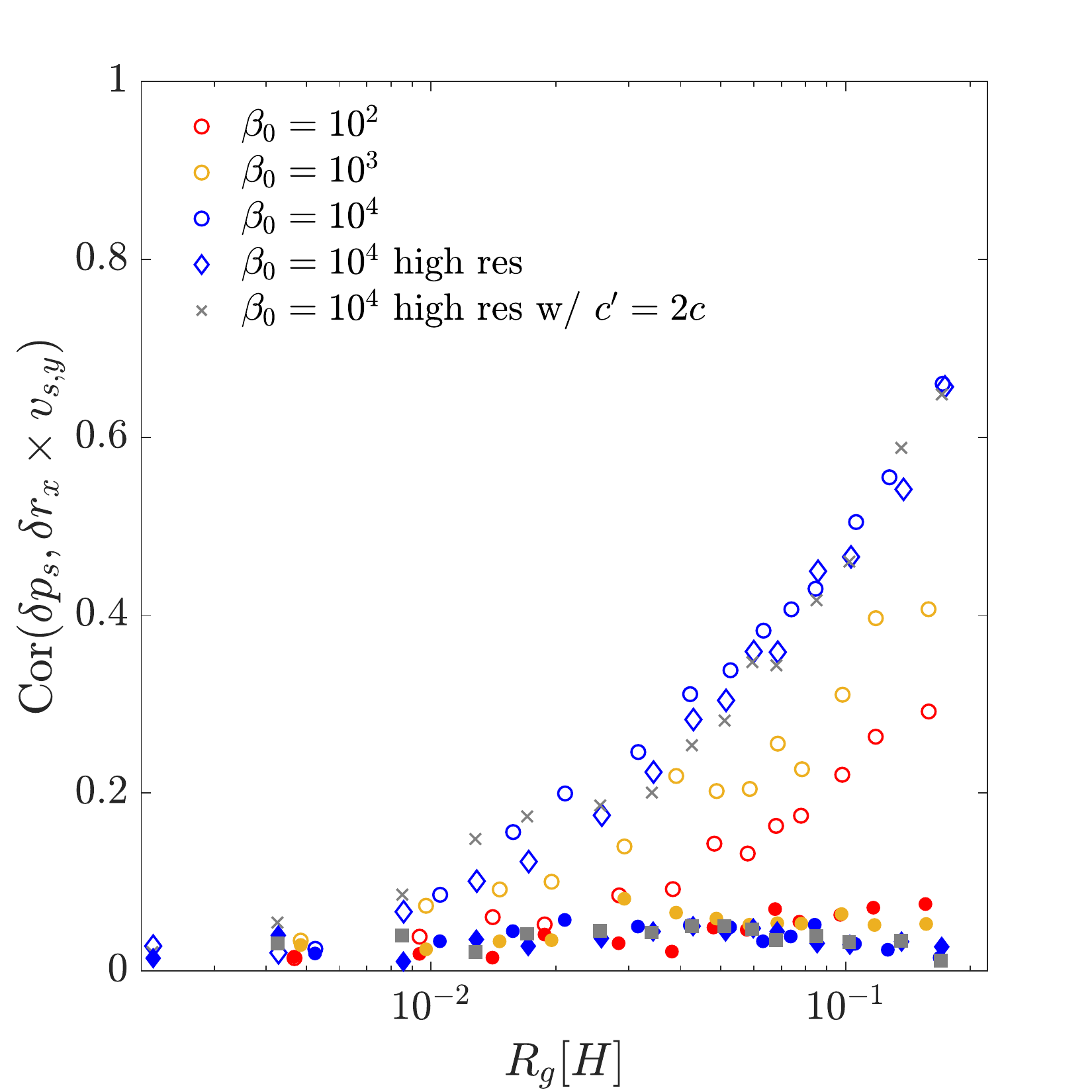}
    \caption{Correlation coefficients between $\delta p_s$ and $\delta r_x \times v_{s,y}$ of relativistic particles, shown as a function of particle gyro radii. Results from different simulations are marked by different colors and symbols (see legend). Hollow markers refer to the results from full MRI turbulence and filled ones are from shear-subtracted turbulence.}
    \label{fig::corr_val_rel}
\end{figure}

In Figure~\ref{fig::corr_traj}, we study such correlation. We choose a time interval $\delta t=2T_0$ to measure individual particle displacements $\delta x$. This interval is chosen so that particles undergo more than one gyro-period to experience some scatterings, but much less than the time it takes to isotropize so that it keeps some initial memory. We show scattered plot of momentum change $\delta p_s$ in shearing frame over this time interval, versus $\delta x \times v_{s,y}$, where $v_{s,y}$ is the azimuthal velocity averaged over the interval. 
All values are normalized to be non-dimensional. We clearly see that no correlation is seen in ``shear-subtracted" turbulence, whereas in full MRI turbulence, such correlation is present, and is more evident for particles with larger $R_g$.
We have also checked that $\delta p_s$ has no correlation with $\delta r_x$ or $v_{s,y}$ individually, and correlation exists only for their product.
More quantitatively, we measure the correlation coefficients Cor$\left( \delta p_s, \delta r_x \times v_{s,y} \right)$ and show its dependence over $R_g$ in Figure~\ref{fig::corr_val_rel}. 
The correlation coefficients in the ``shear-subtracted" turbulence are consistently 0 for all $R_g$, as expected. With shear flows, Cor$\left( \delta p_s, \delta r_x \times v_{s,y} \right)$ increases monotonously with $R_g$ and roughly approximates the ratio between $D_{\rm shear}$ and $D_{\rm MRI}$.
These results confirm the physical picture we outlined earlier as a generic mechanism in turbulent shear flows.

\section{Discussion}
\label{sec::discussion}

\subsection{Particle acceleration in ADAFs}
\label{sec::acc_ADAF}

\begin{figure}
    \centering  
    \includegraphics[width=28em]{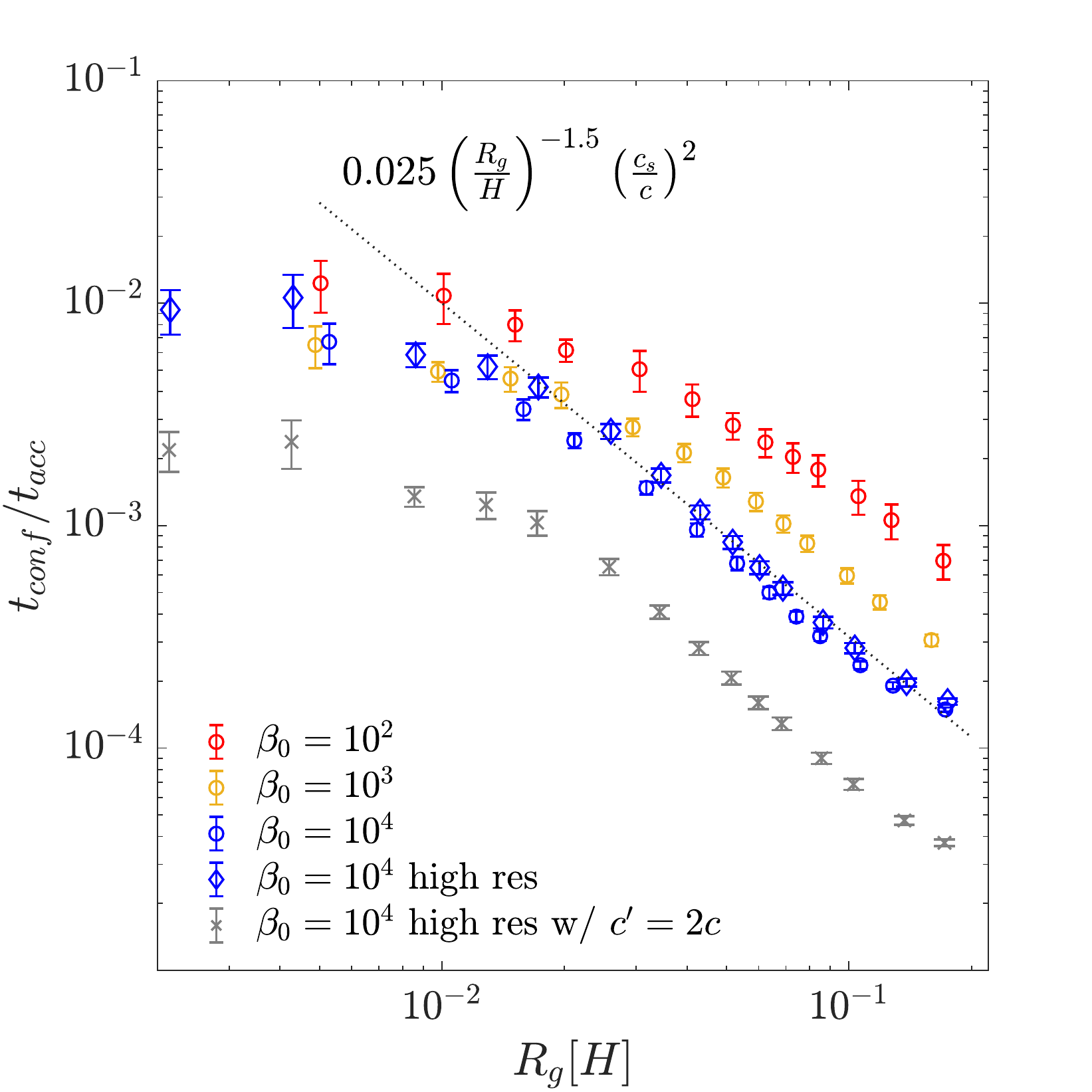}
    \caption{The acceleration efficiency $t_{\rm conf}/t_{\rm acc}$ for ultra-relativistic particles. The colour stands for the $\beta_{0}$ of background MRI simulations, meanwhile the shape represents the resolution.}
    \label{fig::acc_eff}
\end{figure}

We discuss the prospects of accelerating particles, primarily protons, to very high energies in ADAFs, as has been proposed in the recent literature (e.g., \citealp{2015ApJ...806..159K}).
The maximum energy that can be achieved is determined by the timescale of particle acceleration versus the timescale for particles to remain confined in the acceleration site
(e.g., \citealp{2008ApJ...681.1725S,2019ApJ...886..122C,2019MNRAS.485..163K}). Given the local nature of our simulations, particles always stay confined (do not escape). In reality, the confinement time scale $t_{\rm conf}$ can be considered as the time for a particle to diffuse over the characteristic distance of disk thickness $H$ ($\sim 0.3-0.4R$ for ADAF) along the poloidal plane. After traveling such distance, the particle would either escape vertically from the disk, or approach the central object where our assumptions fail, or diffuse to a larger distance where acceleration becomes much slower and hence inefficient. This timescale can be estimated as 
\begin{equation}
    t_{\rm conf} \sim \frac{H^2}{D_{\rm conf,pol}}\ .
\end{equation}
Similarly, at a given momentum, the acceleration timescale $t_{\rm acc}$ 
(neglecting $A_{\rm MRI}$) can be estimated as
\begin{equation}
    t_{\rm acc} \sim \frac{p_s^2}{D_{\rm MRI}}\ .
\end{equation}
The bulk of the CR particle population can be continuously accelerated towards higher energies as long as $t_{\rm acc}<t_{\rm conf}$.

In Figure~\ref{fig::acc_eff}, we show the ratio $t_{\rm conf}/t_{\rm acc}$ as a function of momentum (or $R_g$). We see that for all particle gyro radii considered in our simulations, $t_{\rm conf}/t_{\rm acc}\ll1$, indicating that they are unlikely to be confined before reaching these energies in reality. 
We find that for runs B4 and B3, and for particles with $R_g\gtrsim0.03H$, the ratio of $t_{\rm conf}/t_{\rm acc}$ can be reasonably fitted with a power law with index $\sim(-1.5)$. Results from run B2 is higher, indicating more effective confinement. Given the dependence of $D_{\rm conf,pol}\propto c$ and $D_{\rm MRI}\propto c^{-1}$ on numerical speed of light $c$, we see $t_{\rm conf}/t_{\rm acc}\sim c^{-2}$.
Overall, the following expression reasonably captures the overall trend seen in Figure \ref{fig::acc_eff}:\footnote{One would obtain $t_{\rm conf}/t_{\rm acc}\sim(R_g/H)^{-2}$ following the scalings discussed earlier, but note that scaling relations for $R_g\gtrsim0.03H$ obtained in this regime should be considered only as being empirical and approximate.}
\begin{equation}\label{eq:tratio}
    \frac{t_{\rm conf}}{t_{\rm acc}}\sim0.025 \bigg(\frac{R_g}{H}\bigg)^{-3/2}\bigg(\frac{c_s}{c}\bigg)^{2}\ .
\end{equation}

However, the trend does not continue for particles with $R_g\lesssim0.03H$. We have seen that $D_{\rm conf, pol}\sim$const and tentatively $D_{\rm MRI}\sim p_s^2$ in this regime, and hence $t_{\rm conf}/t_{\rm acc}\sim$const. This trend is seen also Figure~\ref{fig::acc_eff} for $R_g\lesssim0.01H$. A flat dependence on $R_g$ with $t_{\rm conf}/t_{\rm acc}$ suggests that particles would never gain sufficient acceleration before escaping the system. For our fiducial choice of $c_s/c=0.02$, this ratio caps at around $0.01$, and this cap appears to apply to all our simulations (runs B2-B4). To reach order unity, it is thus required that $c_s/c\gtrsim0.1-0.2$. This suggests that only in regions very close to the SMBH, it can be marginally possible to achieve strong particle acceleration. 

Despite the discussion above, we also note that since we do not fully resolve the gyro motion of particles with such small $R_g$, the aforementioned $t_{\rm conf}/t_{\rm acc}$ scaling at small $R_g$ may suffer from potential numerical artifacts and hence large uncertainties. We defer to a future study for particle acceleration in this parameter range. In the following, we consider the most optimistic scenario by assuming that Equation~(\ref{eq:tratio}) is broadly applicable for all $R_g$, as well as for all MRI runs, and only focus on regions very close to the SMBH (so that the assumption is more likely valid). 

By default,
we consider a radius close to the innermost stable circular orbit (ISCO) for a non-spinning black hole, with $r\sim10 r_g$ ($r_g\equiv GM/c^2$ is the gravitational radius), and a disk thickness of (e.g., \citealp{2012MNRAS.426.3241N})
\begin{equation}
    \frac{c_s}{\Omega r} = \frac{H}{r} \sim \frac{1}{3}\ .
\end{equation}
In this case, $c_s/c\sim0.1$, and from Equation~\ref{eq:tratio}, we obtain
$R_g\sim4\times 10^{-3}H$, which is closer yet still below our resolvable scale. The maximum CR energy $\varepsilon_{\rm CR}$ from such $R_g$ is then given by
\begin{equation}\label{eq:ecr0}
    \varepsilon_{CR} \sim 0.085\ er_g \langle B \rangle\bigg(\frac{H}{r}\bigg)^{7/3}\bigg(\frac{r_g}{r}\bigg)^{-1/3}\ .
\end{equation}
It remains to estimate the mean field strength $\langle B\rangle$, which can be conveniently done as follows. Let $M$ be the mass of the central black hole, and $\dot{M}$ be the accretion rate. Steady state accretion yields $\dot{M}\sim2\pi\alpha c_s H\Sigma$, where $\Sigma$ is the surface density. While this relation does not hold in the presence of outflows, as has been identified to play an important role on the large-scale dynamics of ADAFs (e.g., \citealp{2015ApJ...804..101Y, 2021arXiv210203317Y}), the rate of outflow is found to be negligible in the vicinity of the black hole (within a few tens of $r_g$). A useful relation in the MRI turbulence is that angular momentum transport becomes more efficient with stronger magnetization, and in the saturated state, one approximately has \citep{2007ApJ...668L..51P, 2011ApJ...736..144B}
\begin{equation}
    \langle\alpha\rangle\langle\beta\rangle\sim\frac{1}{2}\ .
\end{equation}
This relation is also approximately satisfied in our simulations within a factor of $\sim2$ (see Table \ref{tab::mri}).
Given this relation, together with midplane pressure $P_{\rm mid}\sim\Sigma c_s^2/\sqrt{2\pi}H$ with $\langle B^2\rangle=P_{\rm mid}/\langle\beta\rangle$, we obtain
\begin{equation}
    \langle B^2\rangle \approx 3.2\frac{\dot{M}\Omega}{r}\ .
\end{equation}
We note that the relation is in fact independent of disk thickness $H/r$ and $\alpha$ values.
It is conventional to normalize $\dot{M}$ to Eddington accretion rate, which we define assuming an radiative efficiency of $10\%$ (following \citealp{2014ARA&A..52..529Y}) $\dot{M}_{\rm Edd}=10L_{\rm Edd}/c^2$, where $L_{\rm Edd}=4\pi GMc /\kappa_{\rm es}$, with $\kappa_{\rm es}=0.4 {\rm cm}^2\ {\rm g}^{-1}$ for electron scattering opacity. With these results and definitions, Equation.~(\ref{eq:ecr0}) yields
\begin{equation}
\begin{split}
    \epsilon_{\rm CR}^{\rm max}\approx 5.5{\rm PeV}&\bigg(\frac{M}{4\times10^6M_{\bigodot}}\frac{\dot{M}}{10^{-3}\dot{M}_{\rm Edd}}\bigg)^{1/2}\\
    &\times\bigg(\frac{r}{10r_g}\bigg)^{-11/12}\bigg(\frac{3H}{r}\bigg)^{7/3}\ ,
\end{split}
\end{equation}
where we have plugged in the mass of Sgr A$^*$, $r=10r_g$ and $H/r=1/3$ for reference. 

The present day accretion rate of Sgr A$^*$ is in between $10^{-6}$ and $10^{-7}\dot{M}_{\rm Edd}$ \citep{2003ApJ...598..301Y, 2014ARA&A..52..529Y, 2018ApJ...868..101B}, implying that the maximum energy of $\epsilon_{\rm CR}^{\rm max}\sim$ 10 TeV. On the other hand, there is evidence that Sgr A$^*$ was more active in the past (e.g., \citealp{2013PASJ...65...33R}), and underwent a major outburst a few Myrs ago which created the Fermi bubble \citep{2010ApJ...724.1044S}. The latter is believe to be powered by relic CRs injected during the active accretion phase (e.g., \citealp{2011PhRvL.106j1102C, 2012ApJ...756..182G, 2012ApJ...761..185Y}). We see that so long as accretion rate can reach $\sim10^{-3}M_{\rm Edd}$ (while still being in ADAF state), the galactic center SMBH can become a PeVatron. For the SMBH in the center of M87, with mass of $6.2 \times 10^9 M_{\bigodot}$ and accretion rate varying from $10^{-4} \dot{M}_{\rm Edd}$ to $10^{-7} \dot{M}_{\rm Edd}$ depending on disk accretion mode being SANE or MAD \citep{2019ApJ...875L...5E}, we find $\epsilon_{\rm CR}^{\rm max}\sim 2-60$PeV, significantly larger than that of Sgr A$^*$ thanks to its higher mass.
We note these estimates are overall comparable to those from \citet{2015ApJ...806..159K}, while are higher than those in \citet{2019MNRAS.485..163K} where they estimated escape timescale differently.

We again emphasize that the above estimates are optimistic and require more efficient particle acceleration at smaller $R_g$ than obtained from our direct simulations. If it is not the case, we anticipate fast-spinning SMBHs be more promising candidates because their ISCOs extend closer to the SMBH to achieve larger $c_s/c$ to enhance acceleration.
Nevertheless, corrections from general-relativistic effects also become increasingly important.
More works are also needed to clarify the physics of injection and particle acceleration at such small scales, and to quantify the energy loss and escape processes to predict the particle energy spectrum.

\subsection{Comparison with previous works}
\label{sec::compare}

Our work closely resembles \citet{2016ApJ...822...88K}, who studied particle acceleration by injecting test particles with a handful of gyro-radii in an MRI simulation in shearing-box with $\beta_0=10^4$.
Results from individual particle simulations in their work are in rough agreement with ours when converting to the same units. On the other hand, by employing MRI simulations with 3-6 times higher resolution, as well as explored different net vertical magnetic flux, we cover much more dynamical range, allowing for the observation of a transition of turbulence properties towards smaller scales leading to larger particle mean free path. We have also been careful in limiting simulation time so that particles are sufficiently isotropized and avoid the issue of ``runaway" particles that undermined the results in \citet{2016ApJ...822...88K}. 
Moreover, we have conducted more detailed and systematic measurements and analysis of particle diffusion and acceleration in configuration and momentum space, showing results that are converged with numerical resolution, and
are made ease of use for calculating CR diffusion and acceleration in astrophysical accretion disks. In particular, we have identified the transition in transport properties around $R_g\sim0.01-0.03H$, which likely have important implications when extrapolating to realistic CR energies (whose $R_g\ll H$).

Recently, \citet{2019MNRAS.485..163K} further conducted global MHD simulations of ADAF disks, and studied test particle diffusion in the global context. Over global scale, particles exhibit super-diffusion in the disks, and they have identified a trend of outward bulk motion due to mirroring effect acting on particles. We suspect that the apparent super-diffusive behavior is due to turbulence being non-uniform with outward magnetic gradient in global disks, 
leading to longer mean free path as particles diffuse outwards. The particles in their work have comparable energy to those with $R_g \sim 0.1 H$ in our simulations, but exhibit $D_\text{MRI}(p)\sim p^2$ behavior, rather than $D_\text{MRI}(p)\sim p$ as in our simulations. The cause of this discrepancy is unclear. We notice that the mean free path of such particles is already comparable to global scale in ADAF disks where local simulations may not be a good approximation, while on the other hand our simulations offer much higher numerical resolution, and likely better capture the dynamics of $R_g\ll H$ particles that are more relevant under realistic conditions. We also note that for both their and our works, estimates of the maximum particle energy in ADAFs rely on extrapolation of simulation results to particles with gyro-radii that cannot be properly resolved. Those particles also have smaller mean free path thus less subject to global effects, and we exercise caution that due to a transition in CR transport properties below $R_g\lesssim0.01-0.03H$ that were not noted in their study. Local and global simulations with much higher resolutions are likely needed to help resolve this issue.

The general phenomenology we observe, namely second-order Fermi acceleration, is akin to those observed in recent kinetic simulations of driven turbulence. 
For instance, \citet{2019ApJ...886..122C} and \citet{2020ApJ...893L...7W} found in their kinetic simulations of relativistic driven turbulence that $D_\text{Fermi}(p)\propto p^2$ scaling for energetic particles. Our simulations show a $D_\text{Fermi}(p)\propto p$ scaling at large $p$, but we also find tentative evidence of $D_\text{Fermi}(p)\propto p^2$ for particles with smaller $p$, thus joining the scaling relation obtained towards kinetic scales (despite our turbulence is non-relativistic). Again, higher resolution MHD simulations of the MRI are needed for confirm this scaling.

Finally, although our findings show that shear acceleration is unlikely to dominate in ADAF disks, shear acceleration itself likely plays an important role in other contexts. 
In particular,
relativistic jets from active galactic nuclei (AGN) are characterized by strong transverse velocity gradient off the jet axis, leading to renewed interest in studying particle acceleration in jets with the potential of generating ultra-high energy cosmic rays (e.g., \citealp{2009A&A...506L..41R, 2016ApJ...833...34R, 2017ApJ...842...39L, 2018PhRvD..97b3026K, 2018ApJ...855...31W}).
Most studies are based on theoretical models, while kinetic studies usually focus on abrupt rather than gradual velocity gradient (e.g., \citealp{2013ApJ...766L..19L, 2014NJPh...16c5007A}). Our study offers a first-principle verification of shear acceleration under the specific realization of the MRI turbulence in accretion disks.

\section{Summary and Future Directions}
\label{sec::summary}

In this work, we conduct high-resolution unstratified local shearing box simulations of the MRI in ideal MHD with weak to strong net vertical magnetic flux, and study properties of particle diffusion and acceleration by directly integrating energetic test particles trajectories in simulation snapshots. The test particles can be considered as being injected from kinetic processes in hot accretion flows around compact objects, especially supermassive black holes in the low state, where accretion rates are well below Eddington rate and the plasmas are largely collisionless. Besides studying the acceleration physics, we also aim at assessing the potential of generating very high energy ($\sim$PeV) cosmic rays (CRs), with potential multi-messenger observational signatures.
Our main findings are listed as follows.

\begin{itemize}
\item The MRI turbulence is strongly anisotropic, with toroidal field dominating energy fluctuations. Energy cascade approximately agree with, but show some deviations from critical balance for the range of wave numbers that we probed.

\item Particle diffusion in configuration space is highly anisotropic, with diffusion being most rapid in the azimuthal ($y$) direction. For gyro-radii $R_g\gtrsim0.03H$, diffusion coefficients $D_{{\rm conf},y}\sim30D_{\rm Bohm}$, and $D_{\rm conf,pol}\sim5D_{\rm Bohm}$. At smaller radii, particles approach constant mean free path, which is longer than this scaling.

\item Particle diffusion in momentum space at gyro radii $R_g\gtrsim0.03H$ show $D_\text{Fermi}(p)\propto p$ in the absence of shear, indicating a characteristic acceleration rate $D_\text{Fermi}/p^2$ scale as gyro-frequency. 

\item Particles with smaller gyro radii likely transition towards $D_\text{Fermi}(p)\propto p^2$ scaling, indicating exponential acceleration. Multiple mechanisms (including compressibility) contribute but the process is likely affected by intermittency in the MRI turbulence, and demands further studies.

\item Particles also experience shear acceleration at a rate in agreement with conventional theories. However, it overtakes second order Fermi acceleration only for sufficiently energetic particles whose gyro radii exceed $\sim10\%$ of disk scale height. Particles generally escape from the disk before reaching such energies.

\item Whether strong particle acceleration can be achieved in hot accretion flows crucially depends on the acceleration vs. escape timescales for particles with $R_g\lesssim0.01H$, and favorable conditions may only be marginally realized near the ISCO. In that case, SMBHs with sub-Eddington accretion rates satisfying $(\dot{M}/M_{\rm Edd})(M/M_{\rm SgrA*})\gtrsim10^{-4}$ are likely capable of accelerate particles to PeV energies.

\end{itemize}

The gyro-radii of particles employed in our simulations (as well as other similar works) are larger but close to the gyro-radii of the particles with maximum energy one expects to achieve in ADAF disks. By further increasing the resolution of the MRI simulations by a factor of 4-8, one would be able to comfortably resolve such radii.
Moreover, our measured acceleration and diffusion coefficients bear large uncertainties for particles with small $R_g$ towards grid scale, which we suspect is related to turbulent intermittency associated with numerical dissipation. It is thus desirable to introduce physical dissipation to better control potential numerical artifacts. These considerations point to a dedicated study of particle acceleration from the MRI simulations with even higher resolution and explicit dissipation.

Besides, our studies are idealized in several aspects. The MRI turbulence in (collissionless) ADAF disks tend to develop modest level of pressure anisotropy (e.g., \citealp{2007ApJ...671.1696S}), and it is yet to see how it impacts particle acceleration and transport. Other than the acceleration processes studied in this work, the final CR spectrum and the particularly the flux of CRs from the disk, are further determined by the injection and escape processes. Further kinetic studies \citep{2012ApJ...755...50R,2016PhRvL.117w5101K} are needed to calibrate the outcome of particle injection, while global simulations, such as \citet{2019MNRAS.485..163K} but with higher resolution, are needed to quantify the rate of particle escape. In addition, as particle acceleration is the most effective in the vicinity of the SMBH, general-relativistic (GR) effects become increasingly important, and it is desirable to combine GRMHD simulations (e.g., \citealp{2016ApJS..225...22W}) with GR particle integrators \citep{2019ApJS..240...40B}. Finally, our simulations adopt the MRI snapshots with test particles. Future extensions may co-evolve particles with the MHD fluid under the MHD-PIC framework \citep{2015ApJ...809...55B}. 
It will enable us to further address the influence of temporal fluctuations in turbulence that likely enhances diffusive particle acceleration through resonance broadening and TTD (e.g., \citealp{2020PhRvD.102b3003D}). It will also allow us
to incorporate particle feedback and hence more self-consistently study interplay between thermal and non-thermal plasmas in ADAFs.

\section*{Acknowledgements}

We thank Lev Arzamasskiy, Luca Comisso, Shigeo Kimura, Vahe Petrosian, Brian Reville, Dmitri Uzdensky,
and Mingrui Zhang for useful discussions, particularly during the KITP program on ``Multiscale phenomenon in plasma astrophysics", as well as the anonymous referee for a thorough and constructive report. This research is supported by the National Science Foundation of China under grant No. 11873033, and in part by the National Science Foundation under Grant No. NSF PHY-1748958.

\section*{Data Availability}
 

The data underlying this article will be shared on reasonable request to the corresponding author.


\input{mri_acc.bbl}
\bibliographystyle{mnras}





\bsp	
\label{lastpage}
\end{document}